\documentclass[prb,twocolumn,showpacs,preprintnumbers,amsmath,aps]{revtex4}
\usepackage{graphicx,hyperref}
\usepackage{xcolor}
\usepackage{color}
\usepackage{bm}
\usepackage{subfigure}
\usepackage[utf8]{inputenc}
\usepackage[normalem]{ulem}

\newcommand\vp{\varphi}

\newcommand\DR{\rm{DR}}

\def\nbC{{\mathchoice {\setbox0=\hbox{$\displaystyle\rm C$}%
\hbox{\hbox to0pt{\kern0.4\wd0\vrule height0.9\ht0\hss}\box0}}
{\setbox0=\hbox{$\textstyle\rm C$}\hbox{\hbox
to0pt{\kern0.4\wd0\vrule height0.9\ht0\hss}\box0}}
{\setbox0=\hbox{$\scriptstyle\rm C$}\hbox{\hbox
to0pt{\kern0.4\wd0\vrule height0.9\ht0\hss}\box0}}
{\setbox0=\hbox{$\scriptscriptstyle\rm C$}\hbox{\hbox
to0pt{\kern0.4\wd0\vrule height0.9\ht0\hss}\box0}}}}
\def\nbQ{{\mathchoice {\setbox0=\hbox{$\displaystyle\rm
Q$}\hbox{\raise
0.15\ht0\hbox to0pt{\kern0.4\wd0\vrule height0.8\ht0\hss}\box0}}
{\setbox0=\hbox{$\textstyle\rm Q$}\hbox{\raise
0.15\ht0\hbox to0pt{\kern0.4\wd0\vrule height0.8\ht0\hss}\box0}}
{\setbox0=\hbox{$\scriptstyle\rm Q$}\hbox{\raise
0.15\ht0\hbox to0pt{\kern0.4\wd0\vrule height0.7\ht0\hss}\box0}}
{\setbox0=\hbox{$\scriptscriptstyle\rm Q$}\hbox{\raise
0.15\ht0\hbox to0pt{\kern0.4\wd0\vrule height0.7\ht0\hss}\box0}}}}
\def\nbT{{\mathchoice {\setbox0=\hbox{$\displaystyle\rm
T$}\hbox{\hbox to0pt{\kern0.3\wd0\vrule height0.9\ht0\hss}\box0}}
{\setbox0=\hbox{$\textstyle\rm T$}\hbox{\hbox
to0pt{\kern0.3\wd0\vrule height0.9\ht0\hss}\box0}}
{\setbox0=\hbox{$\scriptstyle\rm T$}\hbox{\hbox
to0pt{\kern0.3\wd0\vrule height0.9\ht0\hss}\box0}}
{\setbox0=\hbox{$\scriptscriptstyle\rm T$}\hbox{\hbox
to0pt{\kern0.3\wd0\vrule height0.9\ht0\hss}\box0}}}}
\def\nbS{{\mathchoice
{\setbox0=\hbox{$\displaystyle     \rm S$}\hbox{\raise0.5\ht0%
\hbox to0pt{\kern0.35\wd0\vrule height0.45\ht0\hss}\hbox
to0pt{\kern0.55\wd0\vrule height0.5\ht0\hss}\box0}}
{\setbox0=\hbox{$\textstyle        \rm S$}\hbox{\raise0.5\ht0%
\hbox to0pt{\kern0.35\wd0\vrule height0.45\ht0\hss}\hbox
to0pt{\kern0.55\wd0\vrule height0.5\ht0\hss}\box0}}
{\setbox0=\hbox{$\scriptstyle      \rm S$}\hbox{\raise0.5\ht0%
\hboxto0pt{\kern0.35\wd0\vrule height0.45\ht0\hss}\raise0.05\ht0%
\hbox to0pt{\kern0.5\wd0\vrule height0.45\ht0\hss}\box0}}
{\setbox0=\hbox{$\scriptscriptstyle\rm S$}\hbox{\raise0.5\ht0%
\hboxto0pt{\kern0.4\wd0\vrule height0.45\ht0\hss}\raise0.05\ht0%
\hbox to0pt{\kern0.55\wd0\vrule height0.45\ht0\hss}\box0}}}}
\def\nbZ{{\mathchoice {\hbox{$\sf\textstyle Z\kern-0.4em Z$}}
{\hbox{$\sf\textstyle Z\kern-0.4em Z$}}
{\hbox{$\sf\scriptstyle Z\kern-0.3em Z$}}
{\hbox{$\sf\scriptscriptstyle Z\kern-0.2em Z$}}}}

\begin{document}

\title{On the breakdown of dimensional reduction and supersymmetry in random-field models}

\author{Gilles Tarjus} \email{tarjus@lptmc.jussieu.fr}
\affiliation{LPTMC, CNRS-UMR 7600, Sorbonne Universit\'e,
4 Pl. Jussieu, 75252 Paris cedex 05, France}

\author{Matthieu Tissier} \email{tissier@lptmc.jussieu.fr}
\affiliation{LPTMC, CNRS-UMR 7600, Sorbonne Universit\'e,
4 Pl. Jussieu, 75252 Paris cedex 05, France}

\author{Ivan Balog} \email{balog@ifs.hr}
\affiliation{Institute of Physics, P.O.Box 304, Bijeni\v{c}ka cesta 46, HR-10001 Zagreb, Croatia}

\date{\today}

\begin{abstract}
We discuss the breakdown of the Parisi-Sourlas supersymmetry (SUSY) and of the dimensional-reduction (DR) property in the random field Ising and 
O($N$) models as a function of space dimension $d$ and/or number of components $N$. The functional renormalization group (FRG) predicts that this 
takes place below a critical line $d_{\rm DR}(N)$. We revisit the perturbative FRG results for the RFO($N$)M in $d=4+\epsilon$ and carry 
out a more comprehensive investigation of the nonperturbative FRG approximation for the RFIM. In light of this FRG description, we discuss the 
perturbative results in $\epsilon=6-d$ recently derived for the RFIM by Kaviraj, Rychkov, and Trevisani.\cite{rychkov_RFIM-II,rychkov_lectures} 
We stress in particular that the disappearance of the SUSY/DR fixed point below $d_{\rm DR}$ arises as a consequence of the nonlinearity of the FRG equations and cannot be found via the perturbative expansion in $\epsilon=6-d$ (nor in $1/N$). We also provide an error bar on the value of the critical dimension $d_{\rm DR}$ for the RFIM, which we find around $5.11\pm0.09$, by studying several successive orders of the nonperturbative FRG approximation scheme.
\end{abstract}


\maketitle

\section{Introduction}

Although having been introduced some 50 years ago\cite{imry-ma75,aharony76,grinstein76,young77}, the random-field Ising model (RFIM) and its 
extension, the random-field O($N$) model (RFO($N$)M), keep stimulating an ongoing research activity and lively debates (for recent papers, see 
[\onlinecite{fytas19,hikami19,angelini20,balog20,rychkov_RFIM-I,tarjus-review,rychkov_RFIM-II,rychkov_lectures,fytas23,cardy23}]). Besides their 
relevance as effective theories for a wide range of situations,\cite{tarjus-review} one of the recurring (and fascinating) questions about the theoretical 
description of random-field systems is the nature of the mechanism by which dimensional reduction and the underlying supersymmetry\cite{parisi79} 
are broken as one lowers the dimension below the upper critical dimension $d=6$. Dimensional reduction (DR) is the property that the critical behavior 
of the RFIM with the disorder strength as control parameter is the same as that of the pure Ising model with temperature as control parameter in 
two dimensions less. It is found at all orders of perturbation theory.\cite{aharony76,grinstein76,young77} The Parisi-Sourlas supersymmetry (SUSY) 
is an emergent symmetry associated with the zero-temperature properties of the model\cite{parisi79} and it entails DR, even at a nonperturbative 
level.\cite{cardy_SUSY,SUSY_klein,tissier12a,tissier12b,rychkov_RFIM-I} A similar behavior is found in the RFO($N$)M.\cite{fisher85} As it was 
proven early on through heuristic\cite{imry-ma75} and rigorous\cite{imbrie84,bricmont87,aizenman89} arguments that DR does not hold in dimensions 
$d=2$ and $d=3$, the search for the process leading to the breakdown of DR and SUSY has enjoyed a continuing interest over the last decades.

In a series of papers since 2004 we have proposed a consistent theoretical explanation of DR and SUSY breakdown as one lowers the 
dimension in the RFIM and RFO($N$)M through a functional renormalization group (FRG) treatment: for a review, see [\onlinecite{tarjus-review}]. The 
breakdown is associated with the appearance of a nonanalytic dependence on the order-parameter fields (a ``cusp'', to be described in detail later on) 
in the cumulants of the renormalized disorder and in the correlation functions at the zero-temperature fixed point that controls the critical behavior of the 
model.\cite{tarjus04,tissier06,tissier11,tissier12b,FPbalog} This cusp in the functional form of the cumulants of the renormalized random field is the 
consequence of the presence of scale-free avalanches in the ground state of the model under infinitesimal changes of an applied source at 
criticality,\cite{tarjus13} avalanches that are indeed observed in computer simulations of the RFIM at zero temperature.\cite{vives_GS,wu-machta_GS,liu_GS} 
The connection between cusp in the functional dependence of the cumulants, avalanches, and breakdown of DR has also been amply demonstrated 
in another disordered model which  describes an elastic manifold pinned in a random 
medium.\cite{fisher86b,BBM,FRGledoussal-chauve,middleton07,doussal-wiese_aval,doussal_annals,wiese_review} The functional character of the 
RG is central in such problems.

There are two different patterns of SUSY and DR breaking in the RFO($N$)M depending on the values of $N$ and $d$. Near the lower critical dimension 
of the RFO($N>2$)M, in $d=4+\epsilon$, the perturbative FRG to 2 loops predicts that the SUSY/DR fixed point which controls the critical point at large $N$ 
first becomes unstable in $N=18.3923\cdots$ when a ``cuspy'' fixed point becomes stable and then vanishes in $N=18$ when it collapses with a SUSY/DR 
unstable  fixed point.\cite{FPbalog} (``Cuspy'' refers to the fact that the cumulants of the renormalized random field have a nonanalytical functional 
dependence in theform of a cusp: note that the functional character of the RG is crucial to capture this effect, even if the FRG is perturbative in $\epsilon=d-4$ 
here.) On the other hand, for the Ising version, $N=1$, the nonperturbative FRG predicts that the SUSY/DR critical fixed point which is present at 
the upper critical dimension $d=6$ disappears around $d_{\rm DR}\approx 5.1$, even before becoming unstable to a cuspy perturbation associated with 
avalanches. There is thus a critical value along the line $d_{\rm DR}(N)$ at which the SUSY/DR fixed point disappears that separates the two different 
patterns and that we have estimated around $N_x\approx 14$ and $d_{\rm DR}(N_x)\approx 4.4$.\cite{FPbalog}

The FRG prediction for the breakdown of SUSY and DR in $d_{\rm DR}\approx 5.1$ for the RFIM  is supported by both 
nonperturbative\cite{tarjus04,tissier06,tissier11,tissier12b,balog20} and $\epsilon=6-d$ perturbative calculations,\cite{tissier_pertFRG} and the results 
are in good agreement with all simulation results in $d=3$, $4$, and $5$.\cite{middleton-fisher02,middleton_4d,fytas13,fytas16,fytas17,fytas18,fytas19}  
In particular it describes very well the DR broken results in $d=4$\cite{middleton_4d,fytas16} and the weak or negligible breaking of SUSY and DR in 
$d=5$.\cite{fytas17,fytas19} Furthermore it has a clear physical interpretation in terms of scale-free avalanches: They are present even in $d\geq d_{\rm DR}$ 
but then only have a subdominant effect at the fixed point (the resulting amplitude of the cusp which comes from the second moment of the avalanches 
is an irrelevant perturbation for $d\geq d_{\rm DR}$), while they dominate the critical behavior when $d<d_{\rm DR}$.\cite{tarjus13} The mechanism of 
the collapse of fixed points and the emergence of a new cuspy fixed point below $d_{\rm DR}$ is very unusual,\cite{FPbalog} which explains the 
corrections to scaling observed in $d=5$ in large-scale simulations.\cite{fytas19,balog20} Finally, the FRG prediction is also compatible with the loop 
expansion around the Bethe lattice\cite{angelini20} and approximate conformal-bootstrap results.\cite{hikami19}

In this paper we revisit this FRG description of SUSY and DR breakdown by first focusing on the recent work of Kaviraj, Rychkov and Trevisani, hereafter 
denoted KRT.\cite{rychkov_RFIM-II,rychkov_lectures} In the latter a perturbative (nonfunctional) RG investigation of the RFIM at 2 loops around the 
Gaussian fixed point in $d=6-\epsilon$ suggests that SUSY breaking operators destabilize the SUSY fixed point in $d\approx 4.2-4.6$. By building on 
our previous FRG analysis  of both the RFO($N$)M in $d=4+\epsilon$\cite{tissier06,tissier06b} and the RFIM in either $d=6-\epsilon$\cite{tissier06,tissier_pertFRG} or nonperturbatively in all dimensions,\cite{tissier06,tissier11,tissier12a,tissier12b,tarjus13,balog_activated,balog20}  
we show that the scaling dimension of these dangerous operators have a counterpart in the eigenvalues that have already been computed within 
the FRG around the SUSY/DR fixed point. Crucially, the functional and nonperturbative character of our approach allows us to address questions 
that cannot be directly answered through the perturbative treatment of KRT. In particular, we show that when the eigenvalue associated 
with the most dangerous (polynomial) operator destabilizing the SUSY/DR fixed point vanishes at a critical dimension $d_{\rm DR}$ (the operator then  becomes marginal), the SUSY/DR fixed point actually {\it disappears} instead of just becoming unstable as predicted in [\onlinecite{rychkov_RFIM-II}] 
and we discuss the mechanism by which this happens. As a result, even by fine-tuning additional control parameters such as the form of the 
bare random-field distribution there is no way to access any SUSY/DR critical point in dimensions below $d_{\rm DR}\approx 5.1$. In computer 
simulations of the RFIM in $d=4$ or $5$ one could therefore only at best observe remnants of SUSY/DR behavior over finite sizes, remnants that 
would disappear in the asymptotic critical regime if large enough system sizes are accessible.

Finally, we  check the robustness of our theoretical prediction that the SUSY/DR fixed point disappears in $d_{\rm DR}\approx 5.1$ for the RFIM. For this 
we study different orders of the nonperturbative approximation scheme that we have previously introduced within the FRG formalism. By considering 
both cruder and improved levels of approximation compared to our previous work,\cite{tissier11,tissier12b,FPbalog,balog20} we find that 
$d_{\rm DR}\approx 5.11\pm 0.09$, thereby providing an error bar for our result.

The paper is organized as follows. In Sec.~\ref{sec_summary} we summarize the recent work of KRT.\cite{rychkov_RFIM-II,rychkov_lectures} 
We next give the key results of our work and put the contribution of Refs.~[\onlinecite{rychkov_RFIM-II,rychkov_lectures}] in the framework of our 
FRG approach.Then, in Sec.~\ref{sec_RFON} we make a detour via the RFO($N>2$)M by reanalyzing its perturbative but functional RG description 
at one- and two-loop orders near the lower critical dimension $d=4$. Here, the study is performed as a function of the number of field components 
$N$ instead of the spatial dimension $d$. We present analytical and numerical results unambiguously showing that when the most dangerous operator 
for destabilizing the SUSY/DR fixed point which is found for large values of $N$ becomes marginal the SUSY/DR fixed point disappears at once by 
collapsing with another (completely unstable) SUSY/DR fixed point. We next come back to the case of the RFIM in Sec.~\ref{sec_back}. We discuss 
the nature and the scaling dimension of the operators that are potentially dangerous for destabilizing the SUSY/DR fixed point as the dimension $d$ 
is decreased. We do so  in the various parametrizations of the replica fields. We also spell out within the nonperturbative FRG the mechanism by which 
the SUSY/DR fixed point disappears in the critical dimension $d_{\rm DR}$ where the most dangerous operator becomes marginal. In 
Sec.~\ref{sec_NPFRG} we check the robustness of the nonperturbative FRG predictions by implementing several successive orders of the 
nonperturbative approximation scheme. The outcome is an apparent rapid convergence for the value of the critical dimension $d_{\rm DR}$. Finally, 
we present some concluding remarks in Sec.~\ref{sec_conclusion} about the physical implications of our findings. In addition, we provide several 
appendices to discuss more technical aspects of our investigation and to further address some of the comments made by 
KRT\cite{rychkov_RFIM-II,rychkov_lectures} on our FRG approach.
\\

\section{The recent work of KRT\cite{rychkov_RFIM-II,rychkov_lectures} in light of the FRG approach}
\label{sec_summary}

\subsection{Summary of the work in [\onlinecite{rychkov_RFIM-II,rychkov_lectures}]}

KRT\cite{rychkov_RFIM-II,rychkov_lectures} start from the replica field-theoretical description of the RFIM in which one considers a bare 
action for replica scalar fields $\phi_a$ with $a=1,\cdots,n$ of the form\cite{cardy_textbook}
\begin{equation}
\begin{aligned}
\label{eq_bare-action}
S[\{\phi_a\}]=& \sum_a\int_{x}\bigg\{\frac{1}{2}[\partial \phi_a(x)]^2+ \frac{r}{2} \phi_a(x)^2 + \frac{u}{4!} \phi_a(x)^4 \bigg\}\\&
- \frac{\Delta}{2}\sum_{a,b}\int_x \phi_a(x)\phi_b(x),
\end{aligned}
\end{equation}
which is obtained after having introduced $n$ replicas of the system and averaged over a Gaussian (bare) random field of zero mean and 
variance $\Delta$. (Here, $ \int_{x} \equiv \int d^d x$.) It then makes use of Cardy's linear transformation of the replica fields.\cite{cardy_transform} 
In $d=6$, when dropping from the action terms that are irrelevant by simple power counting and terms that vanish when the number $n$ of 
replicas go to zero, one ends up with a theory that reproduces the main features of the Parisi-Sourlas SUSY action.\cite{parisi79}
Note that once transformed the fields have different scaling dimensions: $\phi=(1/2)[\phi_1+(\phi_2+\cdots+\phi_n)/(n-1)]$, which is essentially the 
physical (order parameter) field, has a canonical dimension $D_\phi=(d-4)/2$, $\hat\phi=(1/2)[\phi_1-(\phi_2+\cdots+\phi_n)/(n-1)]$, which plays a 
role similar to the ``response'' field,  has a canonical dimension $D_{\hat\phi}=d/2$, and the $(n-2)$ independent ``antisymmetric'' field combinations 
$\chi_i$, which somehow mimic the two anticommuting Grassmannian ghost fields $\bar\psi$ and $\psi$ of the SUSY formalism, have a canonical 
dimension $D_\chi=(d-2)/2$. (Cardy's formalism does not explicitly involves a renormalized temperature, whose dimension would be $-\theta=-2$ in 
the SUSY/DR regime, but the above field dimensions are identical to those obtained introducing a scaling in $1/T$ for $\phi$ and $1/\sqrt{T}$ for 
the $\chi_i$'s.)

The idea followed by KRT\cite{rychkov_RFIM-II} is to study the scaling dimensions of the operators that have not been considered by 
Cardy\cite{cardy_transform} and investigate through a perturbative RG calculation at 2 loops in $\epsilon=6-d$ if these operators can become relevant 
for some value of $\epsilon$. The operators are classified into SUSY-writable, SUSY-null, and SUSY-non-writable, only the last two ones being potentially 
dangerous to destabilize the SUSY fixed point. There are several subtleties that are carefully  handled by KRT. First, Cardy's transformation obscures the 
replica permutational symmetry $S_n$ of the action and one must make sure that only singlets under $S_n$ are retained. Second, these singlets are not 
eigen-operators of scale transformations and must be decomposed into terms of increasing scaling dimensions, the dominant one (of lowest dimension)  
being called the ``leader''.

KRT then find that two (leader) operators, denoted by $(\mathcal F_4)_L$ and  $(\mathcal F_6)_L$, are more dangerous: if extrapolated, their scaling 
dimensions at 2 loops  become relevant and destabilize the SUSY fixed point, in $d\approx 4.6$ for the former and in $d\approx 4.2$ for the 
latter.\cite{rychkov_RFIM-II,rychkov_lectures} There are other operators which appear at first sight even more dangerous because the extrapolated 
$2$-loop expression of their scaling dimension crosses $d$ at a higher spatial dimension than those for $(\mathcal F_4)_L$ and  $(\mathcal F_6)_L$. 
However, the authors argue that this does not take place due to nonperturbative mixing effects that should repel the scaling dimensions of operators 
belonging to the same symmetry class when they approach each other. The claim is that $(\mathcal F_4)_L$ and  $(\mathcal F_6)_L$ are protected 
from this effect but not the others which are then conjectured to never become dangerous.

By construction, the operators under consideration are analytical functions of the replica fields (polynomials). The most dangerous ones have already 
been studied and discussed, albeit in a different framework, by Feldman\cite{feldman02} and later by two of us\cite{tissier06} for both the RFO($N$)M and 
the RFIM. As will be further developed, they are also accounted for in the nonperturbative FRG.\cite{tarjus04,tissier06,tissier11,tissier12a,tissier12b, balog20} 
These operators are 2-replica functions of the form
\begin{equation}
\label{eq_Feldman_op}
\mathcal F_{2p}=\sum_{a,b}[(\phi_a-\phi_b)^2]^p.
\end{equation}
(They are denoted by $A_{p}$ in Ref.~[\onlinecite{feldman02}].) In the RFO($N>2$)M near the lower critical dimension of ferromagnetism, $d=4$, the 
long-distance physics can be described by a nonlinear sigma model and the above operators can be rewritten as 
$\sum_{a,b}[1-\phi_a\cdot\phi_b/(\vert\phi_a\vert\vert\phi_b\vert)]^p$ which can then be represented as linear combinations of random 
anisotropies.\cite{feldman02} 

In the RFIM close to its upper critical dimension $d_{\rm uc}=6$, one finds that these operators have a scaling dimension strictly larger than $d$ at tree 
level but acquire a negative anomalous dimension proportional to $p^2\epsilon^2$ at two loops, so that extrapolations may lead to an intriguing outcome: 
either, as proposed by Feldman,\cite{feldman02} one takes the limit of large $p$ at any fixed small $\epsilon$ and concludes that the operators always 
destabilize the SUSY/DR fixed point or, as done by KRT,\cite{rychkov_RFIM-II} one considers the situation at a fixed $p$ and extrapolate at large 
$\epsilon$ to predict that the operators become relevant at some low enough specific dimension. The two scenarios lead to different pictures of SUSY/DR 
breaking. Our nonperturbative FRG yields yet another picture which dismisses the former scenario and, while having some overlap with 
the latter, also shows some key differences.
 
\subsection{Putting the above results in the framework of our FRG approach}
\label{sub_putting}

\begin{figure}
\includegraphics[width=\linewidth]{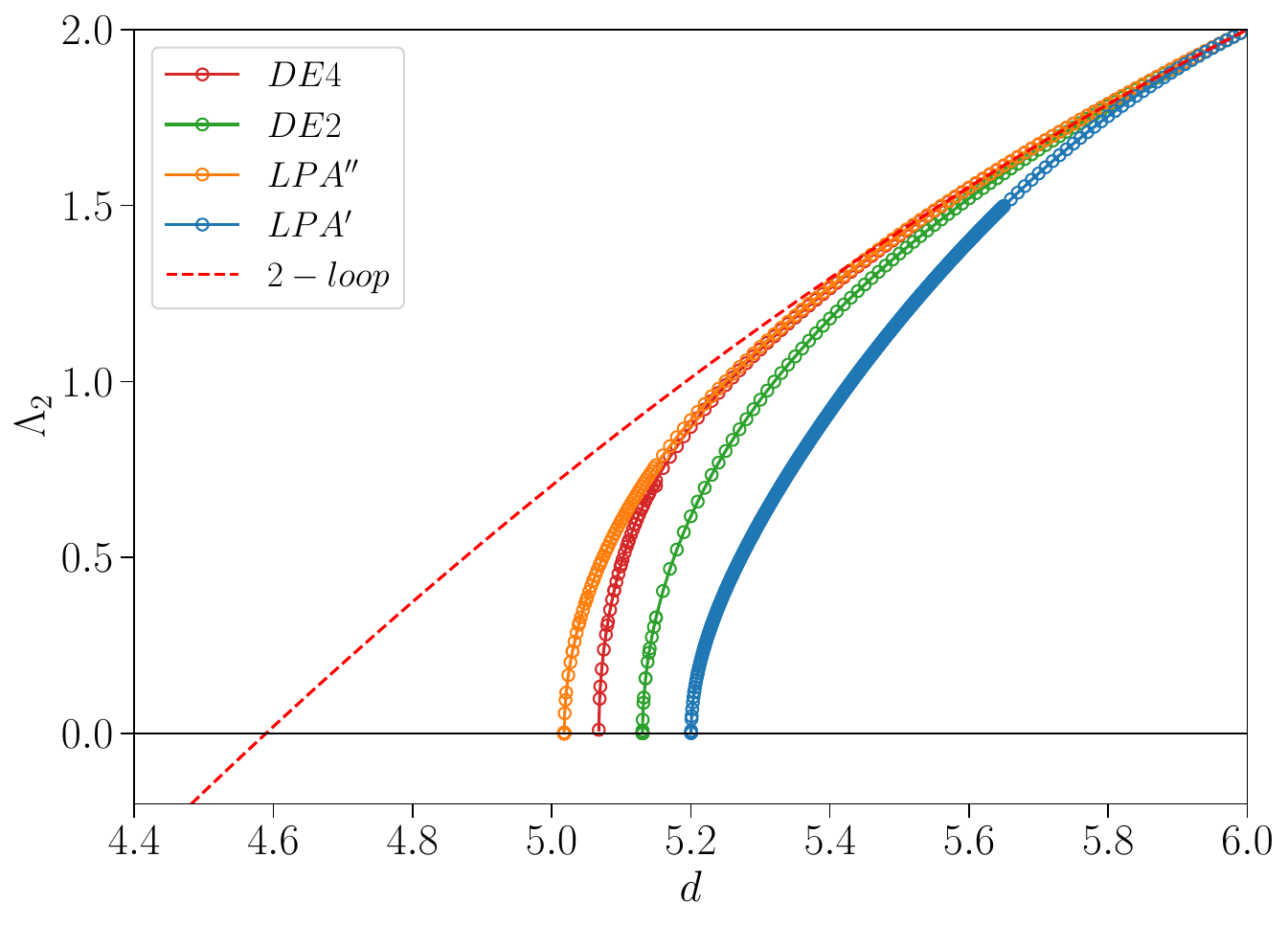}
\caption{Eigenvalue  $\Lambda_2(d)$ corresponding to the most dangerous analytic perturbations around the SUSY/DR fixed point in the RFIM 
and associated with Feldman's operator $\mathcal F_4$. Dashed line: $2$-loop calculation in $d=6-\epsilon$, together with a plausible extrapolation 
(the result coincide with that of KRT\cite{rychkov_RFIM-II,rychkov_lectures}); full lines:  Results of successive levels of the nonperturbative approximation 
scheme of the FRG (LPA', LPA'', DE2, and DE4), which are discussed in Secs.~\ref{sub_RFIM_DE2}, \ref{sec_NPFRG} and Appendix~\ref{app_flow-eqs} 
(full lines). Below $d_{\rm DR}\approx 5.11\pm0.09$ at which $\Lambda_2=0$ the  SUSY/DR fixed point disappears and gives way to a cuspy fixed point 
at which both SUSY and DR are broken. Note that this disappearance which is associated with a square-root singularity in $\Lambda_2(d)$ is out of 
reach of the extrapolated perturbative expansion in $\epsilon$ which only suggests that the eigenvalue becomes negative (relevant) below $d\approx 4.6$.
}
\label{fig_lambda_RFIM}
\end{figure}

The FRG allows one to derive exact flow equations for the cumulants of the effective average action, or scale-dependent Gibbs free-energy functional. 
Because the effective action is the generating functional of all 1-particle irreducible  (1-PI) correlation functions\cite{zinnjustin89}, we will generically call 
the associated cumulants ``1-PI cumulants''. For practical purposes the exact FRG equations can be truncated in a nonperturbative approximation scheme. 
In the case of the RFIM this relies on the combined truncation of the expansions of the effective action in number of field derivatives and order of the 
cumulants. (A perturbative approximation scheme can of course also be used through an expansion in the $\phi^4$ coupling constant of the first cumulant, 
which is marginal at the upper dimension $d=6$, and a subsequent expansion in $\epsilon=6-d$.) Contrary to the perturbative treatment in the coupling 
constant, the nonperturbative scheme provides an account of the functional dependence of the 1-PI cumulants in their field arguments. The truncation is 
chosen such that it does not explicitly break the symmetries and the Parisi-Sourlas SUSY of the theory. Furthermore, the 1-loop perturbative results are 
recovered in the vicinity of $d=6$ (and, for the RFO($N>2$)M, in the vicinity of the lower critical dimension of ferromagnetism, $d=4$).

It should be stressed that all of the operators considered in the approach of KRT [\onlinecite{rychkov_RFIM-II,rychkov_lectures}] have a counterpart in the FRG description, whether the latter makes use of conventional fields,\cite{tarjus04,tissier06} superfields,\cite{tissier11,tissier12a,tissier12b} or is derived within the dynamical formalism.\cite{balog_activated,balog_dynamics} A difference that is worth mentioning is that our FRG formalism 
deals with 1-PI quantities present in the effective action, hence with somehow averaged operators which are functions of the average (replica) fields. On 
the other hand the operators studied by KRT are present in the action and involve the fluctuating (replica) fields. In the FRG the ``averaged'' operators are associated with a Taylor expansion of the functional dependence of the 1-PI cumulants of the renormalized random field.  This can be directly seen for 
Feldman's operators which appear in the expansion of the second 1-PI cumulant  as polynomials in the difference between the two field arguments, 
$(\phi_a-\phi_b)$ [see Eq.~(\ref{eq_Feldman_op})]. (For simplicity we will keep referring to such functions of the average fields appearing in the 1-PI 
cumulants as ``operators'' but one should keep in mind that they are not fluctuating quantities.) Most importantly, $\mathcal F_4$, which corresponds 
to the most dangerous perturbation that may destabilize the SUSY/DR fixed point and which is a SUSY-null operator in the formalism of Refs.~[\onlinecite{rychkov_RFIM-II,rychkov_lectures}]  already played a key role in our FRG treatment because it signals when a cuspless, hence 
SUSY/DR, fixed point can no longer exist. We called the critical spatial dimension at which this happens $d_{\rm DR}$ and found it to be about $5.1$. 
A check of the robustness of the prediction with an estimate of the error bar is provided below in Sec.~\ref{sec_NPFRG}.

A crucial point is that the nonperturbative FRG  is able to show the disappearance of the SUSY/DR fixed point altogether when the most dangerous 
analytic perturbation associated with Feldman's operator $\mathcal F_4$ becomes marginal, {\it i.e.}, when $\Lambda_2=0$. The reason 
is that the nonperturbative FRG in fixed dimension $d<6$ provides a full characterization of the effective action at the SUSY/DR fixed point and of 
the spectrum of eigenvalues (or equivalently of scaling dimensions) around this fixed point.\cite{footnote_SUSY_FP} 
Whereas the latter is determined from the linearization 
of the RG flow equations, the former is obtained via the resolution of fixed-point equations that may be {\it nonlinear} in some coupling constants 
(or rather functions). This should be contrasted with the conventional perturbative RG in which the only nonlinearity concerns the coupling constant 
that is marginal in $d=6$.\cite{footnote_marginal} Then, both the eigenvalues and the characteristics of the fixed-point effective action are derived as 
expansions in powers of this coupling constant (eventually turned into an expansion in $\epsilon=6-d$). If the fixed point disappears in a given dimension 
$d_{\rm DR}<6$ because the (nonlinear) equation describing a specific coupling constant/function  of the fixed-point effective action, in the present 
case that associated with the operator $\mathcal F_4$ which is irrelevant in $d=6$, has no more solution due to the collapse with another fixed point, 
the expansion in $\epsilon$ used in [\onlinecite{rychkov_RFIM-II,rychkov_lectures}] cannot {\it per se} capture this phenomenon. 
We will discuss additional symmetry arguments further below.

We illustrate the outcome of the two frameworks, nonperturbative FRG and conventional perturbative RG, for the eigenvalue $\Lambda_2(d)$ 
associated with the most dangerous perturbation for the SUSY/DR fixed point in Fig.~\ref{fig_lambda_RFIM}. The perturbative calculation of 
KRT up to 2-loop order, which in the present case is also reproduced within the FRG (see below) and was already obtained by Feldman,\cite{feldman02} 
predicts a curve as a function of $\epsilon$ or $d$ that when extrapolated to lower dimension passes through $0$  in $d\approx 4.6$ and then 
becomes negative. On the other hand, the nonperturbative FRG result coincides with the pertubative curve near $d=6$ but strongly deviates from 
it as $d$ decreases and go to $0$ in  $d_{\rm DR}=5.11\pm0.09$ (depending on the level of the nonperturbative approximation scheme: see 
Sec.~\ref{sec_NPFRG}) with a singular square-root behavior.  Below $d_{\rm DR}$ the SUSY/DR fixed point {\it no longer exists}. The extrapolation of 
the perturbative result is of course blind to this feature. As we will also show in the next section, a similar phenomenon takes place in the RFO($N>2$)M 
where an expansion in $1/N$ is structurally unable to detect the disappearance of the SUSY/DR fixed point.

As already pointed out in the Introduction, the FRG not only predicts the disappearance of the SUSY/DR fixed point in $d_{\rm DR}$, but it also 
describes the appearance and the properties of the new fixed point below $d_{\rm DR}$ at which both SUSY and DR are broken. And, much like in 
the case of elastic manifolds in a random environment, it relates this emergence to the appearance of a linear cusp in the functional dependence of 
the 1-PI cumulants of the renormalized random field at the zero-temperature fixed point. This is in turn associated with the fact that the long-distance 
physics at criticality is dominated by avalanches. As also already mentioned but worth stressing again, scale-free avalanches are present in all 
dimensions at criticality. Their effect is subdominant (the associated ``cuspy'' perturbation is irrelevant at the SUSY/DR fixed point) when 
$d\geq d_{\rm DR}$ while they control the long-distance physics when $d< d_{\rm DR}$. However, the disappearance of the SUSY/DR fixed point in 
$d_{\rm DR}$ is not due to the avalanches and the cuspy perturbation {\it per se}. The latter is indeed still irrelevant in $d=d_{\rm DR}$,\cite{FPbalog,balog20} and there is a discontinuity in the associated eigenvalue because the nature of the fixed points (characterized by cuspless 
or cuspy 1-PI cumulants of the renormalized random field) is different above and below $d=d_{\rm DR}$. This is illustrated below in the inset of Fig.~\ref{fig_lambdas_RFIM}. 

To pinpoint $d=d_{\rm DR}$ more accurately, and we agree on this conclusion with KRT, one should instead focus on Feldman's operator 
$\mathcal F_4$ and its scaling dimension (or, equivalently, the eigenvalue $\Lambda_2$). Yet, one should also consider the associated coupling 
constant/function at the fixed point. The critical dimension $d_{\rm DR}$ can then be located in two ways: either looking at the appearance of a cusp 
in the fixed-point function (and a resulting divergence in some properly chosen derivative) as a function of dimension, as we did in our first 
nonperturbative FRG investigations,\cite{tarjus04,tissier06,tissier11,tissier12b} or by studying the vanishing of the eigenvalue $\Lambda_2(d)$, as 
we did in [\onlinecite{FPbalog,balog20}] and in the present paper.

Finally, in their work\cite{rychkov_RFIM-II,rychkov_lectures} KRT also raise concerns about some aspects of our nonperturbative FRG approach. 
These concerns mostly stem from the unusual mechanism by which the SUSY/DR fixed point disappears in $d=d_{\rm DR}$ to give way to a cuspy, 
SUSY/DR broken fixed point and from the peculiarities of a zero-temperature critical fixed point. This will be addressed below.
\\

\section{A detour via the RFO($N$)M}
\label{sec_RFON}

\subsection{SUSY/DR fixed point, nonanalyticities, and dangerous operators}
\label{sub_RFON_FP}

It is instructive to first consider the critical behavior of the RFO($N$)M which also has an underlying SUSY and is naively described by 
DR.\cite{fisher85} (The RFO($N$)M corresponds to the same replica field-theoretical action as in Eq. (\ref{eq_bare-action}) with the fields now 
having $N$ components and the Lagrangian having an O($N$) instead of a $Z_2$ symmetry.) The RFO($N>2$)M  is more directly accessible 
to an analytical treatment when studied near its lower critical dimension $d=4$ where it can be described through a nonlinear sigma model. 
To recall, one starts from an action involving only the unit vectors $\mathbf{n}_a(x)$ describing the orientation of the 
$N$-component replica fields $\boldsymbol{\phi}_a(x)=\phi_0 \mathbf{n}_a(x)$ when their amplitude $\phi_0$ is frozen to a nonzero 
constant,\cite{fisher85,feldman02,tissier06,ledoussal06} 
\begin{equation}
\begin{aligned}
\label{eq_NLSigma_action}
S[\{\mathbf{n}_a\}]= &\int_x \bigg\{\frac{1}{2T}\sum_a \partial_\mu \mathbf{n}_a(x)\partial_\mu \mathbf{n}_a(x)- \\&
\frac 1{2T^2}\sum_{a,b} R(\mathbf{n}_a(x)\cdot\mathbf{n}_b(x)) +\cdots\bigg\}
\end{aligned}
\end{equation}
where the dependence on $\phi_0^2$ has been omitted (one can choose $\phi_0^2\equiv 1$) and the ellipses denote irrelevant terms near 
$d=4$. The function $R(z)$ is the variance of the disorder. At the bare level it is simply equal to $R_B(z)=\Delta z$, which corresponds to the 
random-field disorder in Eq.~(\ref{eq_bare-action}) with the dependence on temperature made explicit. The scalar product of different replica 
fields, $z=\mathbf{n}_a\cdot\mathbf{n}_b$, is dimensionless in $d=4$, so that the whole function $R(z)$ is marginal. One can then proceed 
to a perturbative computation of the effective action in powers of the disorder variance and perform an RG treatment in $d=4+\epsilon$. 
Note that this perturbative RG is {\it functional}: it deals with a function of the fields, the renormalized variance of the disorder $R(z)$ which also corresponds to the second (1-PI) cumulant of the effective action, in place of coupling constants.   
Results have be obtained both at 1-loop\cite{fisher85,feldman02,tissier06,tissier06b,FPbalog} and at 2-loop\cite{tissier06b,ledoussal06,sakamoto06} 
order, so that all statements can be directly proven analytically and rather easily checked numerically.

For the RFO($N$)M the study of Feldman's operators (``operator'' being used here and below in an abuse of language to denote polynomials 
of the average replica fields that are the 1-PI counterpart of fluctuating operators, see above) is equivalent to considering the eigenvalues 
associated with the derivatives of the variance of the renormalized disorder $R(z)$ when the angle between the 2 replica field arguments 
$\mathbf{n}_a$ and $\mathbf{n}_b$ goes to $0$ and its cosine, $z=\mathbf{n}_a\cdot \mathbf{n}_b$, goes to $1$: 
\begin{equation}
R(z)=R(1)+\sum_{p\geq 1} (-1)^p \frac{R^{(p)}(1)}{p!} (1-z)^p.
\end{equation}
One can associate the coupling constant $R^{(p)}(1)$ with Feldman's operator $\sum_{a,b}(1-z_{ab})^p$, which is indeed the analog of 
$\mathcal F_{2p}$ in Eq.~(\ref{eq_Feldman_op}) since $\vert\bm\phi_a-\bm\phi_b\vert^{2p}\propto (1-z_{ab})^p$. 

The eigenvalues corresponding to the $R^{(p)}(1)$'s around the SUSY/DR fixed point are given at one loop in $d=4+\epsilon$ by\cite{tissier06}
\begin{equation}
\begin{aligned}
\label{eq_eigenvalue_RFO(N)}
&\Lambda_{p\geq 2}(N)=\\&
-\epsilon\big [\frac{2p^2-(N-1)p+N-2}{N-2}+p(6p+N-5)R''_*(1)\big ]\\&
=-\frac{\epsilon}{N-2} \big [2p^2-(N-1)p+N-2 \,+\\&
\frac{p(6p+N-5)}{2(N+7)}(N-8-\sqrt{(N-2)(N-18)} \,)\big ],
\end{aligned}
\end{equation}
 where we have used that the fixed point itself is characterized by
\begin{equation}
\begin{aligned}
\label{eq_FP_RFO(N)}
&R'_*(1)=\frac{\epsilon}{(N-2)}, \\&
R''_*(1)=\epsilon\frac{N-8-\sqrt{(N-2)(N-18)}}{2(N-2)(N+7)}. 
\end{aligned}
\end{equation}
The above value of $R'_*(1)$, which determines the critical exponents $\eta$, $\bar\eta=\eta$, and $\nu$, together with the existence of a finite 
$R''_*(1)$ are enough to guarantee that the fixed point indeed satisfies SUSY and DR.\cite{fisher85,tissier06,tissier06b,FPbalog} Note that $R''_*(1)$ 
and $\Lambda_{p\geq 2}(N)$ are real for $N\geq 18$ only (we restrict ourselves to the case $N>2$). The 2-loop calculation leads to similar 
results,\cite{tissier06b} but for our illustration purpose it is sufficient to consider the 1-loop description. 

The eigenvalues are positive, {\it i.e.}, irrelevant, at large $N$ and become negative for some value 
as one decreases $N$. Loosely relating the role of $N$ here with that of $d$ in the RFIM, we observe that the value of $N$ at which the eigenvalue $\Lambda_{p}$ goes 
to zero increases with $p$, {\it i.e.}, Feldman's operators of high $p$ become relevant {\it before} those for small $p$ (with $p=2$ corresponding to 
$\mathcal F_4$), much like what is found for  the RFIM at 2 loops below $d=6$. 
However, without invoking a nonperturbative mechanism of level 
repulsion, one finds that the vanishing of an eigenvalue $\Lambda_{p\geq 3}$ around the SUSY/DR fixed point is cured by a change in the functional 
dependence of the fixed-point cumulant $R_*(z)$, which can then acquire  a nonanalytical dependence of the form $(1-z)^{1+\alpha}$, with 
$\alpha$ real and $>1$, in the vicinity of $z=1$.\cite{tissier06,tissier06b,FPbalog} In the restricted sector associated with $R'_*(1)$ and $R''_*(1)$ 
the fixed point still shows SUSY/DR. All of this is further discussed below. 

\begin{figure}
\includegraphics[width=.9\linewidth]{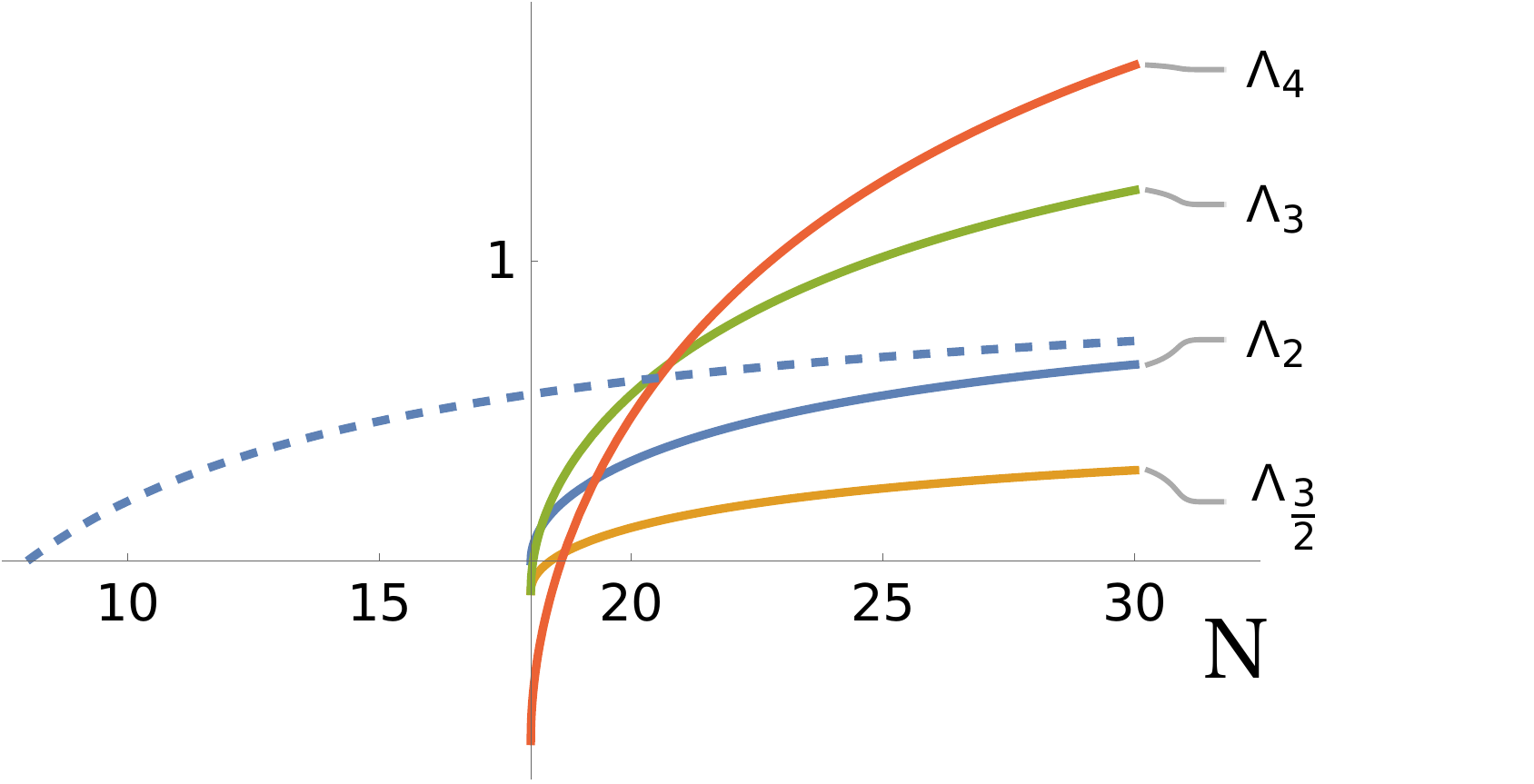}
\includegraphics[width=.9\linewidth]{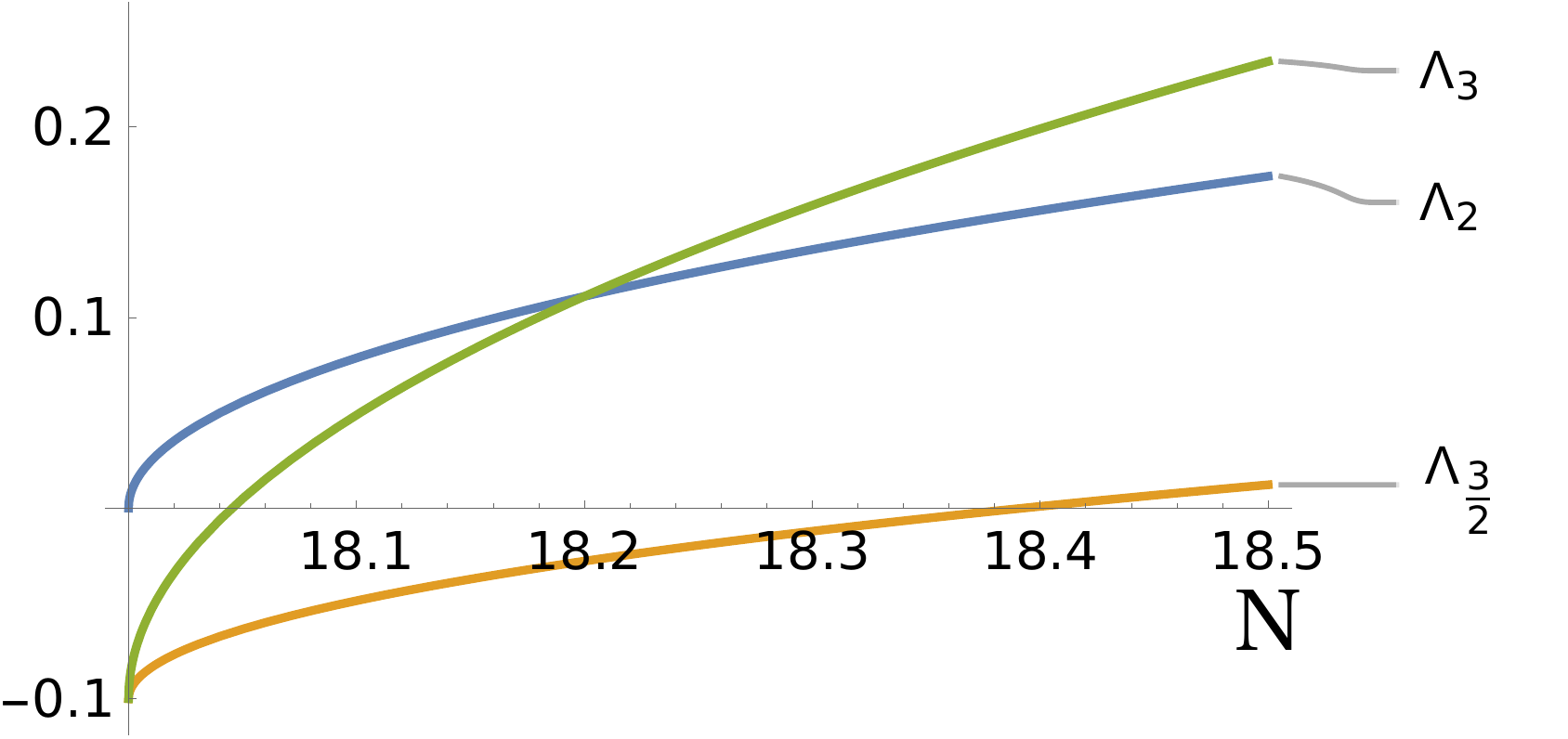}
\caption{Stability of the (most stable)  SUSY/DR fixed point in the RFO($N$)M at 1-loop in $d=4+\epsilon$: Variation with $N$ of the lowest-order 
eigenvalues $\Lambda_p(N)$; $p=2, 3$ and $4$ correspond to Feldman's operators $\mathcal F_4$, $\mathcal F_6$, and $\mathcal F_8$, whereas 
$p=3/2$ corresponds to a nonanalytical perturbation (a cusp in $R'(z)=\Delta(z)$). The eigenvalue $\Lambda_{3/2}$ becomes relevant below 
$N=2(4 + 3\sqrt 3)\approx 18.3923 \cdots$ and $\Lambda_2$ becomes zero in $N=18$. The vertical lines indicate the value $N=18$. 
Also shown is an extrapolation of the large $N$ expression (up to next-to-leading order in $1/N$). This extrapolation vanishes for $N\sim 8$ and is 
clearly blind to the disappearance of the SUSY/DR fixed point in $N=18$. The bottom panel is a zoom in displaying the region near $N=18$.
}
\label{fig_lambda_RFON}
\end{figure}

We also noticed that the expression in Eq.~(\ref{eq_eigenvalue_RFO(N)}) can be analytically continued to noninteger values of $p$, $p=1+\alpha$, 
and that it then controls the RG flow of the amplitude $a_k(\alpha)$ of a nonanalytic term in $(1-z)^{1+\alpha}$ in $R_k(z)$ in the vicinity of $z=1$: 
\begin{equation}
\label{eq_subcusp_amplitude}
\partial_t a_k(\alpha)=\Lambda_{1+\alpha} a_k(\alpha),
\end{equation}
where $t=\ln(k/k_{\rm UV})$ is the RG time with $k$ the running IR cutoff scale and $k_{\rm UV}$ the initial UV scale. The fixed point with 
$a_*(\alpha)=0$ is unstable to a nonanalytical pertubation behaving as $(1-z)^{1+\alpha}$ near $z=1$ when $\Lambda_{1+\alpha}<0$ and a new 
fixed point $R_*(z)$ with $a_*(\alpha)\neq 0$ emerges when $\Lambda_{1+\alpha}=0$. For a given $N$ this occurs for a specific value 
$\alpha_{\#}(N)$ such that $\Lambda_{1+\alpha_{\#}(N)}(N)=0$. (Note that due to the simple form of Eq.~(\ref{eq_subcusp_amplitude}), the value 
of $a_*(\alpha)$ when $\Lambda_{1+\alpha}=0$ is determined by requiring not only that the fixed point be stable but that the function $R_*(z)$ be 
defined over the whole interval of $z$ between $-1$ and $1$; in particular, $R_*(z)$ should be finite when $z=-1$.\cite{sakamoto06}) From the 
work of Sakamoto {\it et al.} on the allowed nonanalytical perturbations,\cite{sakamoto06}  one can conclude that the only acceptable solutions are 
with $\alpha=1/2$, which corresponds to a term in $\sqrt{1-z}$ in the second cumulant of the renormalized random field $\Delta(z)=R'(z)$ and we 
refer to as a ``cusp'', and with $\alpha>1$, which we refer to as ``subcusps''.  We illustrate the appearance of such a subcusp by looking 
at the SUSY/DR fixed point in $N=18$. There, $\Lambda_{5/2}=0$ and a subcusp in $(1-z)^{3/2}$ is expected in $\Delta_*(z)=R'_*(z)$. The DR 
fixed point in $N=18$ is also marginal with respect to $\Delta'(1)=R''(1)$, {\it i.e.} $\Lambda_2=0$, and unstable with respect to the cuspy fixed point, 
with $\Lambda_{3/2}=-\epsilon/10$. However, the fixed-point values of $\Delta'(1)$ and $\Delta''(1)$ are exactly known, and one can numerically find 
the full $z$-dependence of the fixed-point function $\Delta_*(z)$. It unambiguously displays a subcusp in $(1-z)^{3/2}$ when $z\to 1$, as shown 
in detail in Appendix~\ref{app_RFON_subcusp}. 

We also proved the important property that SUSY and DR are {\it only broken in the presence of a cusp} in $\Delta_*(z)=R'_*(z)$. Weaker 
nonanalyticities (subcusps) do not prevent the main scaling behavior from following DR and do not break the SUSY Ward identities. To derive all 
of these results, which carry over to the 2-loop level,\cite{tissier06b,sakamoto_comment,doussal_reply} the functional nature of 
the RG is crucial.

We display in Fig.~\ref{fig_lambda_RFON} the lowest-order eigenvalues $\Lambda_{p}(N)$ for $p=2, 3$ and $4$ (corresponding to Feldman's 
operators $\mathcal F_4$, $\mathcal F_6$, and $\mathcal F_8$), for $p=3/2$ [the ``cuspy'' perturbation to the second cumulant of the random field 
$\Delta_*(z)=R'_*(z)$]. One can see that in the present case the amplitude of the cusp, or equivalently of the $(1-z)^{3/2}$ term in $R_*(z)$, becomes 
relevant when the associated eigenvalue $\Lambda_{3/2}$ changes sign, which takes place in $N=2(4 + 3\sqrt 3)= 18.3923 \cdots$.  The stable fixed 
point is then a cuspy one with a $(1-z)^{3/2}$ nonanalytic dependence in $R_*(z)$ in the vicinity of $z=1$ [and a cusp in $\Delta_*(z)=R_*'(z)$]. This 
happens before Feldman's operator $\mathcal F_4$ becomes marginal, {\it i.e.}, before $\Lambda_2=0$, which takes place in $N=18$. The key point, 
however, is that the SUSY/DR fixed point  {\it no longer exists} below $N=18$, as is easily seen from the expression of $R_*''(1)$ in 
Eq.~(\ref{eq_FP_RFO(N)}). When $N<18$ it only remains the cuspy fixed point. SUSY and DR are broken at this cupsy fixed point. Note that if one 
uses a large-$N$ expansion for the eigenvalue $\Lambda_2$, one finds as illustrated in Fig.~\ref{fig_lambda_RFON} (top) up to the next-to-leading 
order in $1/N$, that $\Lambda_2$ can be extrapolated to $0$ (here, for $N\approx 8$) butthe extrapolated expansion is of course blind to the 
disappearance of the SUSY/DR fixed point in $N=18$.

So, as one decreases $N$, the SUSY/DR fixed point first acquires nonanalytic terms that do not break SUSY/DR, then becomes unstable in 
$N_{\rm cusp}=2(4 + 3\sqrt 3)$ to a cuspy fixed point at which SUSY and DR are broken, and it finally disappears in $N=18$, which is also when 
$\Lambda_2$ vanishes. This coincidence of the vanishing of $\Lambda_2$ and the disappearance of the SUSY/DR fixed point is a central feature 
that can only be found by considering a nonperturbative treatment in $N$ (as opposed to an expansion in $1/N$) and by studying the RG equations 
for the fixed point itself (and not only the linearized version for the determination of the eigenvalue spectrum). It results from an exact property of 
the FRG flow equations (beyond 1- and 2-loop results): Whereas the flow equations for $R^{(p\geq 3)}(1)$ (associated with $\mathcal F_{2p}$) 
is linear, that for $R''(1)$ is nonlinear. At 1-loop order when keeping $R'(1)$ at its SUSY/DR fixed-point value $\epsilon/(N-2)$, one for instance finds
\begin{equation}
\begin{aligned}
\label{eq_nonlinear_RFO(N)}
&\partial_t R''_k(1)=\\& -(N+7) R''_k(1)^2+\frac{\epsilon(N-8)}{N-2}R''_k(1)-\frac{\epsilon^2}{(N-2)^2}  \\&
\end{aligned}
\end{equation}
and, with $R''_k(1)$ kept fixed at its fixed-point value $R''_*(1)$ given in Eq.~(\ref{eq_FP_RFO(N)}), 
\begin{equation}
\begin{aligned}
\label{eq_linear_RFO(N)}
\partial_t R^{(p\geq 3)}_k(1)=\Lambda_p(N) R^{(p\geq 3)}_k(1)+ \mathcal G_{p,k}(N), 
\end{aligned}
\end{equation}
where the $\Lambda_{p\geq 2}(N)$'s are given by Eq.~(\ref{eq_eigenvalue_RFO(N)}) and the $\mathcal G_{p\geq 3,k}(N)$'s are functions of the 
$R^{(q)}_k(1)$'s with $q\leq p-1$.\cite{tissier06}. (We recall that $t=\ln(k/k_{\rm UV})$ is the RG time.) At 2-loop order, the equation for 
$R^{(p\geq 3)}_k(1)$ stays linear and that for $R''_k(1)$ becomes cubic. The degree of the nonlinearity for the latter increases with the loop order, 
so that at $j$-loop order, $\partial_t R''_k(1)=Q_{N,j}(R''_k(1))$ where $Q_{N,j}$ is a polynomial of degree $j$. This nonlinearity is responsible for the 
fact that there is no solution for the SUSY/DR fixed point $R''_*(1)$ below the value of $N$ at which $\Lambda_2(N)$ vanishes. Indeed, the fixed-point 
value is a solution of $Q_{N,j}(R''_*(1))=0$ and $\Lambda_2$ is given by $\Lambda_2(N)=Q_{N,j}'(R''_*(1))$, where $Q_{N,j}'$ is the derivative of 
the polynomial. One expects $\Lambda_2$ to be positive for $N>N_{\rm DR}$ (with $N_{\rm DR}=18$ at one loop) so that $Q_{N,j}'(R''_*(1))>0$. 
When $N=N_{\rm DR}$, $\Lambda_2=Q_{N,j}'(R''_*(1))=0$ while $Q_{N,j}(R''_*(1))=0$. Generically, and in the absence of an additional symmetry, 
this corresponds to the collapse of the SUSY/DR solution with another solution, such that the SUSY/DR solution disappears altogether for  
$N<N_{\rm DR}$: see Appendix~\ref{app_RFON}. Although not a rigorous proof, this strongly supports the fact that {\it the SUSY/DR fixed point 
no longer exists below the value $N_{\rm DR}$ at which the eigenvalue $\Lambda_2$ associated with the operator $\mathcal F_4$ vanishes.}

We stress again that the perturbative expansions in $1/N$ around the large-$N$ limit {\it cannot} capture the annihilation of fixed points even 
if it predicts through an extrapolation the vanishing of the extrapolated eigenvalue $\Lambda_2(N)$.

\subsection{More on the SUSY/DR fixed points and their stability}

In what follows we study in more detail the issue of the stability of the fixed points and the existence of the putative unstable fixed point that 
coalesces and annihilates with the SUSY/DR fixed point when $N=N_{\rm DR}$. 

As we have already mentioned, when $N<N_{\rm cusp}=2(4 + 3\sqrt 3)\approx 18.39$ (at 1-loop order), the stable SUSY/DR fixed point becomes 
unstable to a nonanalytic perturbation associated with a $(1-z)^{3/2}$ term in $R(z)$ [and a cusp in $\Delta(z)=R'(z)$] near $z=1$ and characterized 
by the eigenvalue $\Lambda_{3/2}$. The SUSY/DR fixed point is cuspless but has nonetheless a weak singularity in $(1-z)^{1+\alpha_*(N)}$ with 
$\alpha_*(N)\geq3/2$. 

\begin{figure}
\includegraphics[width=.9\linewidth]{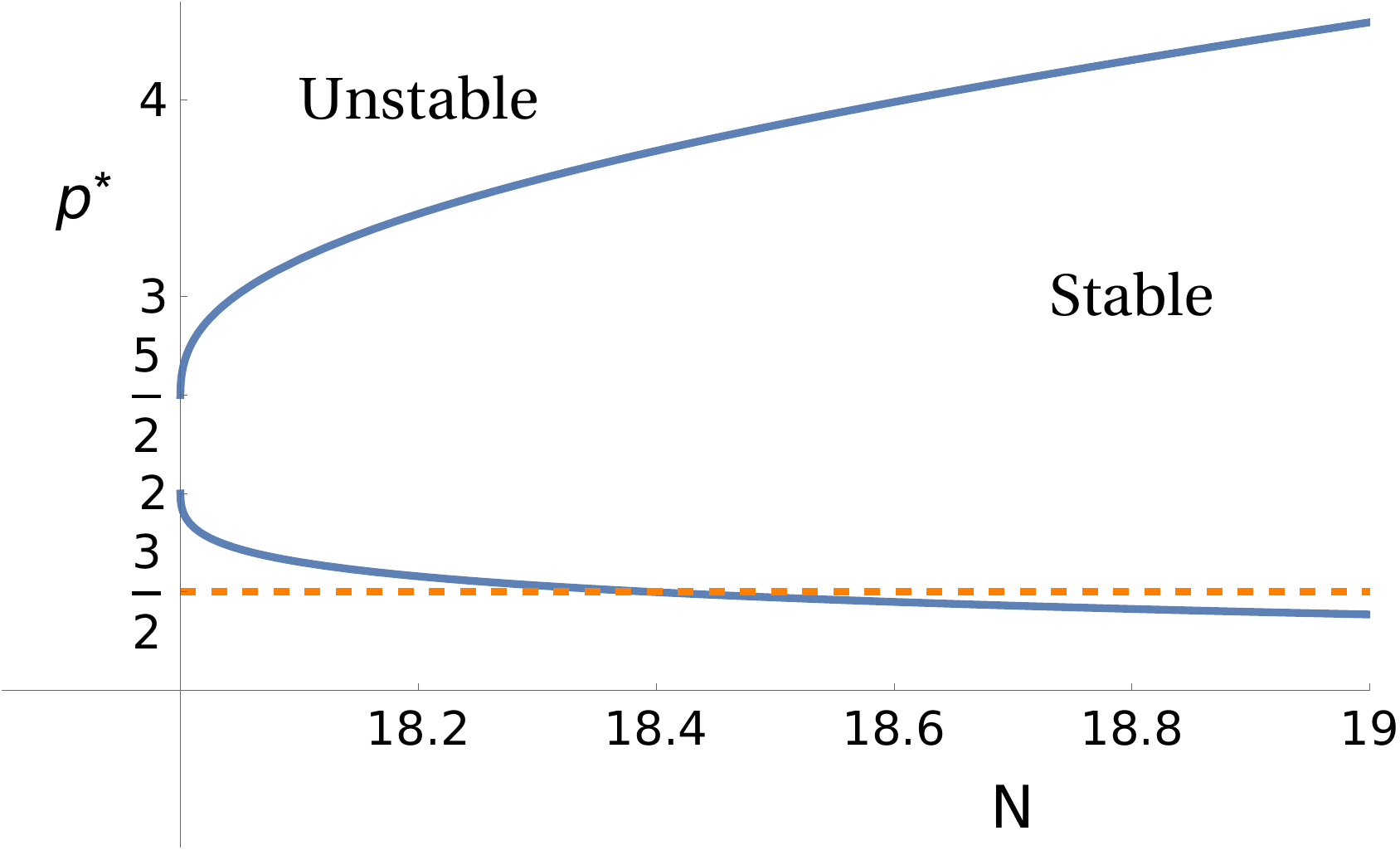}
\includegraphics[width=.9\linewidth]{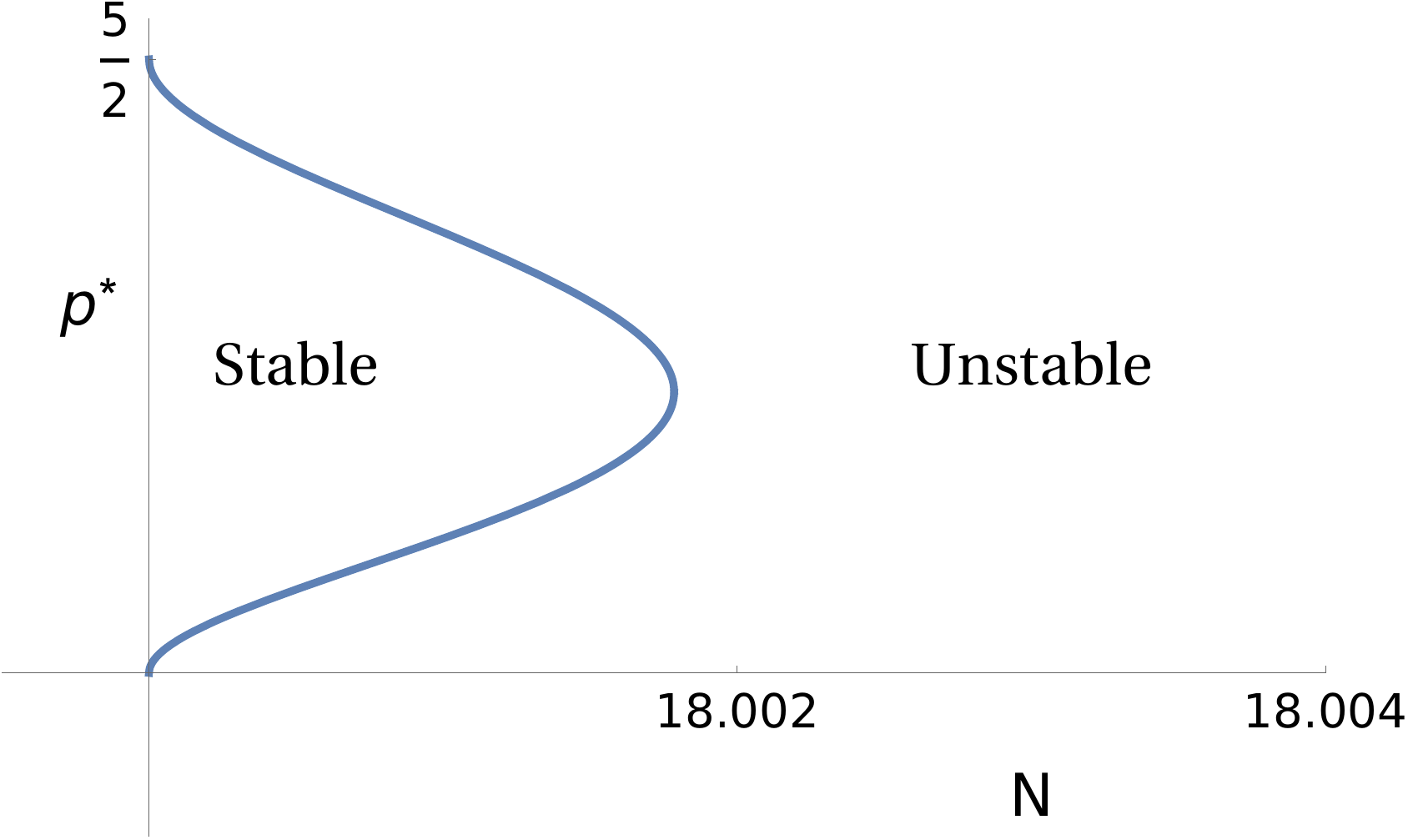}
\caption{Stability of the SUSY/DR fixed points in the RFO($N$)M at 1-loop order in $d=4+\epsilon$. 
Top: Zeros of the eigenvalue $\Lambda_p(N)$, where $p=1+\alpha$ is extended to real values, for perturbations around the (most) stable fixed point. 
The dashed line corresponds to $p=3/2$, {\it i.e.}, to a cusp in $R'(z)=\Delta(z)$, which is the only acceptable value for $p<2$.\cite{sakamoto06,FPbalog} 
The eigenvalue is negative (relevant) above the top (orange) curve $\alpha_*(N)$ and positive (irrelevant) between this curve and the bottom (blue) 
curve $\alpha_-(N)$. Below $N=2(4 + 3\sqrt 3)\approx 18.3923 \cdots$, the fixed point is unstable to the cuspy perturbation, {\it i.e.}, $p=3/2$ or 
equivalently $\alpha=1/2$. 
Bottom: Zeros of the eigenvalue $\Lambda_{p=1+\alpha}(N)$ for perturbations around the unstable fixed point. To the right of the curves the eigenvalues 
are relevant and to the left they are irrelevant. The top and bottom curves giving the two zeros,  $\alpha_*(N)$ and $\alpha_-(N)$, merge in 
$N=18.001785\cdots$.
}
\label{fig_alpha_RFON}
\end{figure}

Consider now the putative unstable fixed point that collapses with the cuspless fixed point when $\Lambda_2(N)=0$. This fixed point is characterized 
by the same value of $R'_*(1)$ as the SUSY/DR most stable fixed point ({\it e.g.}, $\epsilon/(N-2)$ at one loop). It has therefore the same critical 
exponents $\eta$, $\bar\eta$, and $\nu$ because these exponents are obtained from $R'_*(1)$ only.\cite{tissier06,tissier06b} At one loop, the fixed 
point is also specified by $R''_*(1)=\epsilon[N-8+\sqrt{(N-2)(N-18)}]/[2(N-2)(N+7)]$, which is the other solution of Eq.~(\ref{eq_nonlinear_RFO(N)}), and 
it is unstable in the direction $R''(1)$. More generally, the eigenvalue associated with $R^{(p)}(1)$ around this unstable SUSY/DR fixed point is given 
by a simple modification of Eq.~(\ref{eq_eigenvalue_RFO(N)}),
\begin{equation}
\begin{aligned}
\label{eq_eigenvalue_RFO(N)_unstable}
\Lambda_{p\geq 2}(N)=&-\frac{\epsilon}{N-2} \big [2p^2-(N-1)p+N-2+\\&
\frac{p(N-5+6p)}{2(N+7)}(N-8+\sqrt{(N-2)(N-18)} \,)\big ],
\end{aligned}
\end{equation}
which can also be extended to real values $p=1+\alpha$ associated with a nonanalytic perturbation in $(1-z)^{1+\alpha}$ near $z=1$. We note that 
for $N> 18.001785\cdots$ all eigenvalues $\Lambda_{p}(N)$ are strictly negative, meaning that all the associated directions are relevant: see 
Fig.~\ref{fig_alpha_RFON}(b). For $N>18.001785\cdots$, provided the eigenperturbations can be extended over the whole interval $-1\leq z\leq 1$, 
this fixed point therefore appears so unstable that a whole function $R_*(z)$ must be fine-tuned at the start of the FRG flow, even if one restricts the 
initial condition to analytical functions. So, this fixed point is completely unphysical.

The situation is different when $18\leq N\leq  18.001785\cdots$. In this case, there are two zeros for each $N$, $\alpha_-(N)$ and $\alpha_*(N)$, 
with $1<\alpha_-(N)\leq \alpha_*(N)<3/2$: see Fig.~\ref{fig_alpha_RFON}(b). As for the other SUSY/DR fixed point, we expect that the (several times) 
unstable SUSY/DR fixed point acquires a nonanalytical dependence that behaves as $(1-z)^{1+\alpha_-(N)}$ near $z=1$. (Since one is dealing 
with full functions one must check that the corresponding $R_*(z)$ is defined over the full interval $-1\leq z\leq 1$, but this seems 
possible by forming a linear combination with a solution that behaves as $(1-z)^{1+\alpha_*(N)}$ near $z=1$.) The pending question 
is that of the stability of this fixed point in the present domain of $N$. It is unstable in the direction $R''(1)$ and is also unstable in the direction of 
the cusp ($\alpha=1/2$), which implies that it is unstable with respect to both the cuspless SUSY/DR more stable fixed point and to the cuspy 
fixed point that is the most stable for $N<2(4 + 3\sqrt 3)\approx 18.39$. In addition, however, all eigenvalues $\Lambda_{1+\alpha}(N)$ with 
$1<\alpha\leq \alpha_-$ are also negative, {\it i.e.}, relevant, and associated with acceptable eigenfunctions when following the same procedure as 
before.\cite{sakamoto06,FPbalog}  Nonetheless, over the very narrow interval of values of $N$, $18\leq N\leq  18.001785\cdots$, one expects that the 
fixed point can be found by choosing initial conditions of the FRG flow that are restricted to analytical functions $R_0(z)$ and by fine-tuning the two 
remaining relevant directions, $R'(1)$ and $R''(1)$, to their fixed-point values.

\subsection{Recap}
\label{subsec_RFON_recap}

To sum up: Around the SUSY/DR critical fixed point in the RFO($N$)M in $d=4+\epsilon$, Feldman-like operators of the form $(1-z)^p$ near $z=1$ 
become marginal (and potentially relevant) as $N$ decreases from infinity but, provided $p>2$, this is cured by the fact that the SUSY/DR fixed point 
acquires a weak nonanalytical behavior in $(1-z)$, which does not break SUSY nor DR in the main sector of the critical behavior. The most dangerous 
operators are then $\mathcal F_4$, which is characterized by the eigenvalue $\Lambda_2(N)$, and the cuspy perturbation in $(1-z)^{3/2}$ near $z=1$, 
which is characterized by the eigenvalue $\Lambda_{3/2}(N)$. When $\Lambda_{3/2}(N)=0$ the SUSY/DR fixed point becomes unstable to a cuspy 
perturbation and there is an exchange of stability with another fixed point at which the second cumulant of the renormalized random field $\Delta(z)=R'(z)$ 
has itself a cusp and is therefore associated with a breakdown of SUSY and DR. This takes place for $N_{\rm cusp}=2(4 + 3\sqrt 3)\approx 18.39$ at 
$1$-loop order. Below this value the SUSY/DR fixed point continues to exist down to $N=18$, at which $\Lambda_2=0$ ($\mathcal F_4$ becomes 
marginal) and, at the same time, a coalescence with yet another unstable SUSY/DR fixed takes place. Below $N=18$ there are {\it no} SUSY/DR 
fixed points. This phenomenon cannot be found through a perturbative expansion in $1/N$ around the large-$N$ limit. 

\begin{figure}
\includegraphics[width=.9\linewidth]{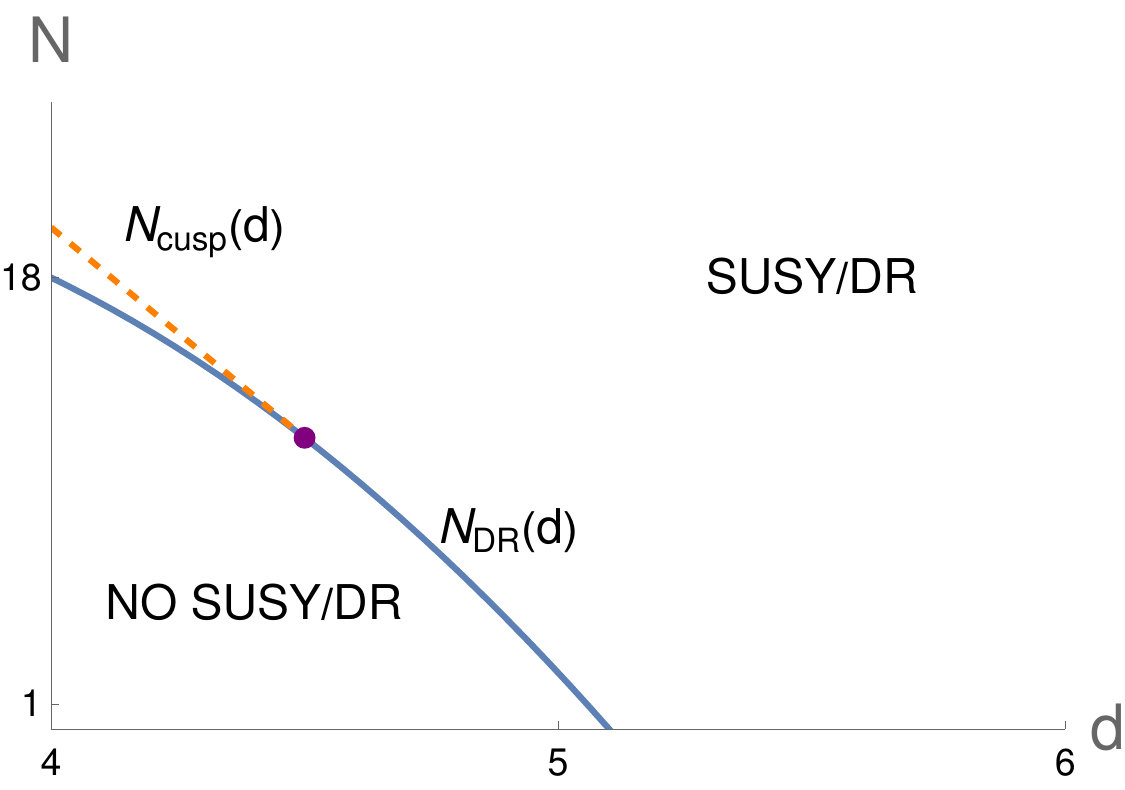}
\caption{Schematic phase diagram of the RFO($N$)M in the ($N$, $d$) plane. The full line, $N_{\rm DR}(d)$, is where the eigenvalue $\Lambda_2$ 
associated with Feldman's operator $\mathcal F_4$ vanishes and the cuspless SUSY/DR fixed point disappears. The dashed line, $N_{\rm cusp}(d)$, 
is where the eigenvalue $\Lambda_{3/2}$ associated with a cusp in $R'(z)$ vanishes and the cuspless SUSY/DR fixed point becomes unstable with 
respect to a cuspy fixed point. The two lines meet around $N_x\approx 14$ and $d_x\approx 4.4$ (estimated from a 2-loop perturbative FRG in 
$d=4+\epsilon$\cite{tissier06b,FPbalog}).For $N<N_x$ and $d>d_x$, which includes the RFIM, the SUSY/DR critical fixed point disappears when it 
is still stable with respect to a cuspy perturbation, {\it i.e.}, $\Lambda_{3/2}>0$ when $\Lambda_{2}=0$.
}
\label{fig_sketch_RFON}
\end{figure}

All of the above is true at both 1- and 2-loop orders, and we have given arguments extending the conclusion to any loop order. One notices from 
the $2$-loop calculation that the two curves associated with $\Lambda_2(N)=0$ and $\Lambda_{3/2}(N)=0$ tend to move toward each other as 
$\epsilon$ increases and, if extrapolated, would cross near $N\sim 14$ and $d\sim 4.4$. At this intersection, $\Lambda_{3/2}$ and $\Lambda_2$ 
vanish together, and for lower values of $N$, which include the RFIM, one expects that $\Lambda_2$ goes to zero before ({\it i.e.}, at a higher $d$ 
than) $\Lambda_{3/2}$. This is sketched in Fig.~\ref{fig_sketch_RFON}. 
\\

\section{Back to the RFIM}
\label{sec_back}

\subsection{Dangerous operators and their scaling dimensions around the SUSY/DR fixed point}
\label{sub_dangerous}

We first address the issue of the dimension of Feldman's operators [given in Eq.~(\ref{eq_Feldman_op})] in the RFIM near $d=6$. The question 
of determining the scaling dimensions at the fixed point, even without invoking nonanalytical field dependences, is unexpectedly rather subtle in 
the RFIM. The problem is found in the usually trivial operation of finding the canonical dimensions of the operators. In the replica approach of 
KRT,\cite{rychkov_RFIM-II,rychkov_lectures} the difficulty comes with the limit $n\to 0$ in the number of replicas. Then, as stressed in 
[\onlinecite{rychkov_lectures}], the 2-point correlation functions of $S_n$ invariant operators vanishes. In principle scaling dimensions can be 
extracted from the operator product expansion (OPE) and/or by considering higher-order correlation functions with other operators but the procedure 
is much far from straightforward. In the FRG approach based on a cumulant expansion and the description of the fixed point as being at zero 
(renormalized dimensionless) temperature,\cite{fisher86b,FRGledoussal-chauve,doussal_annals,tarjus04,tissier12a,tissier12b} the difficulty lies 
in the fact that cumulants of different orders come with different powers of the inverse temperature and that, as a consequence, the limit $T\to 0$ 
must be performed separately for each order: see also Appendix~\ref{app_recap}.

To give examples, near $d=6$, the canonical dimension of the leader of Feldman's operator $\mathcal F_4$, which involves products of 4 
transformed fields, is taken in [\onlinecite{rychkov_RFIM-II}] as $\Delta_{(\mathcal F_4)_L}=2(d-2)$ and that of  $\mathcal F_6$, which involves 
products of 6 transformed fields, as $\Delta_{(\mathcal F_6)_L}=3(d-2)$. In the FRG near the zero-temperature fixed point (see 
also Appendix~\ref{app_recap}), the counterpart of  $\mathcal F_4$ is present in the second 1-PI cumulant and involves 4 replica fields and 
a factor of $1/T^2$. It has a canonical dimension of $2(d-4)+4$, where $(d-4)/2$ is the scaling dimension of the fields and $2$ that of the 
inverse temperature, and this indeed corresponds to $\Delta_{(\mathcal F_4)_L}$ as predicted in [\onlinecite{rychkov_RFIM-II}]. On the other 
hand, the term corresponding to Feldman's $\mathcal F_6$ involves products of 6 replica fields and, also being in the second cumulant, 
a factor of $1/T^2$, so that its canonical dimension is $3(d-4)+4$, which disagrees with the prediction of [\onlinecite{rychkov_RFIM-II}].

Accordingly, the eigenvalues of the linearized flow equations around the (Gaussian) fixed point should be 
$\Lambda_2=\Delta_{(\mathcal F_4)_L}-d=d-4$ for both [\onlinecite{rychkov_RFIM-II}] and the FRG, and 
$\Lambda_3=\Delta_{(\mathcal F_6)_L}-d=2d-6$ for [\onlinecite{rychkov_RFIM-II}] and a less irrelevant value $\Lambda_3=2d-8$ for the FRG. 
(Note that in [\onlinecite{feldman02}] Feldman appears to have used the canonical dimensions for the nonrandom $\phi^4$ theory, {\it i.e.}, 
a dimension of $(d-2)/2$ for the replica fields  which however does not predict the right upper critical dimension in the presence of a random field 
nor the right lower critical dimension for the RFO($N$)M; this choice also gives $\Lambda_3=2d-6$ as in KRT.) 

In the FRG approach, the scaling dimensions are fixed by both casting the flow equations in a dimensionless form such that a fixed point can 
effectively be found and then determining the spectrum of all eigenvalues by solving the linearized FRG equations near this fixed point. 
In the RFIM where, as stressed above, there is an ambiguity in choosing the canonical dimensions of some of the irrelevant operators, 
one may fix the ambiguity by considering the loop contributions to the flow equations (or the fixed-point equations) of the coupling constants/functions associated with these operators. By construction, the eigenvalue equations are indeed linear in the coupling constants/functions themselves (which is a reason for the mentioned ambiguity) but the loop contributions to the flow equations are nonlinear in other coupling constants/functions, which helps determining the proper canonical (engineering) dimensions. The latter should then indeed be fixed by finding a consistent matching between the 
dimensions of the tree-level contributions and those of the loop contributions. We discuss this in detail in Appendix~\ref{app_SUSYequiv}, where we 
illustrate the subtleties coming from Cardy's replica field transformation and from sums over replicas. We also point out the disagreements with KRT.\cite{rychkov_RFIM-II,rychkov_lectures} Further work would nonetheless be needed to settle the issue.
\\

We close this discussion by stressing two points:
\begin{itemize}
\item The problem with the canonical scaling dimensions comes only for operators (or their 1-PI counterparts) that involve sums 
over replicas, such as $\mathcal F_{2p}$ in Eq.~(\ref{eq_Feldman_op}) or the associated leaders given in [\onlinecite{rychkov_RFIM-II}]. For the 
random-field free scalar field theory, which describes the RFIM at the upper critical dimension and has been for instance recently studied in [\onlinecite{trevisani_gaussian}], there is an agreement between the various formalisms concerning the scaling of the correlations functions of 
the physical primary (and composite) fluctuating fields. Sums over replicas are what brings an unusual ambiguity.

\item Most importantly, there is an agreement among all derivations concerning the canonical dimension of Feldman's operator $\mathcal F_4$ 
which turns out to be the most dangerous for the SUSY/DR fixed point below $d=6$. The main conclusions of this paper are therefore not affected 
by the discrepancy about other canonical dimensions, such as those of the $\mathcal F_{2p}$'s with $p\geq 3$.
\end{itemize}

\subsection{Perturbative calculation at 2-loop order of the scaling dimensions/eigenvalues}
\label{sub_RFIM_perturb}

When considering the ``anomalous'' contributions to the scaling dimensions of Feldman's operators at the SUSY/DR fixed point, {\it i.e.}, the 
contributions coming from the Feynman diagrams of the loop expansion beyond the tree level, the outcome is the same whether the calculation 
is performed within the FRG as here and in [\onlinecite{feldman02}] or with Cardy's transformed fields as in [\onlinecite{rychkov_RFIM-II}]. 

For computing the scaling dimension of an irrelevant operator in perturbation theory, the general strategy goes  as follows. One first lists the 
operators (call them $\mathcal O_i$) with the same canonical dimension and the same symmetries. The bare action is then perturbed by a term 
of the form $\sum_i x_i \mathcal O_i$. The calculation consists in computing the beta functions for the coupling constants, $\beta_{x_i}$, at linear 
order in $x_j$. (Note then that the canonical dimensions do not enter the calculation of the anomalous contributions to the scaling 
dimensions.) The scaling dimensions are related to the eigenvalues of the matrix $M_{ij}=\frac{\partial(\beta_{x_i})}{\partial {x_j}}\vert_{x_k=0}$. 
Consider for illustration the eigenvalue $\Lambda_2$ associated with  Feldman's operator $\mathcal F_4$. A major simplification takes place 
because $\mathcal F_4$ does not mix with the other operators of the same dimension. If we attribute the index $1$ to this operator 
($\mathcal O_1=\mathcal F_4$), it can be checked that the associated coupling constant ($x_1$) does not contribute to the RG flows of the other 
coupling constants $x_{i>1}$, at least at two-loop order. The matrix $M_{ij}$ is therefore such that $M_{i1}=0$ for $i>1$. In this situation, the sought 
eigenvalue is simply equal to $M_{11}$. In a second step, it is necessary to consider the diagrams built with one operator $\mathcal F_4$ which 
renormalize the coupling constant $x_1$. The one-loop contribution vanishes, as first observed by Feldman,\cite{feldman02} and we obtain the 
two-loop one as
\begin{equation}
\begin{aligned}
\Lambda_{2}^{\rm 2-loop}(\epsilon)=-\frac{8}{27}\epsilon^2.
\end{aligned}
\end{equation}

More generally, the output of the calculation for the anomalous contribution is
\begin{equation}
\begin{aligned}
\Lambda_{p}^{\rm 2-loop}(\epsilon)=-\frac{p(3p-2)}{27}\epsilon^2,
\end{aligned}
\end{equation}
which coincides with the results of Feldman\cite{feldman02} and KRT.\cite{rychkov_RFIM-II} 
\\

The above results lead to several comments:
\begin{itemize}

\item For $p=2$, and as already stressed, all the derivations agree and one has
\begin{equation}
\begin{aligned}
\label{eq_eigenvalue_RFIM_2}
\Lambda_{2}(\epsilon)=2-\epsilon-\frac{p(3p-2)}{27}\epsilon^2,
\end{aligned}
\end{equation}
which indeed coincides with $\Delta_{(\mathcal F_4)_L}-d$ in  [\onlinecite{rychkov_RFIM-II}]. This is what is plotted in Fig.~\ref{fig_lambda_RFIM}. 
For $p>2$ on the other hand we find
\begin{equation}
\begin{aligned}
\label{eq_eigenvalue_RFIM_TT}
\Lambda_{p}(\epsilon)=2p-2-(p-1)\epsilon -\frac{p(3p-2)}{27}\epsilon^2,
\end{aligned}
\end{equation}
which differs by an additive term $-2(p-2)$ from the result of [\onlinecite{rychkov_RFIM-II}], a discrepancy discussed above.

\item Setting $p=3/2$ in Eq.~(\ref{eq_eigenvalue_RFIM_TT})  gives back our previous result for the eigenvalue $\Lambda_{3/2}$ associated 
with the cuspy perturbation stemming from the presence of scale-free avalanches,\cite{tissier_pertFRG} 
$\Lambda_{3/2}(\epsilon)=1-\epsilon/2 -(5/36)\epsilon^2$. Note that when extrapolated to finite values of $\epsilon$ or low values of $d$, this 
eigenvalue $\Lambda_{3/2}(\epsilon)$ vanishes in $d\approx 4.57$ whereas $\Lambda_{2}(\epsilon)$ which is associated with $\mathcal F_4$ 
vanishes in a slightly higher dimension $d\approx 4.59$, so that the cuspy perturbation is still irrelevant when $\Lambda_2$ vanishes.
(Here in the RFIM, we keep the terminology ``cusp'' and ``cuspy'' to denote a dependence in the second cumulant of the renormalized random field 
$\Delta(\phi_1,\phi_2)$ in $\vert\phi_1-\phi_2 \vert$ when $\phi_1\to\phi_2$. More properly, the cusp arises when the cumulant, which is an even 
function of the field difference, is considered as a function of $(\phi_1-\phi_2)^2$ but we will keep the terminology for simplicity. In a similar vein, 
we will use ``subcusp'' to denote weaker singularities in the field difference.)

\end{itemize}

\subsection{Disappearance of the SUSY/DR fixed point for $d<d_{\rm DR}$: A nonperturbative FRG calculation}
\label{sub_RFIM_DE2}

As already stressed several times, a main difference of interpretation with KRT\cite{rychkov_RFIM-II,rychkov_lectures} is that in our 
nonperturbative FRG calculations the vanishing of $\Lambda_2$ in $d=d_{DR}$ {\it coincides with the disappearance} of the SUSY/DR fixed 
point. This is at odds with their scenario in which the SUSY/DR fixed point becomes unstable but is still present below $d_{\rm DR}$. We 
emphasize that the issue {\it cannot} be directly probed through the perturbative approach that relies on an expansion in the $\phi^4$ coupling 
constant and/or in $\epsilon=6-d$. It instead requires a nonperturbative and functional RG which allows one to investigate not only the stability 
of the SUSY/DR fixed point but also its very existence. This is much like the situation in the RFO($N$)M near its lower critical dimension which 
we have presented in detail above, with the dimension $d$ now playing the role of the number of components $N$;  the shortcomings of the 
$\epsilon=6-d$ expansion are then similar to those of the $1/N$ expansion. Note that symmetry arguments beyond the inconclusive perturbative 
calculation may also be invoked.\cite{rychkov_RFIM-II,rychkov_lectures} Indeed, the coalescence of two fixed points when an operator 
becomes marginal conventionally takes place when the marginal operator does not break the symmetries of the merging fixed points. Whether 
in the present case the nonperturbative prediction of the coalescence of two SUSY/DR fixed points in $d_{\rm DR}$ stems from the unusual, 
functional,   character of the pattern of fixed-point disappearance/appearance or relates the property that the marginal operator $\mathcal F_4$ 
is SUSY-null instead of strictly SUSY-nonwritable (in the language of KRT) remains to be clarified.
\\

The central quantity of the nonperturbative FRG for disordered systems is the scale-dependent effective action (or Gibbs free-energy functional) 
$\Gamma_k[\{\phi_a\}]$ which incorporates fluctuations down to some imposed  IR cutoff $k$ and represents the generating functional of all 
1-PI correlation functions at this (running) scale $k$.\cite{footnote_NPRG} The IR cutoff is implemented through the introduction of regulator 
functions in such a way that they do not explicitly break the Ward identities associated with the Parisi-Sourlas SUSY\cite{tissier11,tissier12a,tissier12b} 
(see also below). This disorder-averaged effective average action depends on the average (or background) fields $\phi_a$ in an arbitrarily large 
number of copies (or replicas) $a=1,2,\cdots$ of the original system, which allows one to generate the 1-PI cumulants with their full functional 
dependence through an expansion in increasing number of free replica sums, \cite{tarjus04,tissier12a,tarjus-review}
\begin{equation}
\begin{aligned}
\label{eq_expansion_cumulants}
\Gamma_k[\{\phi_a\}]= &\sum_a \Gamma_{k1}[\phi_a]-\frac{1}{2}\sum_{a,b}\Gamma_{k2}[\phi_a,\phi_b]+ \\&
\frac 1{3!}\sum_{a,b,c}\Gamma_{k3}[\phi_a,\phi_b,\phi_c]+\cdots \,,
\end{aligned}
\end{equation}
where $\Gamma_{k1}$ is the disorder-averaged Gibbs free energy at scale $k$ and the $\Gamma_{kp}$'s for $p\geq 2$ are essentially the 
cumulants of the renormalized disorder at the scale $k$. The evolution of the 1-PI cumulants $\Gamma_{kp}$ with the IR scale $k$ down to 
$k=0$ is governed by a hierarchy of {\it exact} functional RG flow equations: see also Appendix~\ref{app_recap}. (Note that for 
avoiding the introduction of too many symbols we have used the same notation for the fields in the bare action of Eq.~(\ref{eq_bare-action}) 
and for the background fields involved in the scale-dependent effective action, although the latter are the averages of the former at the scale $k$.)

Truncations are however necessary to turn this exact hierarchy into an operational scheme for studying the FRG flows toward fixed points. An 
efficient ansatz that can capture the long-distance physics, including the influence of avalanches and droplets which we have argued to be central 
to the critical behavior of the RFIM,\cite{tissier06,tissier12b,tarjus13,balog_activated} consists in truncating the expansion in derivatives of the fields 
and at the same time truncating the expansion in cumulants. This must be done in such a way that the Parisi-Sourlas SUSY is not explicitly broken 
(this is of course true as well for all the other symmetries). In our previous studies of criticality in the 
RFIM,\cite{tissier11,tissier12b,FPbalog,tarjus-review,balog20} we have considered the second order of the approximation scheme in which we keep 
the first 1-PI cumulant of at the second order of  the derivative expansion, the second 1-PI cumulant at the local potential approximation level, and 
we neglect higher-order cumulants,
\begin{equation}
\begin{aligned}
\label{eq_ansatz_gammaDE2}
&\Gamma_{k1}[\phi_1]= \int_x \Big [U_k(\phi_1(x))+ \frac 12 Z_k(\phi_1(x))(\partial_{\mu}\phi_1(x))^2 \Big ],\\&
\Gamma_{k2}[\phi_1,\phi_2]=\int_x V_k(\phi_1(x),\phi_2(x)),\\&
\Gamma_{k p\geq 3}=0.
\end{aligned}\end{equation} 
The effective average potential $U_k(\phi_1)$ describes the thermodynamics of the system, $Z_k(\phi_1)$ is the field-renormalization function, and  
$V_k(\phi_1,\phi_2)$ is the $2$-replica effective average potential whose second derivative, $V_k^{(11)}(\phi_1,\phi_2)=\Delta_k(\phi_1,\phi_2)$, is 
the second cumulant of the renormalized random field at zero momentum. $\Delta_k(\phi_1,\phi_2)$ is the key quantity that tracks the effect of 
avalanches and droplets through its functional dependence. We refer to this approximation as DE2 because it involves an expansion up to the second 
order of the derivative expansion (DE) in the first cumulant. We will discuss other levels of the nonperturbative approximation scheme, as well as the 
issues of robustness and convergence in Sec.~\ref{sec_NPFRG}.

Inserting the above ansatz into the exact FRG equations for the cumulants leads to a set of coupled flow equations for 3 functions, $U_k(\phi_1)$,   
$Z_k(\phi_1)$, and $V_k(\phi_1,\phi_2)$: see Refs.~[\onlinecite{tissier11,tissier12b,balog20}]. Fixed points describing scale invariance and the 
spectrum of eigenvalues around them can then be found by casting the resulting FRG flow equations in a dimensionless form. As we are searching 
for zero-temperature fixed points,\cite{villain84,fisher86,cardy_textbook} we define a dimensionless renormalized temperature $\widetilde T_k$ which 
flows to zero as $k \rightarrow 0$ (this is the precise meaning of a ``zero-temperature'' fixed point) and we introduce scaling dimensions such that the 
dimensionful quantities scale as
\begin{equation}
\label{eq_dim1}
x\sim k^{-1},\; T \sim k^{-\theta},\;  Z_{k} \sim k^{-\eta}, \; \phi_a  \sim k^{\frac{1}{2}(d-4+\bar \eta)},
\end{equation}
with $\theta$ and $\bar \eta$ related through $\theta=2+\eta-\bar \eta$. (Note that, formally and as indicated above, the scaling dimension of the 
temperature is $D_T=-\theta$ and is such that for a fixed bare temperature $T$ the dimensionless renormalized temperature 
$\widetilde T_k=k^\theta T$ indeed goes to zero as $k\to 0$, provided of course $\theta>0$.) 
Moreover,
\begin{equation}
\label{eq_dim2}
U_k\sim k^{d-\theta}, \;V_k \sim k^{d-2\theta},
\end{equation}
so that the second cumulant of the renormalized random field $\Delta_k$ scales as $k^{-(2\eta- \bar \eta)}$. Contrary to Cardy's transformed 
fields,\cite{cardy_transform} all the fields $\phi_a$ have the same scaling dimension $D_\phi=(d-4+\bar\eta)/2$, but there is now in addition a 
renormalized temperature with its own scaling dimension: see also Appendix~\ref{app_recap}.

Letting the dimensionless counterparts of $U_k, V_k,\Delta _k,  \phi$  be denoted by lower-case letters, 
$u_k, v_k,\delta _k,  \varphi$,  the resulting FRG flow equations can be symbolically written as
\begin{equation}
\label{eq_flow_dimensionless_u}
\begin{aligned}
\partial_t u'_k(\varphi)=&-\frac 12(d-2\eta_k+\bar\eta_k)u'_k(\varphi)+\frac 12(d-4+\bar\eta_k) \times \\&
\varphi u''_k(\varphi)+ \beta_{u'}(\varphi)
\end{aligned}
\end{equation}
\begin{equation}
\label{eq_flow_dimensionless_z}
\begin{aligned}
&\partial_t z_k(\varphi)=\eta_k z_k(\varphi) +\frac 12(d-4+\bar\eta_k)\varphi z'_k(\varphi)+ \beta_{z}(\varphi)
\end{aligned}
\end{equation}
and
\begin{equation}
\label{eq_flow_dimensionless_delta}
\begin{aligned}
\partial_t \delta_k(\varphi_1,\varphi_2)=&(2\eta_k-\bar\eta_k)\delta_k(\varphi_1,\varphi_2)+ \frac 12(d-4+\bar\eta_k) \times \\&
(\varphi_1\partial_{\varphi_1}+\varphi_2\partial_{\varphi_2})\delta_k(\varphi_1,\varphi_2) +\beta_{\delta}(\varphi_1,\varphi_2)
\end{aligned}
\end{equation}
where $t=\log(k/k_{\rm UV})$ and $k_{\rm UV}$ is a UV cutoff associated with the microscopic scale of the model. The beta functions themselves 
depend on $u_k'$, $z_k$, $\delta_k$ and their derivatives, and they depend as well on the dimensionless IR cutoff function. In addition, the running 
anomalous dimensions $\eta_k$ and $\bar\eta_k$ are fixed by the conditions $z_k(0)=\delta_k(0,0)=1$. All the expressions are given in 
Ref.~[\onlinecite{tissier12b}]. The above flow equations are written for a zero bare temperature. For a nonzero one there are additional terms 
proportional to $\widetilde T_k$ which are however subdominant as they go to zero as $k^{\theta}$ when approaching the fixed 
point.\cite{tissier06,tissier12a,tissier12b} Note finally that the RG is ``functional" as its central objects are functions instead of coupling constants 
and it is ``nonperturbative'' as the approximation scheme does not rely on an expansion in some small coupling constant or function.

Fixed points are studied by setting the left-hand sides of the dimensionless FRG equations in Eqs.~(\ref{eq_flow_dimensionless_u}-\ref{eq_flow_dimensionless_delta}) to zero and the spectrum of eigenvalues, or equivalently of scaling dimensions, around a given fixed point can 
be obtained from the linearization of the equations in the vicinity of this fixed point. The zero-temperature fixed point controlling the critical behavior 
has been determined in previous investigations:\cite{tissier11,tissier12b} Above a critical dimension $d_{\DR}$ close to $5.1$, there exists a stable 
cuspless fixed point (stable, except of course for the relevant direction that corresponds to fine-tuning to the critical point). As already discussed, 
the presence or absence of a cusp now refers to the dependence of $\delta_k(\varphi_1,\varphi_2)$ on the field difference $\varphi_1 - \varphi_2$ 
(the cusp being a square-root dependence on the variable $(\varphi_1 - \varphi_2)^2$ when $\varphi_1\to \varphi_2$). This cuspless fixed point 
entails SUSY and DR.\cite{tissier11,tissier12b}

We stress that the dimensions introduced in Eqs.~(\ref{eq_dim1},\ref{eq_dim2}) have to be chosen such that a fixed point is indeed found for 
the whole functions, $u_k(\varphi)$, $\delta_k(\varphi_1,\varphi_2)$, etc. Near $d=6$ this then fixes the canonical dimensions of all the coupling 
constants obtained by expanding in powers of the fields: See also the discussion in Sec.~\ref{sub_dangerous} and Appendices~\ref{app_recap} 
and \ref{app_SUSYequiv}.

For addressing the presence or absence of a cuspy behavior, it turns out to be convenient to change variable from $\varphi_1$ and $\varphi_2$ 
to $\varphi=(\varphi_1+\varphi_2)/2$ and $\delta\varphi=(\varphi_1-\varphi_2)/2$. The putative cusp is now in the variable $\delta\varphi$. For 
$d\geq d_{\DR}$, the (critical) cuspless fixed point is characterized in the limit $\delta\varphi \rightarrow 0$ by the expansion
\begin{equation}
\label{eq_cuspless_delta}
\delta_*(\varphi,\delta\varphi) = \delta_{*,0}(\varphi) + \frac{1}{2} \delta_{*,2}(\varphi)\delta\varphi^2  + \mathcal O(\vert \delta\varphi \vert^3).
\end{equation}
The signature of the Parisi-Sourlas SUSY is a Ward identity relating the second and the first cumulants,\cite{footnote_SUSY-Ward}
\begin{equation}
\label{eq_cuspless_ward}
\delta_{*,0}(\varphi)=z_*(\varphi)
\end{equation} 
which is satisfied at the cuspless fixed point and implies DR.\cite{tissier11,tissier12a,tissier12b} Indeed, the sector with $\delta\varphi=0$ then 
decouples, and one finds that in this sector the cuspless fixed point in $d$ dimensions is the same as that of the pure $\phi^4$ theory in dimension 
$d-2$ obtained from the same approximation ({\it i.e.}, the second order of the derivative expansion for the effective average action described by 
2 functions, $u'(\varphi)$ and $z(\varphi)$).

We can now fix the functions $u_k'(\varphi)$, $z_k(\varphi)$ and $\delta_{k,0}(\varphi)$ at their SUSY/DR cuspless fixed-point expressions and 
study what happens at and around this fixed point for the terms of the expansion in $\delta\varphi^2$, which are related to the dangerous operators 
$\mathcal F_{2p}$ pointed out by Feldman\cite{feldman02} [see Eq.~(\ref{eq_Feldman_op})]. We focus on the potentially most dangerous one, 
$\mathcal F_{4}$, which corresponds to $\delta_{k,2}(\varphi)$. Here too, the sector in $\delta\varphi^2$ decouples from higher orders in 
$\delta\varphi$ (provided one assumes a regular enough behavior), and one can derive a closed FRG equation for $\delta_{k,2}(\varphi)$, which 
is of the form
\begin{equation}
\label{eq_delta2}
\partial_t \delta_{k,2}(\varphi)= A_*(\varphi)\delta_{k, 2}( \varphi )^2+ L_*(\varphi,\partial_\varphi,\partial^2_\varphi) \delta_{k, 2}( \varphi )+ B_*(\varphi)
\end{equation}
where $L_*$ is a linear operator, $L_*=C_*(\varphi)+D_*(\varphi)\partial_\varphi+E_*(\varphi)\partial_\varphi^2$, and the functions $A_*$, $B_*$, 
$C_*$, $D_*$, $E_*$ are obtained from the known SUSY/DR fixed-point functions and anomalous dimensions, with $A_*(\varphi)\neq 0$. (Note that 
as a result of SUSY, $\bar\eta_*=\eta_*$.) The expressions are given in [\onlinecite{FPbalog}].

From Eq.~(\ref{eq_delta2}) one can (numerically) obtain the fixed point $\delta_{*,2}(\varphi)$ by setting the left-hand side to 0,
\begin{equation}
\label{eq_delta2_FP}
0= A_*(\varphi)\delta_{*, 2}( \varphi )^2+ L_*(\varphi,\partial_\varphi,\partial^2_\varphi) \delta_{*, 2}( \varphi )+ B_*(\varphi),
\end{equation}
and determine the eigenvalues by introducing $\delta_{k,2}(\varphi)=\delta_{*,2}(\varphi)+k^\lambda f_\lambda(\varphi)$ and linearizing the right-hand 
side in the function $f$. The eigenvalue $\Lambda_2$ which is associated with $(\mathcal F_4)_L$ is then the smallest $\lambda$ that satisfies
\begin{equation}
\label{eq_Lambda2_DE2}
\lambda f_\lambda(\varphi)= 2A_*(\varphi)\delta_{*, 2}( \varphi )f_\lambda(\varphi)+ L_*(\varphi,\partial_\varphi,\partial^2_\varphi) f_\lambda(\varphi).
\end{equation}
The key property of the fixed-point equation for $\delta_{*, 2}( \varphi )$ is that it is nonlinear in $\delta_{*,2}(\varphi)$ itself, {\it e.g.}, quadratic at the 
present level of approximation. On the other hand, the FRG equations for all the higher terms $\delta_{k,2p}(\varphi)$ with $p>1$ are linear. As argued 
in detail in the case of the RFO($N$)M near its lower critical dimension (see Sec.~\ref{sec_RFON} and Appendix~\ref{app_RFON}), {\it the nonlinearity 
of the equation is what leads to the disappearance of the fixed point $\delta_{*, 2}( \varphi )$ when the eigenvalue $\Lambda_2$ vanishes}. This is 
precisely what we find when solving the two above equations, Eqs.~(\ref{eq_delta2_FP}) and (\ref{eq_Lambda2_DE2}). (In practice, from the knowledge 
of $u'_{*}(\varphi)$ and $z_{*}(\varphi)$, which are obtained from two coupled equations, we first solve the equation for $\delta_{*, 2} (\varphi)$ and 
then use the input to solve  Eq.~(\ref{eq_Lambda2_DE2}); all partial differential equations are numerically integrated on a one-dimensional grid by 
discretizing the field $\varphi$, and the solution can be studied for any value of $d$.\cite{tarjus13,FPbalog,balog20}) The stable fixed-point function 
$\delta_{*, 2}( \varphi )$ collapses with another unstable fixed-point function in the critical dimension $d_{\rm DR}\approx 5.13$, when the eigenvalue 
$\Lambda_2=0$. 

For $d<d_{\rm DR}$ the SUSY/DR fixed point for the RFIM therefore no longer exists (of course, the DR fixed-point functions $u_*(\varphi)$ and 
$z_*(\varphi)$ are still well defined, but not the fixed-point functions associated with the cumulants of the renormalized disorder for distinct field 
arguments) and $\Lambda_2$, whose equation involves $\delta_{*, 2} (\varphi)$, is no longer defined. A heuristic analytical argument showing that this 
is the case in the absence of a nongeneric cancellation (either accidental or due to an additional symmetry) is given in Appendix~\ref{app_NLFP_DE2}. 
Note that if for $d<d_{\rm DR}$ one solves the FRG equation for $\delta_{k,2}(\varphi)$ in Eq.~(\ref{eq_delta2}) starting from cuspless initial conditions, 
one finds that $\delta_{k,2}(\varphi)$ grows and diverges at a {\it finite} RG scale (which as alluded to in Sec.~\ref{sub_putting}  defines a Larkin length, 
length that diverges as $d\to d_{\rm DR}^-$\cite{FPbalog,balog20}). Accordingly, the running Ward identity associated with the Parisi-Sourlas SUSY 
ceases to be valid at this finite Larkin scale, even at zero temperature when starting from a SUSY-compatible initial condition, and SUSY is then 
broken along the FRG flow.

From the above result it is clear that, despite the fact that the operator $(\mathcal F_4)_L$ which is characterized by the eigenvalue $\Lambda_2$ 
is SUSY-null and not SUSY-nonwritable,\cite{rychkov_RFIM-II} the vanishing of $\Lambda_2$ leads to a breakdown of both SUSY and DR. This is due 
to the coincidence of $\Lambda_2=0$ with the disappearance of a fixed point for $\delta_2(\varphi)$, complemented by the fact that $\delta_2(\varphi)$ 
appears in all other fixed-point equations in the sector of the theory where the replica-field arguments of the renormalized cumulants are not equal. 
Below the dimension $d_{\rm DR}$ in which $\Lambda_2=0$ there is no more {\it bona fide} SUSY/DR fixed point because the existence of such a 
fixed point requires that all dimensionless renormalized 1-PI cumulants be defined, which includes $\delta_2(\varphi)$.  As already mentioned, this 
fixed-point disappearance is different from the situation in the model of an elastic interface in a random environment for which the SUSY/DR fixed point 
is simply the Gaussian one ({\it i.e.}, with $\delta_2(\varphi)=0$) which, although unstable, can be continued below 
$d_{\rm DR}=4$.\cite{FRGledoussal-chauve,wiese_SUSY}

Starting from the FRG flow equations for the higher-order terms $\delta_{k,2p}(\varphi)$ of the expansion of  $\delta(\varphi,\delta\varphi)$ in 
$\delta\varphi$, one can obtain the eigenvalues $\Lambda_{p}(d)$ introduced above. This requires to first find the fixed-point value 
$\delta_{*,2p}(\varphi)$ 
and then to solve the linearized eigenvalue equation around this value. The quality of the result for a given nonperturbative approximation, {\it e.g.}, 
DE2, is expected to deteriorate as $p$ increases. We have therefore not computed the eigenvalues beyond $\Lambda_3$ which is associated to the 
term in $\delta\varphi^4$ in $\delta(\varphi,\delta\varphi)$ and corresponds to Feldman's operator $\mathcal F_6$. The family of eigenvalues can also 
be extended to noninteger values of $p$ and we have calculated the eigenvalue $\Lambda_{3/2}$ which is associated with a cuspy perturbation in 
$\sqrt{\delta\varphi^2}$ in the second 1-PI cumulant of the renormalized random field.

\begin{figure}
\includegraphics[width=\linewidth]{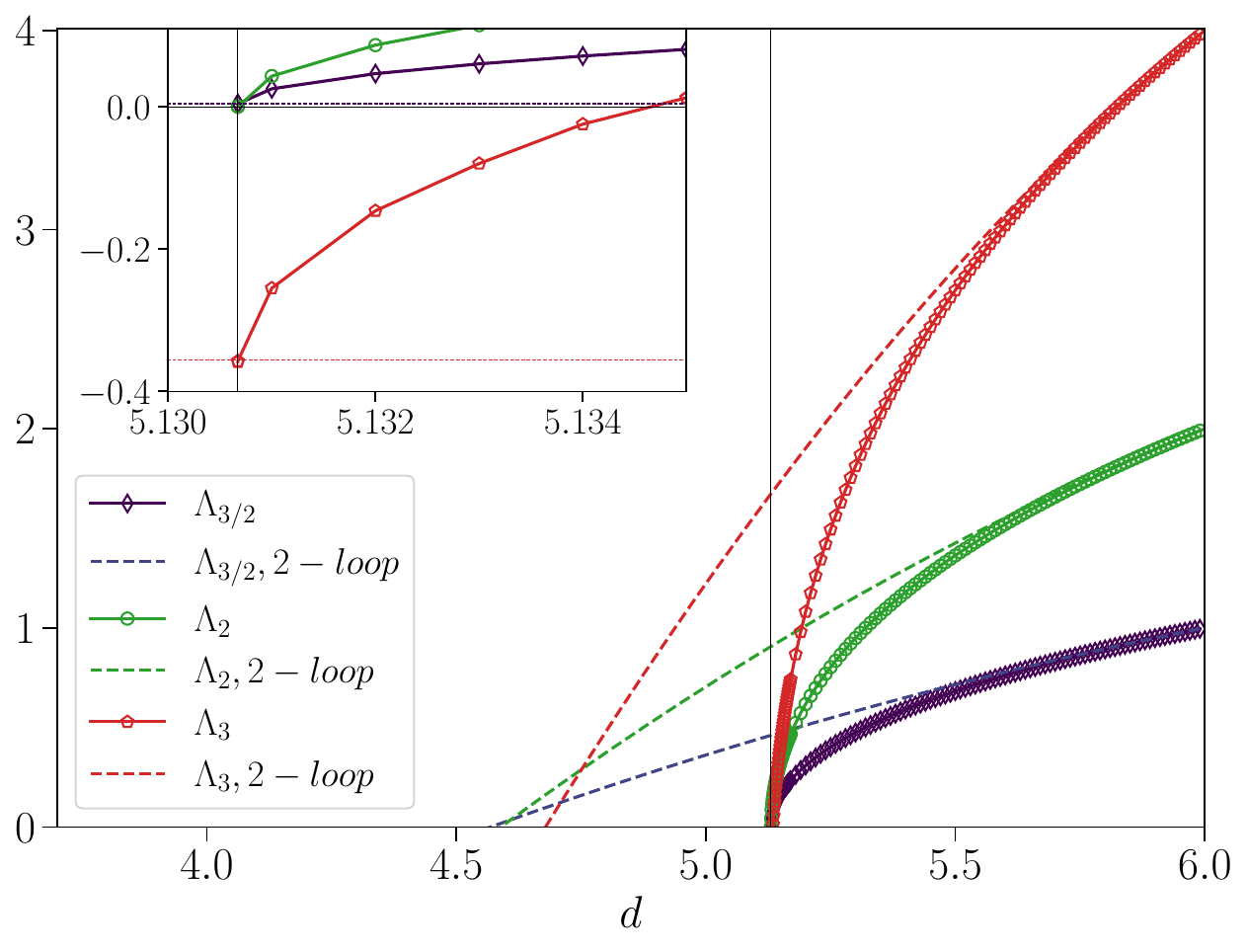}
\caption{Eigenvalues  $\Lambda_2(d)$, $\Lambda_3(d)$, and $\Lambda_{3/2}(d)$ of the most dangerous perturbations around the (most stable) 
SUSY/DR fixed point in the RFIM as calculated from the nonperturbative truncation DE2 discussed in Secs.~\ref{sub_RFIM_DE2}. 
The dashed lines represent the $2$-loop calculation in $d=6-\epsilon$, together with plausible extrapolations.  Inset: Zoom in on the region around 
$d_{\rm DR}$. Note that the cuspy perturbation associated with $\Lambda_{3/2}$ is small but irrelevant when $\Lambda_2$ vanishes in 
$d_{\rm DR}$;  on the other hand, $\Lambda_3$ vanishes at a slightly higher $d$ than $\Lambda_2$ (as also found in the RFO($N$)M: see 
Fig.~\ref{fig_lambda_RFON}). The dashed horizontal lines indicate the values of the three eigenvalues when $d=d_{\rm DR}$.
}
\label{fig_lambdas_RFIM}
\end{figure}

We plot  in Fig.~\ref{fig_lambdas_RFIM} the eigenvalues $\Lambda_2(d)$, $\Lambda_{3}(d)$, and $\Lambda_{3/2}(d)$, as computed  both from 
the  perturbative result of Sec.~\ref{sub_RFIM_perturb} and from our nonperturbative FRG calculation obtained at DE2. The nonperturbative 
calculations are detailed above and  in  Appendix~\ref{app_flow-eqs}. The 2-loop perturbative calculation for $\Lambda_2$ coincides with that of [\onlinecite{rychkov_RFIM-II}] and, as explained in Secs.~\ref{sub_dangerous} and \ref{sub_RFIM_perturb}, the result for $\Lambda_3$ is shifted 
from that of [\onlinecite{rychkov_RFIM-II}] by $-2$. In the present approximation scheme, $\Lambda_3$ vanishes at a slightly higher $d$ than 
$\Lambda_2$. As also found for the RFO($N>2$)M and discussed in Sec.~\ref{sec_RFON}, this does not break SUSY/DR but leads to the 
appearance of a subcusp in $\delta(\varphi,\delta\varphi)$.

As already stressed in~[\onlinecite{tarjus13,FPbalog,balog20}], $\Lambda_{3/2}(d)$ is irrelevant when $\Lambda_2(d)$ vanishes. Physically, this 
means that the presence of scale-free avalanches at criticality does induce cusps in the cumulants of the renormalized random field but that the 
amplitude(s) of these cusps go to zero at the fixed point. The disappearance of the SUSY/DR cuspless fixed point is therefore not due to the effect 
of the avalanches {\it per se}. (On the other hand, scale-free avalanches do control the critical behavior described by the cuspy fixed point below 
$d_{\rm DR}$.) The SUSY/DR cuspless fixed point disappears for $d<d_{\rm DR}$, where $d_{DR}$ coincides with the location of $\Lambda_2(d)=0$. 

Note finally that, as discussed in Sec.~\ref{subsec_RFON_recap}, the fact that $\Lambda_{3/2}>0$ when $\Lambda_2=0$ is in contrast with what 
is observed for the RFO($N$)M near $d=4$ (compare with Fig.~\ref{fig_lambda_RFON}). In the latter case, $d_{\rm cusp}(N)$, or equivalently 
$N_{\rm cusp}(d)$, is given by the location of $\Lambda_{3/2}(d)=0$ which takes place while the SUSY/DR fixed point becomes unstable but still 
exists (with $\Lambda_2>0)$: see the sketch in Fig.~\ref{fig_sketch_RFON} for an illustration. 
\\

We conclude this section by a few additional comments:

\begin{itemize}

\item As already mentioned, our calculation is functional and nonperturbative, and valid for any $d$, but is approximate. It exactly recovers the 
1-loop results in $d=6-\epsilon$ but is not exact at 2-loop order. We compare the approximate results obtained by fitting the data to a quadratic 
polynomial in $\epsilon$ near of $d=6$  (at different orders of the nonperturbative approximation scheme) to the exact perturbative result 
at 2-loop in Sec.~\ref{sub_DE4results} below. 

\item A cruder approximation than the DE2 discussed above is obtained by neglecting the variation on $\varphi$ of $z_k(\varphi)$ and 
$\delta_k(\varphi,\delta\varphi)$ and fixing $\varphi$ to the value of the minimum of the effective potential $u_k(\varphi)$, {\it i.e.}, 
$z_k(\varphi)=z_k(\varphi_{\rm min, k})$ and $\delta_k(\varphi,\delta\varphi)=\delta_k(\varphi_{\rm min, k},\delta\varphi)$ with $u_k'(\varphi_{\rm min, k})=0$. 
We refer to this approximation as the LPA'. It, too,  does not explicitly break SUSY so that the cuspless fixed point exactly satisfies dimensional 
reduction with respect to its LPA' counterpart in the pure $\phi^4$ theory in $d-2$. The equations for $\delta_{*,2}$ and $\Lambda_2$ [see 
Eqs.~(\ref{eq_delta2_FP}) and (\ref{eq_Lambda2_DE2})] become
\begin{equation}
\label{eq_delta2_FP_LPA'}
0= A_*(\varphi_{\rm min,*})\delta_{*, 2}^2+ C_*(\varphi_{\rm min,*}) \delta_{*, 2}+ B_*(\varphi_{\rm min,*})
\end{equation}
and
\begin{equation}
\label{eq_Lambda2_LPA'}
\Lambda_2= 2A_*(\varphi_{\rm min,*})\delta_{*, 2}+  C_*(\varphi_{\rm min,*})
\end{equation}
with $A_*>0$ and $B_*\geq 0$. The situation is now very similar to that encountered in the the FRG of the RFO($N$)M in $d=4+\epsilon$ at 1-loop (see Sec.~\ref{sec_RFON} and Appendix~\ref{app_RFON}), except that all quantities depend on $d$ instead of $N$. One expects a regular behavior with 
$d$ of the coefficients $A_*$, $B_*$, and $C_*$ in the vicinity of $d_{\rm DR}$, so that, barring the occurrence of a nongeneric cancellation, {\it e.g.}, 
due to an additional symmetry, which makes $B_*$ vanish in $d=d_{\rm DR}$, there are two solutions for $\delta_{*, 2}$ above $d_{\rm DR}$: one is 
stable, with an eigenvalue $\Lambda_2>0$, and one is unstable, with an eigenvalue $\Lambda'_2<0$; the two coalesce in $d_{\rm DR}$ where 
$\Lambda_2(d_{\rm DR})=\Lambda'_2(d_{\rm DR})=0$, and there are no real solutions for $d< d_{\rm DR}$. This is indeed what is found numerically 
and is illustrated in the LPA' result of Fig.~\ref{fig_RFIM_dDR}.

\item The validity of the truncation beyond the perturbative regime can be assessed by studying different levels of the nonperturbative approximation 
scheme. This is what we do and detail in the next section. There, we show that the apparent convergence of the approximation scheme is actually fast.

\end{itemize}

\section{RFIM: Robustness and accuracy of the nonperturbative FRG approximation scheme}
\label{sec_NPFRG}

\subsection{Goal}
\label{sub_goal}

Our goal is to check the robustness of the nonperturbative FRG predictions for the explanantion of the SUSY/DR breakdown as a function of 
space dimension. In particular, we focus on the critical dimension $d_{\rm DR}$ at which the SUSY/DR fixed point is predicted to disappear 
and we provide error bars on its value by studying different levels of the nonperturbative approximation scheme.\cite{tissier11,tissier12b,tarjus-review} 
We also want to assess the results for the eigenvalues around the SUSY/DR cuspless fixed point, especially the  eigenvalue $\Lambda_2$ which 
in the work of KRT,\cite{rychkov_RFIM-II,rychkov_lectures} is associated to the leader  $(\mathcal F_4)_L$ of the most dangerous Feldman operator 
that can destabilize the SUSY/DR fixed point.

To obtain $d_{\rm DR}$ and compute the eigenvalues we consider the domain of spatial dimension $d$ in which SUSY and DR are valid at the fixed 
point. The main simplification in the FRG treatment is that the calculation only requires the determination of functions entering the 1-PI cumulants 
when all replica-field arguments are equal. Working with functions of only one field is then much more tractable than the determination of the full 
functional dependence which is needed when $d<d_{\rm DR}$ to capture the cuspy fixed point.\cite{tarjus04,tissier06,tissier12b,tarjus-review,balog20}

In our previous papers,\cite{tissier11,tissier12b,tarjus13,FPbalog,balog20} we have studied the second order of the approximation scheme in which 
we keep the first 1-PI cumulant of at the second order of  the derivative expansion, the second 1-PI cumulant at the local potential approximation level, 
and we neglect higher-order cumulants. This is described in Sec.~\ref{sub_RFIM_DE2} where it is denoted DE2 approximation.

Two cruder approximations can be considered that both avoid an explicit breaking of SUSY: The simplest is LPA' in which one freezes the 
dependence of $Z_k(\phi)$ on $\phi$ and that of $\Delta_k(\phi_1,\phi_2)$ on $\phi=(\phi_1+\phi_2)/2$ (the other independent field variable 
$\delta\phi=(\phi_1-\phi_2)/2$ still being free) to the value at the minimum of the effective average potential, $\phi=\phi_{{\rm min},k}$ with 
$U'_k(\phi_{{\rm min},k})=0$ (see also  above in Sec.~\ref{sub_RFIM_DE2}). An improved approximation, which we call LPA'', consists in still choosing $Z_k(\phi)=Z_k(\phi_{{\rm min},k})$ and $\Delta_k(\phi,\phi)=\Delta_k(\phi_{{\rm min},k},\phi_{{\rm min},k})$ but letting both $\phi_1$ and $\phi_2$ 
unconstrained in $\Delta_k(\phi_1,\phi_2)-\Delta(\phi,\phi)$. More importantly, we also investigate the next level of the nonperturbative approximation 
scheme beyond DE2, which is what we present below.

\subsection{Level DE4 of the nonperturbative FRG approximation scheme}

We now study the next level of approximation, which we call DE4: we keep the first three 1-PI cumulants, the first one being considered at the 
4th order of the derivative expansion (hence the acronym DE4), the second one at the second order of the derivative expansion, and the third 
one at the LPA, {\it i.e.}, explicitly,
\begin{equation}
\begin{aligned}
\label{eq_ansatz_gamma1}
&\Gamma_{k1}[\phi_1]=\\& \int_x \Big [U_k(\phi_1(x))+ \frac 12 Z_k(\phi_1(x))(\partial_{\mu}\phi_1(x))^2 + \frac 12 W_{a;k}(\phi_1(x)) \\&
\times(\partial_{\mu}\partial_\nu\phi_1(x))^2 + \frac 12 W_{b;k}(\phi_1(x))\partial_\mu\partial_\nu\phi_1(x) \partial_{\mu}\phi_1(x)\times \\&
\partial_{\nu}\phi_1(x) + \frac 18 W_{c;k}(\phi_1(x))(\partial_{\mu}\phi_1(x))^2(\partial_{\nu}\phi_1(x))^2 \Big ],
\end{aligned}
\end{equation}
\begin{equation}
\begin{aligned}
\label{eq_ansatz_gamma2}
&\Gamma_{k2}[\phi_1,\phi_2]=\\&\int_x \Big [V_k(\phi_1(x),\phi_2(x))+ X_{a;k}(\phi_1(x),\phi_2(x))\partial_{\mu}\phi_1(x)\partial_{\mu}\phi_2(x)\\&
+ \frac 12 X_{b;k}(\phi_1(x),\phi_2(x))[(\partial_{\mu}\phi_1(x))^2+(\partial_{\mu}\phi_2(x))^2]+ \\&
\frac 12 X_{c;k}(\phi_1(x),\phi_2(x))[(\partial_{\mu}\phi_1(x))^2-(\partial_{\mu}\phi_2(x))^2] \Big ],
\end{aligned}
\end{equation}
\begin{equation}
\label{eq_ansatz_gamma3}
\Gamma_{k3}=\int_x V_{3k}(\phi_1(x),\phi_2(x),\phi_3(x)),
\end{equation} 
and
\begin{equation}
\label{eq_ansatz_gammap}
\Gamma_{kp\geq4}=0,
\end{equation} 
where the functions $V_k$, $X_{a,k}$,  $X_{b,k}$, and $V_{3k}$ are symmetric in the permutations of the arguments while the function 
$X_{c,k}$ is antisymmetric. Recall that the effective average potential $U_k(\phi_1)$ describes the thermodynamics of the system and $Z_k(\phi_1)$ 
is the field-renormalization function. The function $V_k(\phi_1,\phi_2)$ is the $2$-replica effective average potential; its second derivative, 
$V_k^{(11)}(\phi_1,\phi_2)=\Delta_k(\phi_1,\phi_2)$,  is the second cumulant of the renormalized random field at zero momentum and is a  
key quantity that tracks avalanches and droplets through its functional dependence in $(\phi_1-\phi_2)$. Similarly, $V_{3k}(\phi_1,\phi_2,\phi_3)$ 
is the $3$-replica effective average potential whose third derivative, $V_{3k}^{(111)}(\phi_1,\phi_2,\phi_3)=S_{k}(\phi_1,\phi_2,\phi_3)$,  is the 
third cumulant of the renormalized random field at zero momentum. The other functions describe the higher-order momentum dependence of 
the first cumulant and the momentum dependence of the second cumulant.

Inserting the above ansatz into the {\it exact} FRG equations for the cumulants leads to a set of coupled flow equations for 5 functions of one 
field $U_k$, $Z_k$, and $W_{a,b,c;k}$, 4 functions of two fields, $V_k$, $X_{a,b,c;k}$,  and 1 function of three fields, $V_{3k}$. 
It turns out to be more convenient to work with the functions $\Delta_k(\phi_1,\phi_2)=V_k^{(11)}(\phi_1,\phi_2)$, $X_{a;k}(\phi_1,\phi_2)$, 
$X_{b;k}^{(10)}(\phi_1,\phi_2)-X_{c;k}^{(10)}(\phi_1,\phi_2)$, $X_{b;k}^{(11)}(\phi_1,\phi_2)+X_{a;k}^{(11)}(\phi_1,\phi_2)$, and 
$S_{k}(\phi_1,\phi_2,\phi_3)=V_{3k}^{(111)}(\phi_1,\phi_2,\phi_3)$, which in conjunction with the 5 functions of one field present in the first cumulant 
leads to a closed set of FRG equations. Note that we could have alternatively derived these equations by using the FRG in a superfield formalism 
as we did in~[\onlinecite{tissier11,tissier12a,tissier12b}]; the advantage of the latter is to make explicit the Ward identities associated with the 
Parisi-Sourlas SUSY (super-rotations) and the mechanism by which they may break and cease to apply. However, once this information is available, 
it is easier (and fully equivalent) to work with the FRG with conventional replica fields as we do here. 

Finding fixed points that describe scale invariance as well as the spectrum of eigenvalues around them requires casting the resulting FRG flow 
equations in a dimensionless form, as already discussed in Sec.~\ref{sub_RFIM_DE2}. More details are given in Appendix~\ref{app_flow-eqs}. The 
full-blown numerical resolution of the resulting extensive set of coupled partial differential equations is intractable at the present time. However, for 
determining the value of the critical dimension $d_{DR}$ at which DR is broken (and below which SUSY is broken along the FRG flow) 
and to find the spectrum of eigenvalues around the SUSY/DR fixed point, we can restrict ourselves to considering the SUSY/DR fixed point and its 
vicinity when $d \geq d_{\rm DR}$, with $d_{\rm DR}$ yet to be determined.

We have previously shown that the appearance of cusps in the functional dependence of the cumulants of the renormalized random field entails 
the breakdown of SUSY and DR. The cusps appear in the difference between replica fields so that it is convenient to introduce a linear change of field 
arguments: in the second (dimensionless) cumulant, $\varphi=(\varphi_1+\varphi_2)/2$ and $\delta\varphi=(\varphi_1-\varphi_2)/2$ (see 
Sec.~\ref{sub_RFIM_DE2}),  and in the third (dimensionless) cumulant,
\begin{equation}
\begin{aligned}
&\varphi=\frac{\varphi_1+\varphi_2+\varphi_3}{3} \,,\\&
y=\frac{\varphi_1-\varphi_2}{\sqrt 2} \,,\\&
z=\frac{\varphi_1+\varphi_2-2\varphi_3}{\sqrt 6}\,.
\end{aligned}
\end{equation}

An important property of the hierarchy of flow equations for the cumulants of the renormalized random field ($\Gamma_{k2}^{(11)}[\phi_1,\phi_2]$, 
$\Gamma_{k3}^{(111)}[\phi_1,\phi_2,\phi_3]$, etc., and their dimensionless counterparts) when expanded in the differences between replica fields 
is that the sector of equal fields, $\varphi_1=\varphi_2=\varphi_3=\varphi$, $\delta\varphi=y=z=0$, decouples from the sector with nonzero field 
differences, provided that the functional dependence of the cumulants on the latter is regular enough (no cusps). In addition, all functions in the 
sector of equal replica fields can then be related to the functions of the first cumulant as a result of the SUSY Ward identities.\cite{tissier12b} 
As already stressed, the latter are indeed preserved by the present truncation scheme with a choice of IR cutoff functions that satisfy their 
own SUSY Ward identity. In particular, we obtain that if SUSY is valid at the scale $k$,
\begin{equation}
\begin{aligned}
\label{eq_SUSY_ward_DE4_0}
&\delta_{k}(\varphi,\varphi)=z_{k}(\varphi)\\&
x_{a;k}(\varphi,\varphi)=2w_{a;k}(\varphi)\\&
x_{b;k}^{(10)}(\varphi,\varphi)-x_{c;k}^{(10)}(\varphi,\varphi)=w_{b;k}(\varphi)-w'_{a;k}(\varphi)\\&
x_{b;k}^{(11)}(\varphi,\varphi)+x_{a;k}^{(11)}(\varphi,\varphi)=\frac 12 w_{c;k}(\varphi)\\&
s_{k}(\varphi,\varphi,\varphi)=\frac 32 [ w_{b;k}(\varphi)-w'_{a;k}(\varphi)].
\end{aligned}
\end{equation}

With the assumption of a regular enough field dependence, the dimensionless functions can be expanded in the differences between 
replica fields. For the second cumulant, taking into account the symmetries (in particular, $\delta\varphi \to -\delta\varphi$ at constant 
$\varphi$) and the above SUSY related identities, one has
\begin{equation}
\begin{aligned}
\label{eq_DE4_delta}
&\delta_k(\varphi+\delta\varphi,\varphi-\delta\varphi) = z_{k}(\varphi)+ \frac{1}{2} \delta_{k,2}(\varphi)\delta\varphi^2  + \cdots, \\&
x_{a;k}(\varphi+\delta\varphi,\varphi-\delta\varphi)=2w_{a;k}(\varphi) + \frac{1}{2} x_{a;k,2}(\varphi)\delta\varphi^2  + \cdots, \\&
\frac 12[x_{b;k}^{(10)}(\varphi+\delta\varphi,\varphi-\delta\varphi)+x_{b;k}^{(01)}(\varphi+\delta\varphi,\varphi-\delta\varphi)]- \\&
\frac 12[x_{c;k}^{(10)}(\varphi+\delta\varphi,\varphi-\delta\varphi) - x_{c;k}^{(01)}(\varphi+\delta\varphi,\varphi-\delta\varphi)]=\\&
w_{b;k}(\varphi)-w'_{a;k}(\varphi) + \frac{1}{2} x_{e;k,2}(\varphi)\delta\varphi^2 + \cdots, \\&
\end{aligned}
\end{equation}
and
\begin{equation}
\begin{aligned}
\label{eq_DE4_deltabis}
&x_{b;k}^{(11)}(\varphi+\delta\varphi,\varphi-\delta\varphi)+x_{a;k}^{(11)}(\varphi+\delta\varphi,\varphi-\delta\varphi)=\\&
\frac 12 w_{c;k}(\varphi)+ \frac{1}{2} x_{f;k,2}(\varphi)\delta\varphi^2 + \cdots,
\end{aligned}
\end{equation}
when $\delta\varphi\to 0$, where all the functions of $\varphi$ in the right-hand sides are even. For the third cumulant of the random field, 
after taking into account the symmetries (in particular, $y\to -y/2-\sqrt 3 z/2 , z\to \sqrt 3 y/2 - z/2$ at constant $\varphi$,  $y\to -y$ at constant 
$\varphi,z$, and the property that $s_k$ is odd under the global inversion $\varphi\to -\varphi,\,y\to -y,\, z\to -z$), one finds
\begin{equation}
\begin{aligned}
\label{eq_DE4_delta3}
&s_{k}(\varphi +y/\sqrt 2+  z/\sqrt 6,\varphi -y/\sqrt 2+  z/\sqrt 6,\varphi -2  z/\sqrt 6) =\\& 
\frac 32 [ w_{b;k}(\varphi)-w'_{a;k}(\varphi)]+
 \frac{1}{2} s_{k,2}(\varphi)(y^2 +z^2) + \\& \frac 1{3!} s_{k,3}(\varphi)(z^3-3y^2z) + \cdots,
\end{aligned}
\end{equation}
when $y,z \to 0$, where $s_{k,2}$ is an odd function of $\varphi$ and $s_{k,3}(\varphi)$ is even.

The structure of the FRG equations for the cumulants of the random field is such that the sector which is quadratic in the field differences, 
{\it i.e.}, involving the functions $\delta_{k,2}(\varphi)$, $x_{a,e,f;k,2}(\varphi)$, and $s_{k,2}(x=\sqrt 3 \varphi)$, when complemented with the 
cubic term $s_{k,3}(\varphi)$, also decouples from higher orders in field differences. This triangular-like structure of the system of  equations 
likely carries over to all orders of the approximation scheme.

The procedure is then to fix all functions in the sector of equal fields to their fixed-point values. Through the SUSY Ward identities these values 
can all be expressed in terms of the 5 functions, $u_*(\varphi)$, $z_*(\varphi)$, $w_{a,b,c;*}(\varphi)$, which, through the ensuing DR, coincide 
with those computed at the DE4 for the pure $\phi^4$ theory in dimension $d-2$. We next consider the set of coupled flow equations for the 6 
functions $\delta_{k,2}(\varphi)$, $x_{a,e,f;k,2}(\varphi)$, $s_{k,2}(\varphi)$, and $s_{k,3}(\varphi)$:
\begin{equation}
\label{eq_flow_dimensionless_d2DE4}
\begin{aligned}
\partial_t \delta_{k,2}(\varphi)=&
(d-4+2\eta_*) \delta_{k,2}(\varphi)\\&+\frac 12(d-4+\eta_*)\varphi\delta'_{k,2}(\varphi) +\beta_{\delta_2}(\varphi),
\end{aligned}
\end{equation}
\begin{equation}
\label{eq_flow_dimensionless_x_aDE4}
\begin{aligned}
\partial_t x_{a;k,2}(\varphi)=& (d-2+2\eta_*)x_{a;k,2}(\varphi) \\&+ \frac 12(d-4+\eta_*)\varphi x'_{a;k,2}(\varphi) +\beta_{x_{a;2}}(\varphi),
\end{aligned}
\end{equation}
\begin{equation}
\label{eq_flow_dimensionless_x_eDE4}
\begin{aligned}
\partial_t x_{e;k,2}(\varphi)=& \frac 12(3d-8+5\eta_*)x_{e;k,2}(\varphi) \\&+ \frac 12(d-4+\eta_*)\varphi x'_{e;k,2}(\varphi) +\beta_{x_{e;2}}(\varphi),
\end{aligned}
\end{equation}
\begin{equation}
\label{eq_flow_dimensionless_x_fDE4}
\begin{aligned}
\partial_t x_{f;k,2}(\varphi)=& (2d-6+3\eta_*)x_{f;k,2}(\varphi) \\&+ \frac 12(d-4+\eta_*)\varphi x'_{f;k,2}(\varphi) +\beta_{x_{f;3}}(\varphi),
\end{aligned}
\end{equation}
\begin{equation}
\label{eq_flow_dimensionless_s2DE4}
\begin{aligned}
\partial_t s_{k,2}(\varphi)=& \frac 12(3d-8+5\eta_*) s_{k,2}(\varphi) \\&+\frac 12(d-4+\eta_*)\varphi s'_{k,2}(\varphi) +\beta_{s_2}(\varphi),
\end{aligned}
\end{equation}
\begin{equation}
\label{eq_flow_dimensionless_s3DE4}
\begin{aligned}
\partial_t s_{k,3}(\varphi)=&(2d-6+3\eta_*) s_{k,3}(\varphi) \\&+\frac 12(d-4+\eta_*)\varphi s'_{k,3}(\varphi) +\beta_{s_3}(\varphi),
\end{aligned}
\end{equation}
where the beta functions themselves depend on $\delta_{k,2}(\varphi)$, $x_{a,e,f;k,2}(\varphi)$, $s_{k,2}(\varphi)$, $s_{k,3}(\varphi)$, their 
derivatives, and on the DR fixed-point functions. Through this dependence, the above equations are coupled nonlinear second-order partial 
differential equations with a nonlinearity that can be up to cubic in the functions. The equations given above correspond to a zero bare 
temperature for which the subdominant terms in O($\widetilde T_k$) are absent. (The expressions for the beta functions are too long to be 
shown here but they can be systematically and straightforwardly derived with the help of Mathematica.)

As done at the DE2 level (see Sec.~\ref{sub_RFIM_DE2}), we consider both the fixed-point equations obtained by setting the left-hand sides 
of the FRG flow equations to zero and the eigenvalue equations obtained by linearizing the flow equations. Again, we stress that the validity 
of the scaling dimensions used to cast the FRG equations in a dimensionless form is guaranteed by the fact that a {\it bona fide} fixed point can 
actually be found with, in particular, fixed-point solutions for the functions $\delta_{k,2}(\varphi)$, $x_{a;k,2}(\varphi)$,..., $s_{k,3}(\varphi)$. Within 
the spectrum of eigenvalues, $\Lambda_2$ should be the one that starts in $2-\epsilon$ around the upper critical dimension and $\Lambda_3$ 
that starting in $4-2\epsilon$: see Sec.~\ref{sub_RFIM_perturb}. We have also computed the eigenvalue $\Lambda_{3/2}$ associated with cuspy 
perturbations in the second and the third 1-PI cumulants of the renormalized random field.

\begin{figure}
\includegraphics[width=\linewidth]{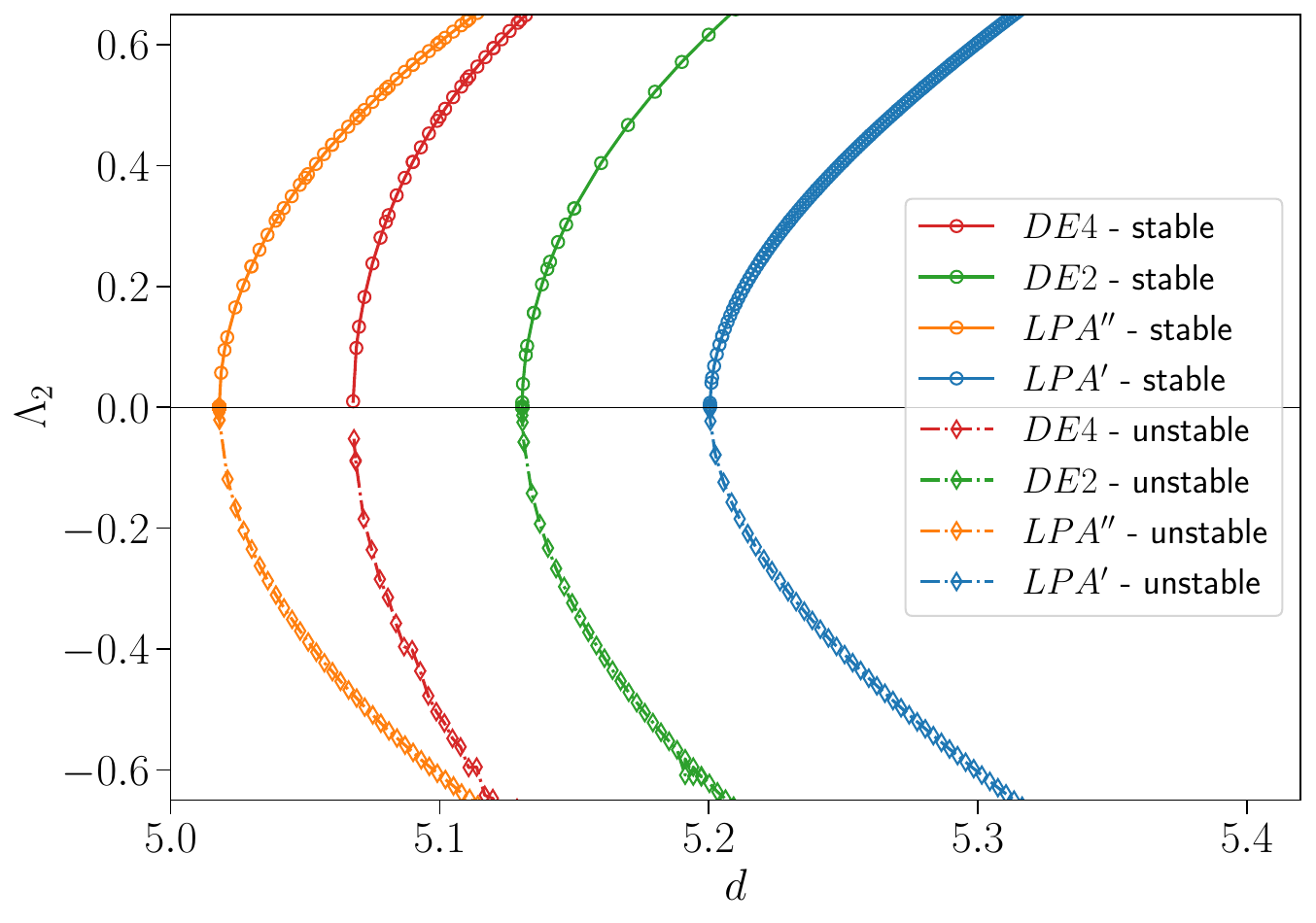}
\caption{Determination of the critical dimension $d_{\rm DR}$ in the RFIM from the vanishing of the eigenvalue $\Lambda_2(d)$ associated with 
the operator $\mathcal F_4$ for the successive nonperturbative FRG approximations LPA', LPA'', DE2, and DE4. The (positive) eigenvalue 
$\Lambda_2(d)$ vanishes as a square-root and collapses with the (negative) eigenvalue $\Lambda'_2(d)$ associated with an unstable SUSY/DR 
fixed point. Below $d_{\rm DR}$, there are no SUSY/DR fixed points. Note that as shown in Fig.~\ref{fig_lambda_RFIM} all the curves for 
$\Lambda_2$ converge to the same value with the same slope when $d\to 6$. 
}
\label{fig_RFIM_dDR}
\end{figure}

Finally, as in [\onlinecite{FPbalog,balog20}] the critical dimension $d_{\rm DR}$ is determined by looking at the vanishing of the eigenvalue 
$\Lambda_2(d)$. It vanishes with a square-root behavior and collapses with the eigenvalue found for a SUSY/DR unstable fixed point. This allows a 
crisp determination of $d_{\rm DR}$: see Fig.~\ref{fig_RFIM_dDR}. Alternatively, $d_{\rm DR}$ can be located as the dimension below which, 
{\it e.g.}, $\delta_{k,2}(\varphi)$ diverges in a finite RG time. As this RG time, which is associated with the Larkin length discussed in 
Sec.~\ref{sub_RFIM_DE2}, diverges when $d\to d_{\rm DR}-$, this procedure is however less accurate than that using the vanishing of $\Lambda_2(d)$.

A heuristic argument extending to the present DE4 approximation why the cuspless SUSY/DR fixed point disappears below the dimension at 
which $\Lambda_2$ vanishes as a result of the nonlinearity of the flow equations in 
Eqs.~(\ref{eq_flow_dimensionless_d2DE4}-\ref{eq_flow_dimensionless_s3DE4}) is given in Appendix~\ref{app_NLFP_DE4}. We stress again that 
the DR fixed point restricted to the 1-replica (first cumulant) sector is of course always present  because it corresponds to the Wilson-Fisher 
fixed point in dimension $d-2$ and can be continued below $d_{\rm DR}$. However, for the RFIM at criticality, all the cumulants with their full 
functional dependence must reach a fixed point. The global RFIM SUSY/DR fixed point then disappears below $d_{\rm DR}$ because it no longer 
exists in the sector associated with the cumulants of order 2 and higher and for field arguments that do not coincide. As already discussed, this 
property is missed by the conventional perturbation approach of KRT.\cite{rychkov_RFIM-II,rychkov_lectures}
\\

\subsection{Results}
\label{sub_DE4results}

We combine the results already obtained at the DE2 level with those that we have presently calculated for the cruder approximation LPA' 
discussed in Sec.~\ref{sub_RFIM_DE2} and for the improved truncation DE4 introduced above. We also consider the improved LPA' approximation, 
which we have called LPA'' (see Sec.~\ref{sub_goal}), where we recall that $z_k(\varphi)$ and $\delta_{k}(\varphi,\varphi)$ are fixed at their 
value in the running minimum of the potential, $\varphi_{\rm{min},k}$, but the $\varphi$-dependence of 
$\delta(\varphi+\delta\varphi,\varphi-\delta\varphi)-\delta_{k}(\varphi,\varphi)$ is not constrained. So defined, this LPA'' does not explicitly 
break SUSY so that, as the other approximation levels, it leads to DR so long as the cuspless fixed point exists. We have already shown the data 
for the eigenvalue $\Lambda_2(d)$ corresponding to all the studied levels of the approximation scheme in Fig.~\ref{fig_lambda_RFIM}(a).

As explained before, we determine the critical dimension $d_{\rm DR}$ by looking at the location where the most dangerous eigenvalue 
$\Lambda_2(d)$ vanishes. The latter does so with a square-root singularity which is associated with the collapse of the stable SUSY/DR fixed 
point with another unstable SUSY/DR fixed point (see above) which we have been able to track in the vicinity of $d_{\rm DR}$. The results are 
displayed in Fig.~\ref{fig_RFIM_dDR}. This allows us to extract the value of $d_{\rm DR}$.  We find $d_{\rm DR}\approx 5.2005$ for the lowest 
order approximation LPA',  $d_{\rm DR}\approx 5.0180$ for the LPA'', and $d_{\rm DR}\approx 5.0678$ for the highest order DE4.

For the previously studied DE2 level of the approximation scheme,\cite{balog20}  we obtain $d_{\rm DR}\approx 5.1307$ through the same procedure. 
We can therefore conclude that the results are robust with respect to the approximation order. We are also able to provide an estimate of 
$d_{\rm DR}$ with, for the first time, an error bar accounting for the 4 levels of approximation: 
\begin{equation}
\label{eq_error-bar}
d_{\rm DR}\approx 5.11 \pm 0.09.
\end{equation}
In addition, it should be noted that the values obtained by increasing the order of the approximation scheme appear {\it to oscillate around 
$5.11$}.\cite{footnote_perturb-fit}

Note that the above values obtained for $d_{\rm DR}$ at the different levels of the nonperturbative approximation scheme are determined with 
a high precision (at least 5 digits) as the location where the eigenvalue $\Lambda_2$ vanishes: see Fig.~\ref{fig_RFIM_dDR}. On the other hand, 
the solution of the FRG equations at each level of approximation somewhat depends on the detailed form of the (dimensionless) regulator functions 
that are introduced to implement the IR cutoff on the functional RG flows: see Sec.~\ref{sub_RFIM_DE2} and Refs.~[\onlinecite{berges02,dupuis_review}]. 
In all our calculations we have used an exponential regulator with a prefactor that is optimized in a dimension near but strictly above $d_{\rm DR}$ 
(separately at each approximation level) according to the principle of minimum sensitivity.\cite{dupuis_review} From the detailed work of 
[\onlinecite{depolsi20}] we expect that the effect of the regulator at each approximation level is within the global error bar obtained from comparing 
different levels [given in Eq.~(\ref{eq_error-bar})].

As already mentioned, the present calculations are nonperturbative but approximate. The robustness and apparent convergence of the predictions 
of the successive orders of the approximation scheme is a strong support for the approach. We have also compared the results obtained in the 
vicinity of $d=6$ to the exact 2-loop calculation. We have fitted our numerical data for $\Lambda_2$, $\Lambda_3$, and $\Lambda_{3/2}$ to a 
quadratic polynomial in $\epsilon=6-d$ near $d=6$:
\begin{equation}
\begin{aligned}
&\Lambda_2(d)=2-\epsilon - a_{2}\epsilon^2 + {\rm O}(\epsilon^3) \\&
\Lambda_3(d)=4-2\epsilon - a_{3}\epsilon^2 + {\rm O}(\epsilon^3) \\&
\Lambda_{3/2}(d)=1-\epsilon/2 - a_{3/2}\epsilon^2 + {\rm O}(\epsilon^3),
\end{aligned}
\end{equation}
where the $\epsilon^2$ coefficients $a_2$, $a_3$ and $a_{3/2}$ are given in Table~\ref{table1} with error bars due to the fitting procedure. We 
find that the numerical values approach the exact ones as the approximation order increases and that they oscillate around the latter (as for the 
value of the critical dimension $d_{\rm DR}$). The relative error at DE4 is less than $20\%$ for $\Lambda_2$ and $\Lambda_3$ and about 
$30\%$ for $\Lambda_{3/2}$.
\\

\begin{table}[]
\caption{$\epsilon^2$ coefficient of the eigenvalues $\Lambda_2$, $\Lambda_3$ and $\Lambda_{3/2}$ from successive orders of the 
nonperturbative FRG approximation scheme, together with the exact 2-loop result.}
\label{table1}
\begin{tabular}{llllllllllllll}
& \\
&Order  \;\;\;\;\;\;\;\;\;\;\;\;\;\;\;\;$\Lambda_2$  \;\;\;\;\;\;\;\;\;\;\;\;\;\;\;\;\;\;\;$\Lambda_3$   \;\;\;\;\;\;\;\;\;\;\;\;\;\;\;\;\;\;\;$\Lambda_{3/2}$ \\&
\\&
  \;LPA'    \;\;\;\;\;\;\;\;$-1.03\pm0.07$                                                         \\&
  \;LPA''   \;\;\;\;\;\;\;$-0.26 \pm 0.01$            					 \;\;\;\;$-0.69\pm0.09$                        \\&
  \;DE2  \;\;\;\;\;\;\;\;\;$-0.40\pm0.02$             					\;\;\;\;$-0.99\pm0.03$                    \;\;\;\;$-0.19\pm 0.01$        \\&
  \;DE4  \;\;\;\;\;\;\;\;\;$-0.25\pm0.02$                				 \;\;\;\;$-0.65\pm0.02$                            \;\;\;\;$-0.10\pm 0.01$    \\&
  Exact \;\;\;\;\;\;\;\,$-8/27\approx 0.30$                      			\, \;\;$-7/9\approx 0.78$                           \;\; \;$-5/36\approx0.14$
\end{tabular}
\end{table}

\section{Conclusion}
\label{sec_conclusion}

By first revisiting the perturbative FRG results for the RFO($N$)M in $d=4+\epsilon$ and then carrying out a more comprehensive 
investigation of the nonperturbative approximation scheme to the FRG of the RFIM, we have put the perturbative results recently 
derived by KRT,\cite{rychkov_RFIM-II,rychkov_lectures} which involve a comprehensive and more rigorous development of Feldman's 
ideas\cite{feldman02} recast within Cardy's parametrization of the RFIM field theory,\cite{cardy_SUSY} in light of our 20-year-old FRG 
description of the breakdown of SUSY and dimensional reduction (DR) in random-field systems.\cite{tarjus04,tarjus-review}

There are two main differences with the treatment of KRT which illustrate the power of the nonperturbative FRG. First, the latter is able to 
describe what happens when SUSY and DR are broken. It indeed predicts a non-SUSY fixed point at which the 1-PI cumulants of the renormalized 
random field display a nonanalytical (``cuspy'') dependence on their field arguments and it provides a physical picture emphasizing the role of 
scale-free collective phenomena that appear in the form of avalanches (at zero temperature) and droplets (at nonzero temperature) at criticality. All 
of this is well supported by state-of-the-art computer simulations. Second, the nonperturbative calculations show that in the critical dimension where 
the eigenvalue associated with the (Feldman) operator which is most dangerous for destabilizing the SUSY/DR fixed point vanishes, there is an 
annihilation of fixed points that leads to the disappearance of the SUSY/DR fixed point below this dimension. This disappearance stems from 
the nonlinear nature of the associated fixed-point equation and is not captured through the perturbative RG and the $\epsilon=6-d$ expansion. 
\\

As the critical change of behavior is predicted to take place near $d_{\rm DR}\approx 5.1$, what are the observable consequences of the 
different scenarios beyond a general compatibility with the main critical behavior obtained in simulation results in $d=4,\,5,\, 6$?

Within the FRG approach of the RFIM we predict that scale-free avalanches are present but have a subdominant effect for $d\geq d_{\rm DR}$ 
and become central to the critical behavior for $d<d_{\rm DR}$. A physical argument relies on comparing the fractal dimension $d_f$ of the 
largest system-spanning avalanches at criticality (in a large but finite system) and the scaling dimension of the spontaneous magnetization 
(times the volume of the system), {\it i.e.}, $d-(d-4+\bar\eta)/2=(d+4-\bar\eta)/2$.\cite{tarjus13} Our prediction, which is substantiated by the 
nonperturbative FRG calculations,\cite{tarjus13,FPbalog,balog_activated,balog20} is that $(d+4-\bar\eta)/2-d_f=0$ when $d<d_{DR}$ and 
$(d+4-\bar\eta)/2-d_f>0$, which explains the subdominant effect of the avalanches, when $d>d_{DR}$. The difference between $(d+4-\bar\eta)/2$ 
and $d_f$ in the latter case is due to the fact that the number of system-spanning critical avalanches scales with the system size as $L^\lambda$ 
with $\lambda=(d+4-\bar\eta)/2-d_f$.\cite{tarjus13} The exponent $\lambda$ precisely coincides with the eigenvalue associated with a cuspy 
perturbation around the SUSY fixed point, $\lambda\equiv \Lambda_{3/2}$.\cite{tarjus13,FPbalog,balog_activated,tissier_pertFRG,balog20} 
From the solution of the mean-field RFIM, one finds that $d_f=4$ and $\lambda=1$,\cite{tarjus13,tarjus-review} and one expects that the same 
values are derived from the field-theoretical version of the RFIM at the upper critical dimension $d_{\rm uc}=6$. Our suggestion is then to carry 
out a simulation of the ground state of the RFIM in $d=5$ and $d=6$ in the presence of an applied field, as in 
[\onlinecite{vives_GS,wu-machta_GS,liu_GS}], and measure the statistics of the system-spanning avalanches near criticality. Doing this, one has 
access to the fractal dimension $d_f$ and, with more difficulty, to the exponent $\lambda= \Lambda_{3/2}$ characterizing the number of these 
avalanches. This type of determination has been for instance attempted for the athermally driven RFIM near its (out-of-equilibrium) critical 
point.\cite{perkovic99}

On the other hand, KRT have suggested that the SUSY fixed point could be reached below the dimension where it becomes unstable, say, 
in $d=4$, by fine-tuning the distribution of the random field.\cite{rychkov_RFIM-II,rychkov_lectures} In contrast, we have shown that, as a 
consequence of the nonlinear nature of the fixed-point equation associated with the dangerous operator $\mathcal F_4$, the SUSY/DR fixed 
point disappears exactly when it becomes unstable. No SUSY/DR fixed point should therefore be found in $d=4$ (nor in $d=5$ but this is too 
close to $d_{\DR}$ to bring any decisive conclusion.\cite{balog20}) This issue is nonetheless hard to settle in computer simulations due to the 
finite-size effects. Indeed, for $d<d_{\rm DR}\approx 5.1$, we find that, even if one starts the FRG flow with initial conditions that are compatible 
with a SUSY/DR fixed point (in a restricted sector of the theory), conditions that involve cuspless cumulants of the random field, a cusp must 
appear along the flow at a scale which by analogy with random manifolds in a disordered environment we associated with a 
``Larkin length''.\cite{wiese_review} This length diverges rapidly as $d\to d_{\rm DR}^-$\cite{FPbalog,balog20} but is finite in $d=4$. It is clear 
that by fiddling with the initial distribution of the quenched disorder one can vary the Larkin length and make it larger, which would allow the RG 
flow to first describe a behavior resembling that predicted by DR while eventually evolving toward the proper cuspy (SUSY- and DR-broken) fixed 
point. The system sizes required for reaching the asymptotic critical behavior may however be out of reach of present-day simulations. 
(A different issue is whether there exists below $d_{\rm DR}$ another, unstable, ``cuspy'' fixed point, at which both SUSY and DR 
break down, on top of the one that we have already found numerically and that describes the critical behavior of the RFIM; we have not carried out 
extensive computations to search for it and we therefore cannot {\it a priori} exclude its presence.)

Finally, as we have previously advocated,\cite{FPbalog,balog_activated,baczyk_LR3D,balog_LR1D} it would be interesting to study by computer 
simulation the long-range RFIM because one may have a direct access to the critical change of behavior between SUSY/DR and non SUSY/DR 
fixed points in $d=3$\cite{baczyk_LR3D,FPbalog} or between cuspless and cuspy fixed points in $d=1$\cite{balog_LR1D} by continuously varying 
the range of the interactions and of the bare random-field correlations. The disappearance or not of the SUSY/DR or cuspless fixed point when it 
is predicted to become unstable could then be more crisply probed.

\acknowledgements{We thank A. Kaviraj, S. Rychkov, and E. Trevisani for useful exchanges and discussions. IB acknowledges the support of 
the Croatian Scientific Foundation grant HRZZ-IP-2022-9423.}

\appendix

\begin{figure}[!t]
\includegraphics[width=.8\linewidth]{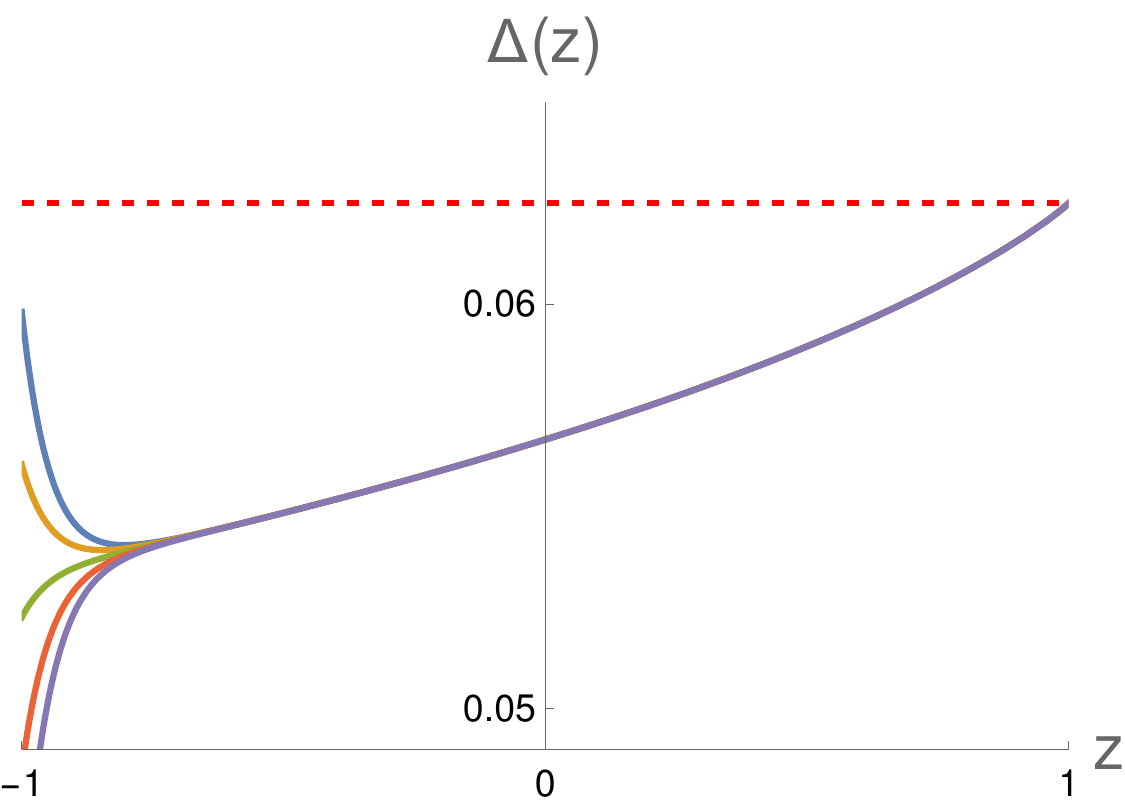}
\includegraphics[width=.8\linewidth]{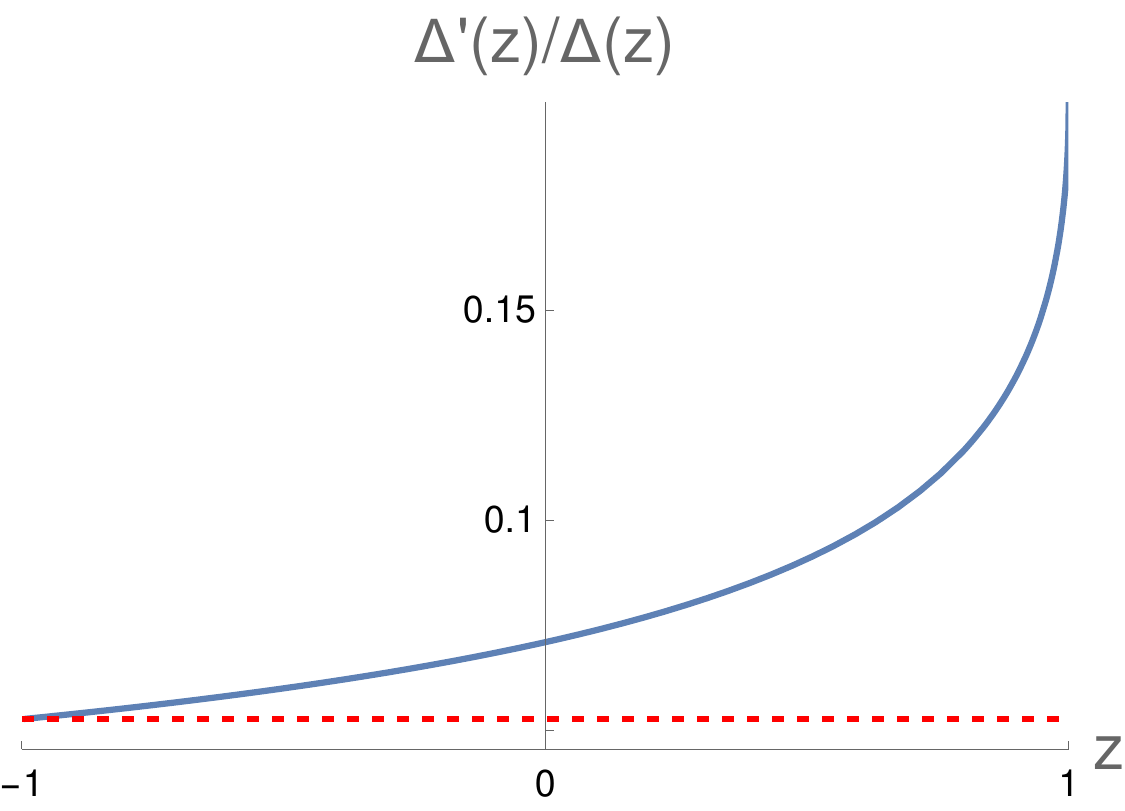}
\includegraphics[width=.8\linewidth]{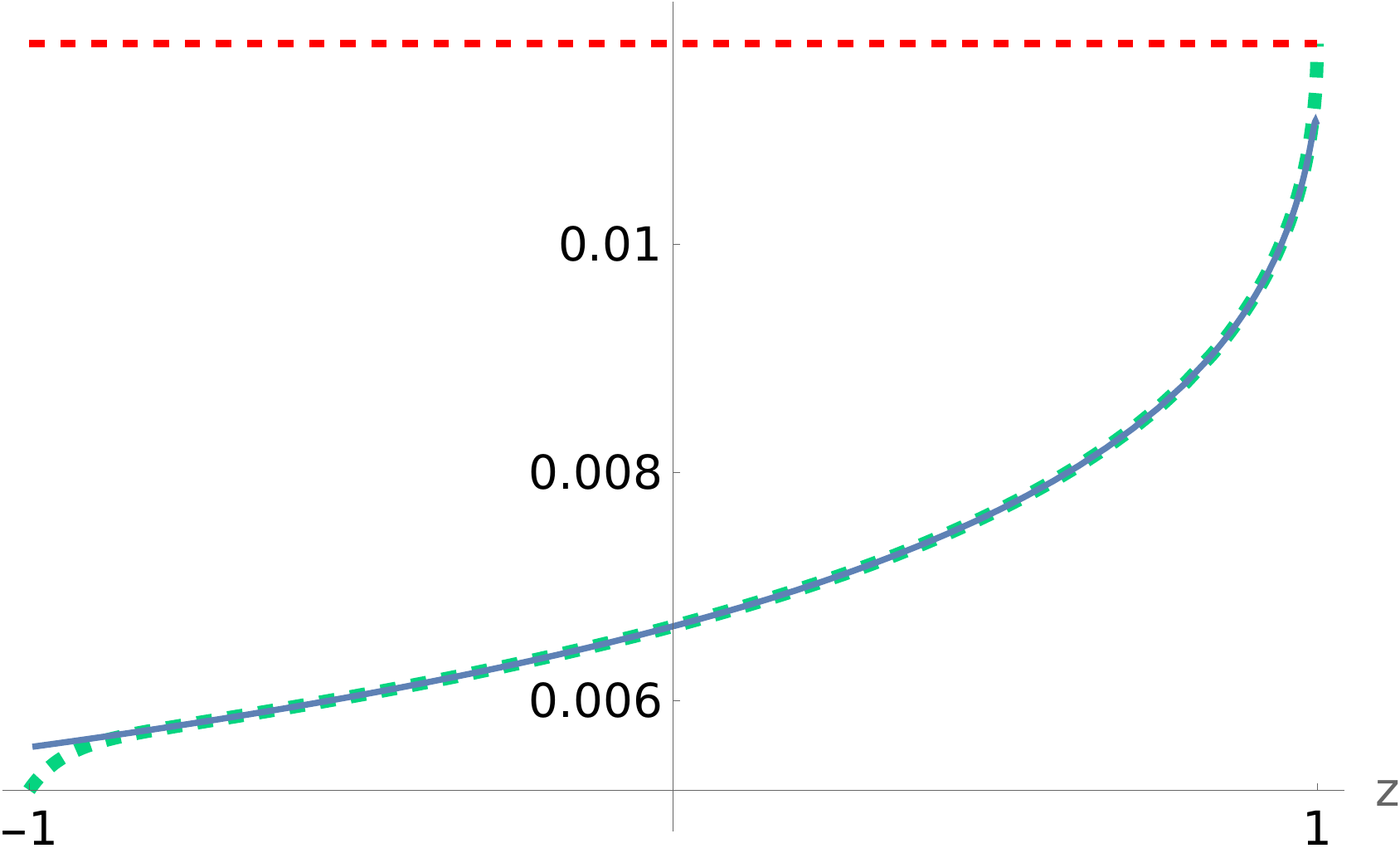}
\caption{Evidence for a subcusp in the SUSY/DR fixed point $\Delta_*(z)$ of the RFO($N$)M at one loop for $N=18$. Top: Function $\Delta(z)$ obtained 
from a Taylor expansion up to $(1-z)^n$ for $n$ between $30$ and $40$ by steps of $2$ (from bottom to top near $z=-1$). The value 
$f(1)=0.011763907083777499$ is optimized for the best apparent convergence near $z=-1$ for $n=34$.
Middle: $\Delta'_*(z)/\Delta_*(z)$ over the interval $[-1,1]$ as obtained from the expansion up to $n=40$ with the same value of $f(1)$ as in (a). The 
horizontal line is the exact value of the ratio $\Delta'_*(-1)/\Delta_*(-1)=1/19$. The curve appears to converge to the exact value but deviates and sharply 
drops as one approaches $z=-1$ because the fine-tuning of $f(1)$ is not precise enough at the level of the 15th digit. 
Bottom: $(\Delta_*(z)-[ \frac 1{16}-\frac 1{80}(1-z)])/(1-z)^{3/2}$ versus $\sqrt{1-z}$ over the whole interval $-1\leq z\leq 1$. The (red) horizontal dashed 
line is equal to $f(1)=0.01176\cdots$, which confirms the presence of a subcusp. The dashed green curve is the result of the expansion (up to order $40$) 
about $z=1$: It coincides with the numerical solution (full line) except when approaching $z=-1$.
}
\label{fig_subcusp_N18}
\end{figure}

\section{Illustration of the presence of a subcusp at the SUSY/DR fixed point of the RFO($N$)M at one loop}
\label{app_RFON_subcusp}

The simplest illustration is to look at the SUSY/DR fixed point in $N=18$. There, the expected subcusp  in the cumulant of the renormalized random 
field $\Delta_*(z)=R'_*(z)$ is in $(1-z)^{3/2}$ when $z\to 1$ (as $\Lambda_{5/2}=0$). We work at 1-loop order and we set $\epsilon\equiv 1$. In 
$N=18$ the two SUSY/DR fixed points coincide ($\Lambda_2=0$) and are unstable with respect to a cuspy perturbation ($\Lambda_{3/2}=-1/10$). 
However, this does not prevent us from finding the fixed point as we fix $\Delta_k(1)$ and $\Delta'_k(1)$ to their known DR fixed-point values, 
{\it i.e.}, for $N=18$ and $\epsilon=1$, $\Delta_k(1)=1/16$ and $\Delta'_k(1)=1/80$. We therefore look for a solution of the form
\begin{equation}
\Delta_*(z)= \frac 1{16}-\frac 1{80}(1-z)+(1-z)^{3/2} f(z) + (1-z)^2 g(z)
\end{equation}
where $f(z)$ and $g(z)$ are regular function of $z$ in the vicinity of $z=1$ [they have a Taylor expansion in powers of $(1-z)$]. The value of 
$f(1)$ is not fixed and should be determined by requiring that the functions $f(z)$ and $g(z)$ are regular over the whole interval of definition 
$[-1,1]$; in particular, the functions should be finite in $z=-1$. As we will see this requirement indeed selects a unique solution.

We study the fixed-point equation in two stages: 

First, we solve the fixed-point equations for $f(z)$ and $g(z)$ in an expansion around $z=1$ to rather high orders of $n$ up to $40$. Solving 
the set of equations at order $n$ gives expressions of all derivatives up to $f^{(n+1)}(1)$ and $g^{(n)}(1)$. They are polynomials of $f(1)$ which 
is the only unknown. The polynomial for $f^{(n+1)}$ is of degree $2(n+1)+1$ and that for $g^{(n)}(1)$ of degree $2(n+1)$. We observe that to 
keep the polynomials ``small enough'' so that the expansions of $f(z)$ and $g(z)$ in powers of $(1-z)$ have a chance to converge over a 
large enough interval one needs to choose the value of $f(1)$ in a very narrow range that drastically shrinks as $n$ increases. For 
$30\leq n\leq 40$ we find that the expansions have apparently converged to unique curves for $z\gtrsim -0.8$ but the last segment down to 
$z=-1$ is sensitive to the 15th significant digit of $f(1)=0.01176390708377\cdots$, as illustrated in Fig.~\ref{fig_subcusp_N18}(a). 
To improve the results one can also include information coming from the exact behavior of $\Delta_*(z)$ near $z=-1$. The value $\Delta_*(-1)$ 
itself is not analytically known but one for instance finds that $\Delta'_*(-1)/\Delta_*(-1)=1/19$. Fig.~\ref{fig_subcusp_N18}(b) shows how 
the approximation of $\Delta'_*(z)/\Delta_*(z)$ for $n=40$ behaves as one approaches $z=-1$ and finally deviates from the exact value. To 
do better one would need to fine-tune the value of $f(1)$ even more precisely.

Next, we numerically solve the fixed-point equation for $\Delta_*(z)$ which is a second-order differential equation. The resolution is performed 
with Mathematica with a working precision of $30$ digits. For boundary values, it is better not to use conditions in $z=1$ which have a 
peculiar character, and we instead consider conditions in $z=0$. We choose $\Delta_*(0)$ and $\Delta'_*(0)$ within $10^{-7}-10^{-9}$ of 
the values obtained through the previous procedure ($z=0$ is well within the region where the expansion has apparently converged) and 
we further fine-tune them so that the function $\Delta_*(z)$ is well-behaved down to $z=-1$. The outcome has a clear subcusp in $(1-z)^{3/2}$, 
as shown in Fig.~\ref{fig_subcusp_N18}(c).
\\

\section{Disappearance of the SUSY/DR fixed point in the RFO($N$)M}
\label{app_RFON}

We consider the mechanism by which the SUSY/DR fixed point in the RFO($N$)M near $d=4$ disappears (or not) at any given loop 
order at the value $N_{\rm DR}$ for which the eigenvalue $\Lambda_2$ associated with Feldman's operator $\mathcal F_4$ vanishes. 
(At 2 loops in $d=4+\epsilon$, $N_{\rm DR}=18- (49/5)\epsilon$.\cite{tissier06b}) We thereby complement the discussion of 
Sec.~\ref{sub_RFON_FP} below Eq.~(\ref{eq_nonlinear_RFO(N)}).

We simplify the notations by replacing $R''(1)$ by $X$ and the polynomial $Q_{N,k}$ by $Q_N$. The fixed-point value $X_*(N)$ is 
then solution of $Q_N(X_*(N))=0$ and the eigenvalue controlling the stability is $\Lambda_2(N)=Q'_N(X_*(N))$. We are interested by 
the behavior in the vicinity of $N_{\DR}$, defined by $\Lambda_2(N_{\rm DR})=0$, when $\delta N=N-N_{\DR} \to 0$.

The coefficients of the polynomial $Q_N(X)$, whose degree depends on the number of loops in the expansion in $\epsilon=4-d$,  are 
expected to be regular in $\delta N$ around $\delta N=0$ (this is for instance verified at 1- and 2-loop levels), with
\begin{equation}
Q_N(X)=Q_{N_{\rm DR}}(X)+ R_{N_{\rm DR}}(X) \delta N+{\rm O}(\delta N^2),
\end{equation}
Furthermore, $Q_N(X)$ can be Taylor expanded as well around $X_{*0}=X_*(N_{\rm DR})$,
\begin{equation}
\begin{aligned}
Q_N(X)=& Q_{N}(X_{*0})+Q'_N(X_{*0})(X-X_{*0})+ \\&
\frac 12 Q''_N(X_{*0}) (X-X_{*0})^2 + {\rm O}((X-X_{*0})^3).
\end{aligned}
\end{equation}
When $\delta N\to 0$, after defining $\delta X_*=X_*(N)-X_{*0}$ and using that by construction 
$Q_{N_{\rm DR}}(X_{*0})=Q'_{N_{\rm DR}}(X_{*0})=0$, we find from the two above expansions that
\begin{equation}
\begin{aligned}
\label{eq_sol}
Q_N(X_*(N))=0=& R_{N_{\rm DR}}(X_{*0}) \delta N+ R'_{N_{\rm DR}}(X_{*0}) \delta N \delta X_*  + \\&
\frac 12Q''_{N_{\rm DR}}(X_{*0})  \delta X_*^2 
+ {\rm O}(\delta N^2,\delta N \delta X_*^2).
\end{aligned}
\end{equation}
For a given (small) $\delta N$ the above equation has two real solutions $\delta X_*(\delta N)$ provided the discriminant
\begin{equation}
\begin{aligned}
D= R'_{N_{\rm DR}}(X_{*0})^2 \delta N^2 -2 R_{N_{\rm DR}}(X_{*0}) Q''_{N_{\rm DR}}(X_{*0}) \delta N 
\end{aligned}
\end{equation}
is positive. The two solutions merge and $D=0$ when $\delta N=0$. When $\delta N\to 0$ the sign of $D$ is generically given by 
the second term of the right-hand side which then changes sign when $\delta N$ changes sign. This implies that the 2 real solutions 
of Eq. (\ref{eq_sol}) annihilate in $N=N_{\rm DR}$ and that there are no real solutions for $N<N_{\rm DR}$. For this not to happen, 
one must have $R_{N_{\rm DR}}(X_{*0}) Q''_{N_{\rm DR}}(X_{*0})=0$, and this may then correspond to a crossing of real solutions 
(depending on the higher-order terms in $\delta N$ and $\delta X_*$) and an exchange of stability of the associated fixed points. The 
condition, however, has no reason to be satisfied in the absence of an additional symmetry. {\it Mutatis mutandis} (replacing $N$ by $d$), 
such a symmetry exists for instance at the SUSY/DR Gaussian fixed point of an elastic interface in a random environment (which then 
corresponds to $X_{*0}=0$). This fixed point then crosses with another fixed point at the upper critical dimension $d=4$ but is still 
present albeit unstable below. However, this nongeneric phenomenon is absent in the RFO($N$)M in $d=4+\epsilon$. One can easily 
check that $R_{N_{\rm DR}}(X_{*0}) Q''_{N_{\rm DR}}(X_{*0}) \neq 0$ at both 1-loop and 2-loop orders.
\\

\section{Brief recap of the FRG formalism for the RFIM}
\label{app_recap}

The starting point of the FRG formalism for the RFIM is an exact RG equation for the scale-dependent effective action (or 
Gibbs free-energy functional) $\Gamma_k[\{\phi_a\}]$ which generates the 1-PI correlation functions when fluctuations are incorporated 
from the microscopic (UV) scale down to an IR cutoff $k$. This equation reads\cite{wetterich93},
\begin{equation}
\begin{aligned}
\label{eq_erge}
\partial_k\Gamma_k\left[\{ \phi_a\}\right ]= \frac{1}{2}\sum_{a,b}  \int_q  \partial_k R_{k,ab}(q^2) \big (\big[ \bm \Gamma _k^{(2)}+ \bm R_k\big]^{-1}\big)_{q,-  q}^{ab},
\end{aligned}
\end{equation}
where  $\bm \Gamma_k^{(2)}$ is the matrix formed by the second functional derivatives of $\Gamma_k$ with respect to the replica fields  
and the operator $\bm P_k[\{ \phi_a\}] \equiv [ \bm\Gamma _k^{(2)}+ \bm R_k]^{-1}$ is the exact propagator at the scale $k$. In the case 
of the RFIM,\cite{tissier11,tissier12a,tissier12b} the IR regulator $R_{k,ab}(q^2)$ is taken as 
$R_{k,ab}(q^2)=\widehat{R}_k(q^2)\delta_{ab} +\widetilde{R}_k(q^2)$. The functions $\widehat{R}_k(q^2)$ and $\widetilde{R}_k(q^2)$ are 
chosen to provide an IR cutoff on the fluctuations, which enforces a decoupling of the low- and high-momentum modes at the scale $k$. 
The function $\widehat{R}_k(q^2)$ adds a mass $\sim k^{2-\eta}$ to replica-field modes with $q^2<k^2$ and decays rapidly to zero for 
$q^2>k^2$, whereas the function $\widetilde{R}_k(q^2)$ (which must be chosen proportional to $-\partial_{q^2}\widehat{R}_k(q^2)$ to avoid 
an explicit SUSY breaking \cite{tissier11,tissier12a,tissier12b}) reduces the fluctuations of the bare random field.

The 1-PI cumulants $\Gamma_{kp}[\phi_{a1},\cdots,\phi_{ap}]$ are generated by the expansion of $\Gamma_k[\{\phi_a\}]$ 
in increasing number of free replica sums which is given in Eq.~(\ref{eq_expansion_cumulants}). When inserted in Eq.~(\ref{eq_erge}) this 
provides an infinite hierarchy of exact functional RG equations for the 1-PI cumulants, whose first equations are for instance given in 
Appendix C of [\onlinecite{tissier12b}] and not reproduced here. 

To study scale invariance at criticality and describe the associated fixed point one needs to introduce scaling dimensions 
and dimensionless quantities. As stressed in several places in this manuscript, the RFIM fixed point is at zero dimensionless renormalized 
temperature, {\it i.e.}, the latter flows to zero as one approaches the fixed point: $\widetilde T_k\sim k^\theta T$ with $\theta>0$ as $k\to 0$ 
and $T$ the bare temperature. Temperature is thus irrelevant, albeit dangerously so, because one cannot simply set it to zero in the 
effective action. Each 1-PI cumulant indeed scales differently with temperature,
\begin{equation}
\label{eq_rescaled-cumulant}
\Gamma_{kp}[\phi_{a1},\cdots,\phi_{ap}]=\frac 1{\widetilde T_k^p} \bigg[\gamma_{kp}[\varphi_{a1},\cdots,\varphi_{ap}] +
{\rm O}(\widetilde T_k)\bigg ] ,
\end{equation}
where the $\gamma_{kp}$'s have a finite limit when $\widetilde T_k \to 0$. In the above expression the replica fields are rescaled as 
$\varphi_a(\tilde x)=k^{-(d-2+\eta)/2}\widetilde T_k^{1/2}\phi_a(x)$ $\forall a$, with $\tilde x=k x$ [see Eq.~(\ref{eq_dim1})]. One can   
check that this leads to a consistent hierarchy of dimensionless exact FRG equations for the 1-PI cumulants 
$\gamma_{kp}[\varphi_{a1},\cdots,\varphi_{ap}]$ that allows one to describe the zero-temperature fixed point of the critical RFIM.

As discussed in Sec.~\ref{sub_dangerous}, near the upper critical dimension $d=6$, the canonical dimensions of the coupling 
constants and the associated operators are obtained by first expanding in number of free replica sums  [Eq.~(\ref{eq_expansion_cumulants})] 
and then expanding the resulting cumulants in powers of the replica fields or of the difference between replica fields as in Feldman's operators.  
The canonical dimensions follow from the above, with the anomalous dimensions $\eta=\bar\eta=0$ and, as a result, $\theta=2$. One counts 
the power of the inverse temperature with dimension $\theta=2$, which depends on the cumulant that is considered [see 
Eq.~(\ref{eq_rescaled-cumulant})],  hence in the number of free replica sums involved, and the power of the replica fields or replica field 
differences with dimension $(d-4)/2$.
\\

\section{Multi-copy formalism and relation to the operator classification in~[\onlinecite{rychkov_RFIM-II}] }
\label{app_SUSYequiv}

\subsection{Averaging over disorder}

There are several ways to handle the quenched disorder in the RFIM in order to generate an effective  disorder-averaged theory. One is 
the conventional replica formalism in which one introduces $n$ replicas of the original system, averages over the disorder, and takes the limit 
$n\to 0$. Another one is the Parisi-Sourlas SUSY method that starts from the functional minimization equation describing the ground state 
of the system at zero temperature. (Both methods have their limitations, {\it e.g.}, if  replica permutational symmetry or SUSY are  
broken.) There are also ways to access the cumulants of the random free-energy functionals by introducing replicas or copies of the original 
system which are coupled to different applied sources (replica symmetry is then explicitly broken). This  can be combined with the 
Boltzmann-Gibbs distribution at equilibrium in a replica field theory,\cite{tarjus04,tissier06} a minimization equation at zero temperature 
in a superfield theory,\cite{tissier11,tissier12a,tissier12b} and a Langevin equation describing the dynamics in a dynamical field 
theory.\cite{balog_activated,balog_dynamics} In the following, we discuss the replica field formalism (which we abbreviate as ``rep-f'' below), 
the Parisi-Sourlas SUSY formalism (abbreviated simply as ``SUSY'') and the replica superfield theory (in the limit which was 
called ``Grassmannian ultralocality'' in [\onlinecite{tissier11,tissier12a,tissier12b}]) which we denote by ``rep-sf''.

For the bare $\phi^4$ RFIM theory, the corresponding actions read
\begin{equation}
\begin{aligned}
S_{\rm{rep-f}}=&\sum_a\int_x \big[\frac 12 (\partial\phi_a)^2+\frac 12 r\phi_a^2+\frac 1{4!} g\phi_a^4 \big] \\&
-\frac 12 \Delta \sum_{ab}\int_x \phi_a\phi_b,
\end{aligned}
\end{equation}
where  $\phi_a$ is a (replica) field depending on the space coordinate $x$, and
\begin{equation}
\begin{aligned}
S_{\rm{SUSY}}=\int_{x \theta\bar\theta}\big[-\frac 12 \Phi\Delta_{\rm{SUSY}}\Phi+\frac 12 r\phi^2+\frac 1{4!} g\Phi^4 \big] ,
\end{aligned}
\end{equation}
\begin{equation}
\begin{aligned}
S_{\rm {rep-sf}}= & \sum_a \int_{x \theta_a\bar\theta_a}\big[\frac 12 (\partial_\mu \Phi_a)^2+\frac 12 r\Phi_a^2
+\frac 1{4!} g\Phi_a^4 \big] \\&-\frac \Delta2\sum_{ab}\int_{x \theta_a\bar\theta_a\theta_b\bar\theta_b}\Phi_a\Phi_b ,
\end{aligned}
\end{equation}
where $\Phi$ is a superfield that depends on $x$ and two Grassmann coordinates $\theta$ and $\bar\theta$ and $\Phi_a$ is 
a (replica) superfield that depends on $x$ and two Grassmann coordinates $\theta_a$ and $\bar\theta_a$.  

The superfields can be expanded in their Grassmann coordinates,
\begin{equation}
\begin{aligned}
\label{eq_expand}
&\Phi=\phi(x)+\bar\theta\psi(x)+\bar\psi(x)\theta+\bar\theta\theta\hat\phi(x)\\&
  \Phi_a=\phi_a(x)+\bar\theta_a\psi_a(x)+\bar\psi_a(x)\theta_a+\bar\theta_a\theta_a\hat\phi_a(x).
\end{aligned}
\end{equation}
Finally, the supersymmetric Laplacian $\Delta_{\rm{SUSY}}=\partial_\mu^2+\Delta\partial_\theta\partial_{\bar\theta}$ involves 
the standard Euclidean Laplacian and derivatives with respect to the Grassmann coordinates.

Starting from the replica field approach, Cardy found a linear transformation of the fields that allows a diagonalization of the 
quadratic part in the limit $n\to 0$. The new fields introduced by Cardy are
\begin{equation}
\begin{aligned}
&\hat\phi=\frac 12(\phi_1+\frac1{1-n}\sum_{a=2}^n\phi_a),\\&  
\phi=\frac 12(\phi_1-\frac1{1-n}\sum_{a=2}^n\phi_a),
\end{aligned}
\end{equation}
as well as $(n-2)$ fields $\chi_i$, $i=3,\cdots,n$, which have no component along the $\phi_1$ field and which 
are orthogonal to $\phi$ and $\hat\phi$. In order to simplify the final expressions, it is often convenient to introduce an extra field 
$\chi_{2}$ (so that we now have $(n-1)$ $\chi_i$  field) which satisfies $\sum_{i=2}^{n}\chi_i=0$. The sum over the $(n-1)$ indices 
$i$ is denoted by $\sum'_i=\sum_{i=2}^{n}$.

In this new set of variables, the relation with the Parisi-Sourlas approach is striking: The fields $\phi$ and $\hat\phi$ closely resemble 
the fields $\phi$ and $\hat\phi$ of the SUSY formalism and, in the limit $n\to 0$, a loop of the $(n-2)$ bosonic fields $\chi_i$ has the 
same contribution asa pair of Grassmann ghost fields $\psi$ and $\bar\psi$. We stress that Cardy's approach is just a rewriting of the 
replica formalism in the limit $n\to 0$, so that the outcome should be the same as with a conventional definition of the replica fields.

\subsection{Dimensionless quantities}

We now discuss the different ways of introducing dimensionless quantities. For simplicity, but without altering the main point to be 
made, we do not consider anomalous dimensions.

\subsubsection{Parisi-Sourlas SUSY}

The coordinates are rescaled as
\begin{equation}
\begin{aligned}
  x=k^{-1}\tilde x\qquad
  \theta=k^{-1}\tilde \theta\qquad
  \bar\theta=k^{-1}\tilde{\bar\theta},
\end{aligned}
\end{equation}
so that the measure of integration and the Laplacian change to
\begin{equation}
\begin{aligned}
&\int_{x\theta\bar\theta}=k^{-d+2}\int_{\tilde x\tilde{\theta}\tilde{\bar\theta}}\\&
  \Delta_{\rm{SUSY}}=k^2\tilde\Delta_{\rm{SUSY}}.
\end{aligned}
\end{equation}
The scaling dimension of the field is then fixed by requiring that the kinetic term remains equal to unity, which imposes
\begin{equation}
\begin{aligned}
  \Phi=k^{\frac {d-4}2}\tilde \Phi .
\end{aligned}
\end{equation}
This implies that the various fields appearing in the decomposition of Eq.~(\ref{eq_expand}) transform as
\begin{equation}
\begin{aligned}
  \phi=k^{\frac {d-4}2}\tilde \phi \qquad  \psi=k^{\frac {d-2}2}\tilde \psi \qquad  \bar\psi=k^{\frac {d-2}2}\tilde \bar\psi \qquad  \hat\phi
  =k^{\frac {d}2}\tilde{\hat\phi}.
\end{aligned}
\end{equation}
The rescaling of the coupling constants therefore comes as
\begin{equation}
\begin{aligned}
  \tilde r=r k^{-2},\qquad  \tilde g=g k^{d-6}.
\end{aligned}
\end{equation}
Note that the variance of the random field, $\Delta\equiv 1$, is unchanged under rescaling.

\subsubsection{Replica superfields}

Not much needs to be said in this case. The rescaling of the coordinates and fields is the same as in the Parisi-Sourlas SUSY 
formalism. We observe that the variance of the random field comes with 2 integrals over Grassmann coordinates and 2 Grassmann 
derivatives in the SUSY formalism, while it appears with 4 Grassmann integrals in the replica superfield  formalism. But, scalingwise, 
this is the same.

\subsubsection{Replica fields}

In the replica field formalism, it was understood a long time ago that there exists a dangerously irrelevant variable. The most 
convenient way to account for this is to rewrite the action by multiplying the 1-replica part by one power of an inverse temperature 
$\beta$, the 2-replica part by two powers, etc.,
\begin{equation}
\begin{aligned}
S_{\rm {rep-f}}=&\beta\sum_a\int_x \big [\frac 12 (\partial\phi_a)^2+\frac 12 r\phi_a^2+\frac 1{4!} g\phi_a^4\big ] \\&
-\frac {\beta^2}{2} \Delta \sum_{ab}\int_x \phi_a\phi_b .
\end{aligned}
\end{equation}

In the vicinity of the fixed point, the running dimensionless inverse temperature $\tilde\beta_k$ behaves as $k^{-2}$, {\it i.e.}, the inverse 
temperature has scaling dimension $D_\beta=2$ (and temperature a dimension of $D_T=-2$). It is then necessary to rescale the field with 
powers of the (inverse) temperature. The fields transform as
\begin{equation}
\begin{aligned}
 \phi_a=  k^{\frac{d-2}2} \sqrt{\tilde\beta_k} \,\tilde \phi_a ,
\end{aligned}
\end{equation}
which yields
\begin{equation}
\begin{aligned}
  \phi_a=k^{\frac{d-4}2}\tilde \phi_a .
\end{aligned}
\end{equation}
This coincides with the scaling dimension of the fields $\phi$ in the SUSY and the replica superfield approaches. Comparing with the 
latter formalisms, the inverse temperature plays the role of two Grassmann integrations or two Grassmann derivatives.

\subsubsection{Cardy's parametrization}

The quadratic part of the action now reads
\begin{equation}
\begin{aligned}
  \int_x 2\partial_\mu\hat\varphi\partial_\mu\vp+\frac 12\sum_i'(\partial_\mu\chi_i)^2-2\Delta \hat\varphi^2+ 
  r[2\hat\varphi \vp+\frac 12\sum_i'\chi_i^2]
\end{aligned}
\end{equation}
where we recall that $\sum_i'=\sum_{i=2}^{n}$. This form enables one to determine the scaling dimension of the fields as
\begin{equation}
\begin{aligned}
\vp=k^{\frac{d-4}2}\tilde{\vp}\qquad  
\chi=k^{\frac{d-2}2}\tilde\chi\qquad  
\hat\varphi=k^{\frac{d}2}\tilde{\hat\varphi} .\qquad  
\end{aligned}
\end{equation}
These dimensions coincide with those found in the SUSY formalism. Similarly, one can rewrite the $\phi_a^4$ interaction in terms 
of Cardy's fields and one finds
\begin{equation}
\begin{aligned}
\sum_a\phi_a^4=& 8\hat\varphi\vp(\vp^2+\hat\varphi^2)+6(\vp-\hat\varphi)^2\sum_i'\chi_i^2
\\& -4(\vp-\hat\varphi)\sum_i'\chi_i^3+2\sum_i'\chi_i^4 .
\label{eq_phi4_cardy}
\end{aligned}
\end{equation}
Not all terms have the same dimension but consideration of the leading ones indicates that one must rescale the coupling 
constant according to $\tilde g=g k^{d-6}$.

\subsection{Feldman's operators}

\subsubsection{Microscopic realization}

For any integer $p>0$, we now consider a coupling to quenched disorder that, in the action, takes the form
\begin{equation}
\begin{aligned}
  S_{\rm{dis}}=-\sum_{i=1}^{2p} \int_x \sigma_i(x) \phi^i(x)
\end{aligned}
\end{equation}
where the $\sigma_i$'s are Gaussian-distributed quenched random variables with zero mean and variances given by
\begin{equation}
\begin{aligned}
  \overline {\sigma_i(x)\sigma_j(y)}=\Delta_{2p} (-1)^iC_{2p}^i\delta(x-y)\delta_{i+j,2p}.
\end{aligned}
\end{equation}
Repeating the construction of the replica action leads to terms in the 2-replica part of the form
\begin{equation}
\begin{aligned}
  S_{2p}=-\frac 12 \Delta_{2p}\sum_{ab}\int_x(\phi_a-\phi_b)^{2p},
\end{aligned}
\end{equation}
which are precisely the operators considered by Feldman:\cite{feldman02} see Eq.~(\ref{eq_Feldman_op}). (Note that so defined the 
variance matrix has a positive determinant but is not positive definite. In a proper treatment this should be corrected by higher-order terms.)

If we implement the Parisi-Sourlas construction in this case, we find that no terms are generated, except for $p=1$ which corresponds 
to the random field (up to a coupling to a random temperature which is here of no interest to us). The terms with $p>1$ can thus be 
associated with the SUSY null and SUSY nonwritable terms of ~[\onlinecite{rychkov_RFIM-II}]. In what follows, we focus on these terms. 
Before averaging over disorder, one has contributions proportional to the auxiliary fields that read
\begin{equation}
\begin{aligned}
  e^{\hat\phi \sum_{i=0}^{2p} \ i \sigma_i\phi^{i-1}+ \sum_{i=1}^{2p}\  i (i-1)\bar\psi\psi\sigma_i\phi^{i-2}} .
\end{aligned}
\end{equation}
After averaging, one obtains for $p>1$ terms proportional to $\hat\phi^2$,
\begin{equation}
\begin{aligned}
&\hat\phi^2  \sum_{i=1}^{2p-1}i(2p-i)\phi^{i-1}\phi^{2p-i-1}(-1)^i C_{2p}^i \\&
=\hat\phi^2 \phi^{2p-2}2p(2p-1)\sum_{j=0}^{2p-2}(-1)^{j+1}C_{2p-2}^j \\&
=-\hat\phi^2 \phi^{2p-2}2p(2p-1)(1-1)^{2p-2} =0 .
\end{aligned}
\end{equation}
There is also a term in $\hat\phi\bar\psi\psi$ which vanishes for the same reason and a term in $(\bar\psi\psi)^2$ which is zero 
due to the anti-commuting property.

In the replica superfield construction, we obtain 
\begin{equation}
\begin{aligned}
S_{2p}=-\frac 12 \Delta_{2p}\int_{x\theta_a\bar\theta_a\theta_b\bar\theta_b}\sum_{ab}(\Phi_a-\Phi_b)^{2p} .
\end{aligned}
\end{equation}
Performing the Grassmann integrals then yields
\begin{equation}
\label{eq_Sp}
\begin{aligned}
& S_{2p}= -p(2p-1) \Delta_{2p}(\phi_a-\phi_b)^{2p-4}\int_{x}\sum_{ab}\Big[-\hat\phi_a\hat\phi_b \,\times \\&
(\phi_a-\phi_b)^{2} +2(p-1)\bar\psi_a\psi_a\hat\phi_b(\phi_a-\phi_b)-2(p-1)\bar\psi_b\psi_b\hat\phi_a  \\&
\times (\phi_a-\phi_b) +2(p-1)(2p-3)\bar\psi_a\psi_a\bar\psi_b\psi_b\Big]   .
\end{aligned}
\end{equation}
We observe that if we put all replica fields $\phi_a$ equal and all replica fields $\psi_a$ equal, the expression in Eq.~(\ref{eq_Sp}) 
vanishes. This is why the Parisi-Sourlas SUSY construction cannot describe such contributions.

Finally, in terms of Cardy's fields, one finds contributions that start with the $2p$th powers of $\chi$. In particular,
\begin{equation}
\begin{aligned}
S_4=-\frac {\Delta_4}2\int_x \Big [&6(\sum'_i\chi_i^2)^2-16\hat\varphi\sum'_i\chi_i^3+48\hat\varphi^2\sum'_i\chi_i^2 \\& 
-32\hat\varphi^4 \Big ]
\end{aligned}
\end{equation}
and 
\begin{equation}
\begin{aligned}
S_6=&-\frac {\Delta_6}2\int_x \Big [30(\sum'_i\chi_i^4)(\sum'_i\chi_i^2)-20(\sum'_i\chi_i^3)^2 \\&
- 24\hat\varphi\sum'_i\chi_i^5+120\hat\varphi^2\sum'_i\chi_i^4 -320\hat\varphi^3\sum'_i\chi_i^3 \\&
+ 480\hat\varphi^4\sum'_i\chi_i^2-128\hat\varphi^6 \Big ].      
\end{aligned}
\end{equation}

\subsubsection{Dimensionless coupling constants $\tilde\Delta_p$}

In the replica field formalism, $\Delta_{2p}$ appears at the 2-replica level, {\it i.e.}, the second cumulant, which comes with a factor  
$\beta^2$. We deduce that
\begin{align}
  \tilde \Delta_{2p}&=\frac{\Delta_{2p}}{\tilde\beta_k^2} k^{-d}\left[k^{\frac{d-4}2}\right]^{2p}\\
  &=\Delta_{2p} k^{\left(d-4\right)(p-1)}.
  \label{eq_our_scaling}
\end{align}
In the replica superfield formalism, we define the rescaled variable as
\begin{align}
  \tilde\Delta_{2p}&=\Delta_{2p} k^{4-d}k^{p(d-4)}\\
  &=\Delta_{2p} k^{(d-4)(p-1)} ,
  \label{eq_our_scaling2}
\end{align}
which is compatible with the previous result.

Note that in the two above methods, all terms multiplying the coupling constant $\Delta_{2p}$ have the same dimension. This is no 
longer true in Cardy's field parametrization. Keeping only the leading term in $p>1$, we conclude that
\begin{align}
  \tilde\Delta_{2p}&=\Delta_{2p} k^{-d}k^{(d-2)p}\\
  &=\Delta_{2p} k^{d(p-1)-2p} ,
  \label{eq_their_scaling}
\end{align}
which does not coincide with the previous result, except for $p=2$. This is surprising because Cardy's formalism is just a rewriting of the 
replica action. As already pointed out, we observe that, in the replica superfield formalism, the terms with $p\geq2$ vanish if we put all 
the replica fields equal, which is, in some sense what is done in Cardy's formalism.

\subsection{3-replica operator}

For later use, it is interesting to consider a generalization of Feldman's operators that involves 3 replicas,
\begin{equation}
  \tilde S_6=w_6\sum_{abc}\int_x (\phi_a-\phi_b)^2(\phi_b-\phi_c)^2(\phi_c-\phi_a)^2 ,
\end{equation}
which can be rewriten in Cardy's formalism as
\begin{equation}
\begin{aligned}
&\tilde S_6=-6w_6\int_x \Big [(\sum_i'\chi_i^2)^3-4\hat\varphi (\sum_i'\chi_i^3)(\sum_i'\chi_i^2)\\&
+4\hat\varphi^2\big[\sum_i'\chi_i^4+3(\sum_i'\chi_i^2)^2\big]-16\hat\varphi^3\sum_i'\chi_i^3
  +16\hat\varphi^4\sum_i'\chi_i^2 \Big ].
\end{aligned}
\end{equation}

When introducing dimensionless coupling constants in the replica field formalism, we must do it such that
\begin{align}
  \tilde w_6=k^{2d-6}w_6 .
\end{align}
We then obtain exactly the same scaling in Cardy's approach. It is interesting to observe that, in the latter, $\Delta_6$ and $w_6$ 
have the same dimension because the leaders have both 6 powers of $\chi$ while, in the replica field approach, they differ by a 
power of $k^2$ because one is a 2-replica interaction and the other a 3-replica one.

\subsection{Examples of Feynman diagrams}

We now discuss a consistency check for determining scaling dimensions through the Feynman diagrams. To illustrate our argument, 
we look at Feynman diagrams that contribute to the renormalization of $\Delta_6$. Of course, $\Delta_6$ is an irrelevant coupling constant 
that does not need to be renormalized in order to make the theory finite. Our point is to show how different formalisms (replica field and 
Cardy) treat them.

In the replica field theory, one can build a diagram with three 4-point vertices of the 1-replica action (the term proportional to $g$). This
diagram contributes to the renormalization of $\Delta_6$ if one uses two disconnected propagators and one connected one.

\begin{figure}
\includegraphics[width=.3\linewidth]{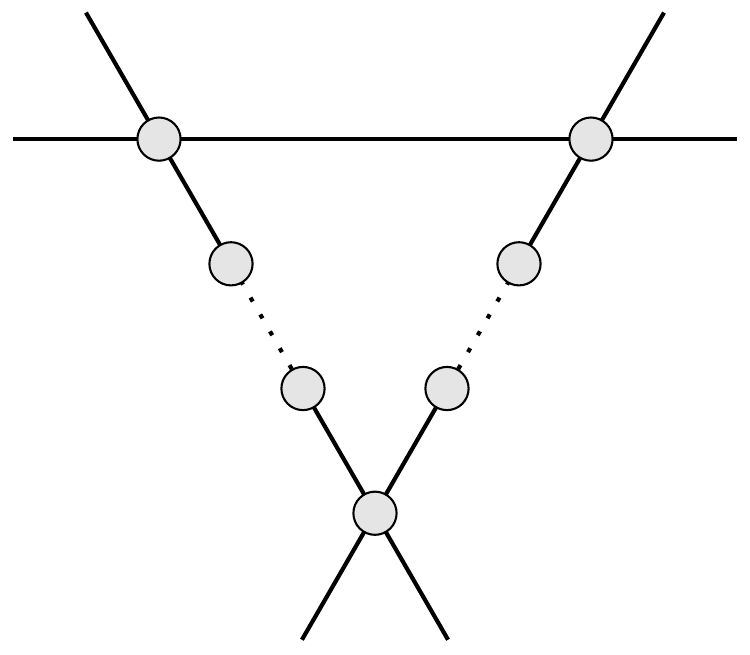}
\includegraphics[width=.3\linewidth]{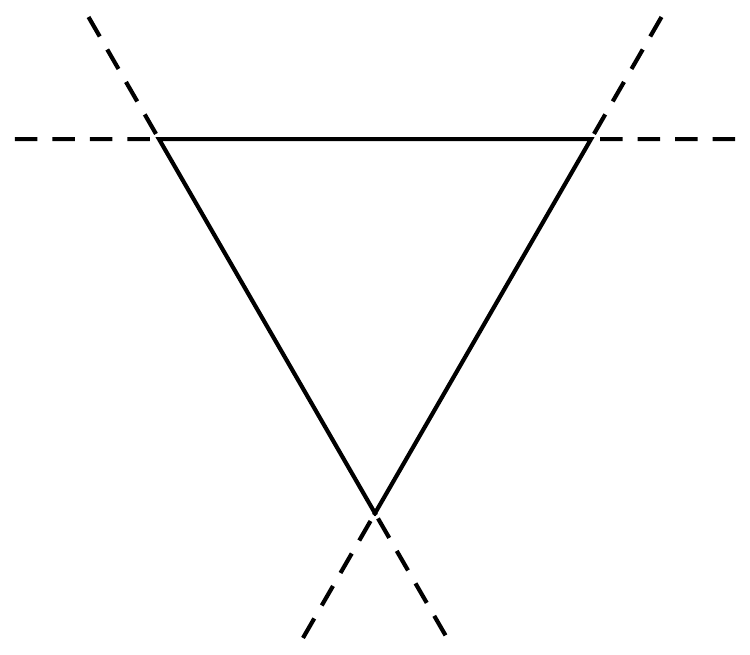}
\includegraphics[width=.3\linewidth]{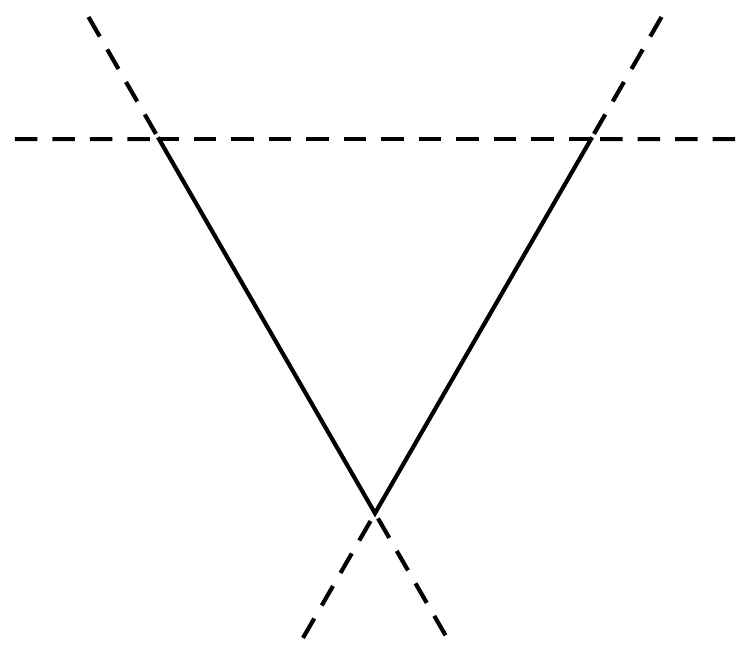}
\caption{One-loop Feynman diagrams with three 4-point vertices possibly contributing to the renormalization of $\Delta_6$ in different 
formalisms. Left: Our calculation with explicit replica symmetry breaking which scales as $k^{8-2d}$; the full line is a connected propagator 
and the two other edges involving a dashed line and two full lines correspond to disconnected propagators.
Middle: The calculation of KRT with Cardy's fields~[\onlinecite{rychkov_RFIM-II}]; the incoming dashed segments correspond to $\chi$ 
fields and a full line correspond to a $\langle\varphi\varphi\rangle$ propagator; this diagram scales as $k^{6-2d}$ but actually does not 
contribute to the renormalization of $\Delta_6$ but to that of $w_6$ which is a 3-replica and not a 2-replica quantity.
Right: The proper diagram expressed with Cardy's fields that contributes to the renormalization of $\Delta_6$; it has two 
$\langle\varphi\varphi\rangle$ propagators and one $\langle\chi_i\chi_j\rangle$ propagator and scales as $k^{8-2d}$, as the diagram on 
the left. (Because $\langle\chi_i\chi_j\rangle$ is not purely diagonal, the diagram also contributes to the renormalization of $w_6$ but 
it is subdominant compared to the middle diagram.)
}
\label{fig_diagrams}
\end{figure}

This diagram when evaluated in zero external momenta reads
\begin{equation}
  g^3 \int_q\frac 1{(q^2+r)^5}\, .
\end{equation}
Note that the integral is both IR and UV finite.  After introducing dimensionless variables, one finds that this diagram scales as
$k^{3(6-d)}k^d k^{-10}$. The first contribution corresponds to the scaling of the 3 powers of $g$, the second to the scaling of 
the integration measure, and the last to the propagators. One therefore obtains a factor of $k^{2(4- d)}$ which coincides with 
the scaling found in the replica field approach for the operator coupling constant $\Delta_6$, see Eq.~(\ref{eq_our_scaling}).

We now reproduce the calculation with Cardy's fields. In order to renormalize $\Delta_6$, one needs to draw a Feynman diagram with 
only$\chi$ external legs. The leading terms in the 1-replica 4-point interaction comes with two powers of $\chi$ and 2 powers of $\varphi$. 
Wetherefore expect that the relevant diagram has three $\langle\varphi\varphi\rangle$ propagators. In total, once dimensionless
coupling constants are introduced, the diagram scales as $k^{3(6-d)}k^d k^{-12}=k^{6-2d}$, which corresponds to the 
scaling in Eq.~(\ref{eq_their_scaling}).

The previous argument, however, has a flaw because, in the Feynman diagram considered above, all external $\chi$ legs arising from the
different 4-point interactions have independent indices. Stated otherwise, this particular diagram renormalizes something proportional to 
$(\sum_i'\chi_i^2)^3$, which corresponds to $w_6$ and not to $\Delta_6$. To renormalize the Feldman operator, one needs to
connect the indices of the external $\chi$ legs. This is possible if one uses subleading terms in Eq.~(\ref{eq_phi4_cardy}), and 
more specifically the term in $\varphi\sum_i'\chi_i^3$, in two of the three interaction terms appearing in the diagram. For the third interaction 
term one can use the leading contribution, $\varphi^2\sum_i'\chi_i^2$. The diagram now has one $\langle\chi_i\chi_j\rangle$ propagator, 
which contains a piece with $\delta_{ij}$ and therefore renormalizes $\Delta_6$, and two $\langle\varphi\varphi\rangle$ propagators. After 
introducing the dimensionless variables, the diagram is found to scale as $k^{3(6-d)}k^d k^{-10}=k^{4-2d}$, which now coincides with the 
scaling of Eq.~(\ref{eq_our_scaling}) in the replica field approach.

This is illustrated in Fig.~\ref{fig_diagrams}. 

The above development shows that care should be exerted when dealing with operators involving sums over replicas (and their 
associated coupling constants). The scaling obtained in the FRG, with the replica field or superfield approach and the 
introduction of a running dimensionless temperature that flows to zero, for casting the functional flow equations in a dimensionless form  
appears fully consistent, either nonperturbatively or via an expansion in Feynman diagrams. Is it possible, on the other hand, that the use of 
Cardy's field parametrization with no explicit reference to a renormalized temperature might run into difficulties? The jury is still out, as there 
may be subtleties coming from accidental cancellations, the mixing of leaders and followers, and the possible difficulty to directly 
compare results obtained in the 1-PI formalism with those obtained at the level of the renormalized action.
\\

\section{Disappearance of the SUSY/DR fixed point in the RFIM in the NP-FRG approximation DE2}
\label{app_NLFP_DE2}

We start from the fixed-point equation for $\delta_{*,2}$ in Eq.~(\ref{eq_delta2}) which we reproduce below, making explicit the 
dependence on $\epsilon=d-d_{\rm DR}$. We assume that there is a $d_{\rm DR}$ in which $\Lambda_2=0$, so that
\begin{equation}
\label{eq_delta2_FP_app}
0= A_*(\varphi;\epsilon)\delta_{*, 2}( \varphi;\epsilon)^2+ L_*(\varphi,\partial_\varphi,\partial^2_\varphi;\epsilon) \delta_{*, 2}( \varphi;\epsilon)
+ B_*(\varphi;\epsilon),
\end{equation}
where we recall that the functions $u''_k(\varphi)$, $z_k(\varphi)$, $\delta_{k,0}(\varphi)$, and the anomalous dimensions are fixed 
at their SUSY/DR fixed-point expressions.

We also consider the eigenvalue equation, Eq.~(\ref{eq_Lambda2_DE2}), in $\epsilon=0$:
\begin{equation}
\begin{aligned}
\label{eq_lambda2_app}
&\lambda_0 f_{\lambda_0}(\varphi)= \\&
2A_*(\varphi;0)\delta_{*, 2}( \varphi;0)f_{\lambda_0}(\varphi)+ L_*(\varphi,\partial_\varphi,\partial^2_\varphi;0) f_{\lambda_0}(\varphi).
\end{aligned}
\end{equation}
By construction, all eigenvalues are $>0$ (irrelevant) except one, equal to $\Lambda_2$ that vanishes; the corresponding eigenfunction 
is then denoted $f_0(\varphi)$. Because $\delta_2(\varphi)$ is an even function we restrict ourselves to even eigenfunctions.

As mentioned in the main text, $ A_*(\varphi;\epsilon)$, $B_*(\varphi;\epsilon)$, and the linear operator 
$L_*(\varphi,\partial_\varphi,\partial^2_\varphi;\epsilon)$ are regular function of $\epsilon$,
\begin{equation}
\label{eq_delta2_FP_app}
A_*(\varphi;\epsilon)=A_*(\varphi;0) + \epsilon \dot A_*(\varphi;0) + {\rm O}(\epsilon^2),
\end{equation}
etc., where a dot denotes a derivative with respect to $\epsilon$. Note that $A_*\neq0$.

We now expand the fixed-point function $\delta_{k,2}(\varphi;\epsilon)$ around $\delta_{k,2}(\varphi;0)$. We do so by using the basis formed 
by the eigenfunctions in $\epsilon=0$, {\it i.e.},
\begin{equation}
\label{eq_FP_expansion_app} 
\delta_{*, 2}( \varphi;\epsilon)=\delta_{*, 2}(\varphi;0) + c_0(\epsilon)f_0(\varphi)+\sum_{\lambda_0>0}  c_{\lambda_0}(\epsilon)f_{\lambda_0}(\varphi),
\end{equation}
where $c_0(\epsilon)\to 0$ and $c_{\lambda_0}(\epsilon)\to 0$ as $\epsilon\to 0$. It is expected that the coefficients associated with the 
irrelevant directions in $\epsilon=0$ are regular at small $\epsilon$, $c_{\lambda_0}(\epsilon)=\epsilon \tilde c_{\lambda_0} +\cdots$ for 
$\lambda_0>0$, while the coefficient along the zero mode may behave in a singular way as $\epsilon\to 0$.

Inserting the above results and definitions into Eq.~(\ref{eq_delta2_FP_app}) leads to
\begin{equation}
\label{eq_delta2_FP_expand}
\begin{aligned}
-\epsilon F(\varphi)=&\epsilon \sum_{\lambda_0>0}  \lambda_0 \tilde c_{\lambda_0}f_{\lambda_0}(\varphi) + 
A_*(\varphi;0)c_0(\epsilon)^2 f_0(\varphi)^2 \\& +{\rm O}(\epsilon^2,\epsilon c_0(\epsilon)).
\end{aligned}
\end{equation}
where the function $F(\varphi)=\dot A_*(\varphi;0)\delta_{*, 2}( \varphi;0)^2
+\dot L_*(\varphi,\partial_\varphi,\partial^2_\varphi;0) \delta_{*, 2}( \varphi;0)+\dot B_*(\varphi;0)$ is supposed to be known and 
$A_*(\varphi;0)\neq 0$. Generically, one expects the solution of the above equation for $\epsilon >0$, {\it i.e.}, $d<d_{\rm DR}$, to have $c_0(\epsilon)^2\propto \epsilon$, which implies a square-root behavior. This behavior however cannot carry over to $\epsilon<0$ and, 
$c_0$ being continuous in $\epsilon$, the solution therefore disappears when $d<d_{\rm DR}$.

The above argument is not rigorous but is indicative of what the generic behavior should be.
\\

\section{NP-FRG flow equations at DE4 approximation level}
\label{app_flow-eqs}

We start from the exact FRG flow equations for the first cumulants which are given in Appendix C of [\onlinecite{tissier12b}] and 
we insert the DE4 truncation defined in Eqs.~(\ref{eq_ansatz_gamma1}-\ref{eq_ansatz_gammap}). This provides a closed set of partial 
differential equations for the functions $U_k(\phi_1)$, $Z_k(\phi_1)$, $W_{a,b,c;k}(\phi_1)$, $V_k(\phi_1,\phi_2)$, $X_{a,b,c;k}(\phi_1,\phi_2)$, 
and $V_{3k}(\phi_1,\phi_2,\phi_3)$. It is actually more convenient to work with the cumulants of the renormalized random field, 
$\Delta_k(\phi_1,\phi_2)=V_k^{(11)}(\phi_1,\phi_2)$ and $S_k(\phi_1,\phi_2,\phi_3)=V_{3k}^{(111)}(\phi_1,\phi_2,\phi_3)$, which also 
lead to a closed set of flow equations.

The next step is to introduce scaling dimensions and dimensionless quantities, along the procedure described in and around 
Eqs.~(\ref{eq_dim1}) and (\ref{eq_dim2}). The resulting dimensionless FRG flow equations can then be symbolically written as
\begin{equation}
\label{eq_flow_dimensionless_uDE4}
\begin{aligned}
\partial_t u'_k(\varphi)=&-\frac 12(d-2\eta_k+\bar\eta_k)u'_k(\varphi)\\&+\frac 12(d-4+\bar\eta_k)\varphi u''_k(\varphi)+ \beta_{u'}(\varphi) 
+{\rm O}(\widetilde T_k),
\end{aligned}
\end{equation}
\begin{equation}
\label{eq_flow_dimensionless_zDE4}
\begin{aligned}
\partial_t z_k(\varphi)&=\eta_k z_k(\varphi) \\&+\frac 12(d-4+\bar\eta_k)\varphi z'_k(\varphi) + \beta_{z}(\varphi) +{\rm O}(\widetilde T_k),
\end{aligned}
\end{equation}
\begin{equation}
\label{eq_flow_dimensionless_waDE4}
\begin{aligned}
\partial_t w_{a;k}(\varphi)&=(2+\eta_k)w_{a;k}(\varphi) \\&
+\frac 12(d-4+\bar\eta_k)\varphi w'_{a;k}(\varphi)+ \beta_{w_{a}}(\varphi) +{\rm O}(\widetilde T_k),
\end{aligned}
\end{equation}
\begin{equation}
\label{eq_flow_dimensionless_wbDE4}
\begin{aligned}
\partial_t w_{b;k}(\varphi)&=\frac 12(d+2\eta_k+\bar\eta_k)w_{b;k}(\varphi) \\&
+\frac 12(d-4+\bar\eta_k)\varphi w'_{b;k}(\varphi)+ \beta_{w_{b}}(\varphi) +{\rm O}(\widetilde T_k),
\end{aligned}
\end{equation}
\begin{equation}
\label{eq_flow_dimensionless_wcDE4}
\begin{aligned}
\partial_t w_{c;k}(\varphi)&=(d-2+\eta_k+\bar\eta_k)w_{c;k}(\varphi) \\&
+\frac 12(d-4+\bar\eta_k)\varphi w'_{c;k}(\varphi)+ \beta_{w_{c}}(\varphi) +{\rm O}(\widetilde T_k),
\end{aligned}
\end{equation}
\begin{equation}
\label{eq_flow_dimensionless_deltaDE4}
\begin{aligned}
&\partial_t \delta_k(\varphi_1,\varphi_2)=(2\eta_k-\bar\eta_k)\delta_k(\varphi_1,\varphi_2) +\frac 12(d-4+\bar\eta_k)\\&
\times (\varphi_1\partial_{\varphi_1}+
\varphi_2\partial_{\varphi_2})\delta_k(\varphi_1,\varphi_2) +\beta_{\delta}(\varphi_1,\varphi_2) +{\rm O}(\widetilde T_k),
\end{aligned}
\end{equation}
\begin{equation}
\label{eq_flow_dimensionless_x_abcDE4}
\begin{aligned}
&\partial_t x_{a,b,c;k}(\varphi_1,\varphi_2)=(2+2\eta_k-\bar\eta_k)x_{a,b,c;k}(\varphi_1,\varphi_2) \\&
+ \frac 12(d-4+\bar\eta_k)(\varphi_1\partial_{\varphi_1}+
\varphi_2\partial_{\varphi_2})x_{a,b,c;k}(\varphi_1,\varphi_2) \\&+\beta_{x_{a,b,c}}(\varphi_1,\varphi_2) +{\rm O}(\widetilde T_k),
\end{aligned}
\end{equation}
and
\begin{equation}
\label{eq_flow_dimensionless_sDE4}
\begin{aligned}
&\partial_t s_k(\varphi_1,\varphi_2,\varphi_3)=\frac 12 (d+6\eta_k-3\bar\eta_k)s_k(\varphi_1,\varphi_2,\varphi_3) \\&+
\frac 12(d-4+\bar\eta_k)(\varphi_1\partial_{\varphi_1}+\varphi_2\partial_{\varphi_2}+\varphi_3\partial_{\varphi_3})
s_k(\varphi_1,\varphi_2,\varphi_3)\\& +\beta_{s}(\varphi_1,\varphi_2,\varphi_3) +{\rm O}(\widetilde T_k),
\end{aligned}
\end{equation}
where, we recall, $t=\log(k/k_{\rm UV})$ and $k_{\rm UV}$ is a UV cutoff associated with the microscopic scale of the model. The beta 
functions themselves depend on $u_k'$, $z_k$, $w_{a,b,c;k}$, $\delta_k$, $x_{a,b,c;k}$, $s_k$ and their derivatives, and they depend 
as well on the (dimensionless) regulator functions that are introduced to implement the IR cutoff on the functional RG flows. In addition, 
the running anomalous dimensions $\eta_k$ and $\bar\eta_k$ are fixed by the conditions $z_k(0)=\delta_k(0,0)=1$. 
The O($\widetilde T_k$) terms are subdominant when one approaches the fixed point as $\widetilde T_k$ goes to zero as 
$k^\theta$. They are also exactly zero when the bare temperature $T$ is set to zero. The resulting zero-temperature beta functions are sufficient 
to study the fixed point and the spectrum of eigenvalues around it. The expressions for these beta functions, and if needed for the 
subdominant terms  proportional to $\widetilde T_k$, are obtained via Mathematica. They are too long to be displayed in this appendix but 
a Mathematica notebook can be made available for anyone interested in using the equations.

Note also that the SUSY Ward identities for equal replica fields that are discussed in the main text are preserved by the above flow 
equations, provided (i) they are satisfied in the initial condition, (ii) one works at zero bare temperature, and (iii) the 6 functions whose flows 
are given in Eqs.~(\ref{eq_flow_dimensionless_d2DE4}-\ref{eq_flow_dimensionless_s3DE4}) indeed reach a fixed point.
\\

\section{Disappearance of the SUSY/DR fixed point in the RFIM at the DE4 NP-FRG approximation level}
\label{app_NLFP_DE4}

We group the 6 functions $\delta_{k,2}(\varphi)$, $x_{a;k,2}(\varphi)$, $x_{e;k,2}(\varphi)$, $x_{f;k,2}(\varphi)$, $s_{k,2}(\varphi)$, and 
$s_{k,3}(\varphi)$ in a vector ${\bf X}_k(\varphi)$ and we fix all the functions $u_k''(\varphi)$, $z_k(\varphi)$, $w_{a,b,c;k}(\varphi)$, and, 
via SUSY [see Eq.~(\ref{eq_SUSY_ward_DE4_0})],  $\delta_{k}(\varphi,\varphi)$, $x_{a;k}(\varphi,\varphi)$, 
$x_{b;k}^{(10)}(\varphi,\varphi)-x_{c;k}^{(10)}(\varphi,\varphi)$, $x_{b;k}^{(11)}(\varphi,\varphi)+x_{a;k}^{(11)}(\varphi,\varphi)$, and 
$s_{k}(\varphi,\varphi,\varphi)$, as well as the anomalous dimensions of the field, to their SUSY/DR fixed-point expressions. Then, making 
explicit the linear and nonlinear parts of Eqs.~(\ref{eq_flow_dimensionless_d2DE4}-\ref{eq_flow_dimensionless_s3DE4}), 
one can rewrite the flow equations in the vicinity of the putative critical dimension $d_{\rm DR}$ as
\begin{equation}
\begin{aligned}
\label{eq_flowX}
&\partial_t  X_{\alpha,k}(\varphi;\epsilon)=C_{\alpha}(\varphi;\epsilon)+
L_{\alpha\beta}(\varphi,\partial_\varphi,\partial_\varphi^2;\epsilon)X_{\beta,k}(\varphi;\epsilon)+\\&
A_{\alpha\beta\gamma}(\varphi;\epsilon)X_{\beta,k}(\varphi;\epsilon)X_{\gamma,k}(\varphi;\epsilon) 
 + B_{\alpha\beta\gamma\delta}(\varphi;\epsilon)X_{\beta,k}(\varphi;\epsilon) \times \\&
 X_{\gamma,k}(\varphi;\epsilon) X_{\delta,k}(\varphi;\epsilon),
\end{aligned}
\end{equation}
where $\alpha=1,\cdots, 6$, summation over repeated indices is implied, $\epsilon=d-d_{\rm DR}$, 
$L_{\alpha\beta}(\varphi,\partial_\varphi,\partial_\varphi^2;\epsilon)$ is a linear operator, and $A_{\alpha\beta\gamma}(\varphi;\epsilon)>0$. 
The cubic term actually only appears in the equation for $X_6\equiv s_3$. 

From Eq.~(\ref{eq_flowX}) one obtains the fixed-point equations by setting the left-hand side to zero and the eigenvalue equations by 
linearizing the equations for a small perturbation around the fixed point. When $\epsilon=0$ and with 
$X_{\alpha,k}(\varphi;0)=X_{\alpha,*}(\varphi;0)+k^{\lambda_0} f_{\lambda_0,\alpha}(\varphi;0)$, this leads to
\begin{equation}
\begin{aligned}
\label{eq_eigenvalueX0}
&\lambda f_{\lambda_0,\alpha}(\varphi)=L_{\alpha\beta}(\varphi,\partial_\varphi,\partial_\varphi^2;0)f_{\lambda_0,\alpha}(\varphi)
+[A_{\alpha\beta\gamma}(\varphi;0) + \\&
A_{\alpha\gamma\beta}(\varphi;0)] X_{\beta,*}(\varphi;0)f_{\lambda_0,\beta}(\varphi)  + [B_{\alpha\beta\gamma\delta}(\varphi;0)
+B_{\alpha\beta\delta\gamma}(\varphi;0) \\&
+B_{\alpha\delta\gamma\beta}(\varphi;0)] X_{\beta,*}(\varphi;0)X_{\gamma,*}(\varphi;0)f_{\lambda_0,\delta}(\varphi),
\end{aligned}
\end{equation}
where by definition of $d_{\rm DR}$ ({\it i.e.}, $\epsilon=0$), one eigenvalue is zero and the others are strictly positive.

We work in the limit of vanishingly small $\epsilon$. As for the DE2 case (see Appendix~\ref{app_NLFP_DE2}) we expand the difference 
between the fixed-point functions in $\epsilon$ and those in $\epsilon=0$ in the basis formed by the eigenfunctions of the linearized equations 
in $\epsilon=0$. We expect that $L_{\alpha\beta}$, $A_{\alpha\beta\gamma}$, $B_{\alpha\beta\gamma\delta}$, $C_{\alpha}$ are regular 
functions of $\epsilon$ around $\epsilon=0$, with $L_{\alpha\beta}=L_{\alpha\beta}(\epsilon=0)+\epsilon \dot L_{\alpha\beta}(\epsilon=0)+\cdots$, 
and similarly for the other functions. We also expect that the coefficients of the expansion of ${\bf X}_{*}(\varphi;\epsilon) -{\bf X}_{*}(\varphi;0)$ 
along the eigenfunctions with strictly positive (irrelevant) eigenvalues are also regular in $\epsilon$, {\it i.e.},
\begin{equation}
\begin{aligned}
X_{\alpha,*}(\varphi;\epsilon)& - X_{\alpha,*}(\varphi;0) = \\& 
\epsilon \sum_{\lambda_0>0} \tilde c_{\lambda_0,\alpha\beta}f_{\lambda_0,\beta}(\varphi) + c_{0,\alpha\beta}(\epsilon)f_{0,\beta}(\varphi),
\end{aligned}
\end{equation}
where the $c_{0,\alpha\beta}$'s go to $0$ in a possibly singular way as $\epsilon \to 0$.

The equations for the fixed point then become, for $\alpha=1,\cdots, 6$,
\begin{equation}
\begin{aligned}
\label{eq_FPX}
&-\epsilon \big [ F_\alpha(\varphi)-\sum_{\lambda_0>0} \lambda_0 \tilde c_{\lambda_0,\alpha\beta}f_{\lambda_0,\beta}(\varphi) \big ]= \\&
A_{\alpha\beta\gamma}(\varphi;0)c_{0,\beta\beta'}(\epsilon)c_{0,\gamma\gamma'}(\epsilon)
f_{0,\beta'}(\varphi)f_{0,\gamma'}(\varphi)
\end{aligned}
\end{equation}
up to a ${\rm O}(\epsilon c_0(\epsilon), c_0(\epsilon)^3)$, with
\begin{equation}
\begin{aligned}
&F_\alpha(\varphi)=\\&{\dot L}_{\alpha\beta}(\varphi,\partial_\varphi,\partial_\varphi^2;0)X_{\beta,*}(\varphi;0)+
{\dot A}_{\alpha\beta\gamma}(\varphi;0)X_{\beta,*}(\varphi;0)  \\& \times
X_{\gamma,*}(\varphi;0) + {\dot B}_{\alpha\beta\gamma\delta}(\varphi;0)X_{\beta,*}(\varphi;0)X_{\gamma,*}(\varphi;0) X_{\delta,*}(\varphi;0) 
\\&+\dot C_{\alpha}(\varphi;0) 
\end{aligned}
\end{equation}
a known function. Generically, in the absence of accidental (or symmetry-induced) cancellation, one expects that the solution of the above 
set of equations which is supposed to exist for $\epsilon\geq 0$ entails that the $c_{0,\alpha\beta}(\epsilon)$'s behave as $\sqrt{\epsilon}$ 
when $\epsilon\to 0$. This however cannot apply when $\epsilon<0$ and the solution should then generally disappear.
\\

\section{Digressions on the peculiarities of a zero-temperature critical fixed point in the presence of disorder and on the unusual mechanism 
by which the SUSY/DR fixed point disappears}

\subsection{On nonanalytic operators at a zero-temperature fixed point with disorder}

In their paper KRT\cite{rychkov_RFIM-II}  raise a criticism to the work in which we study the relevance of the avalanches at the critical 
point of the RFIM by a 2-loop perturbative FRG in $\epsilon=6-d$\cite{tissier_pertFRG}. To do so we considered a functional {\it nonanalytic} 
perturbation around the SUSY (zero-temperature) fixed-point effective action and computed its dimension in perturbation. This perturbation 
is generated by the presence of scale-free avalanches and creates a cusp in the second 1-PI cumulant of the renormalized random field. 
Because we work with the effective action, {\it i.e.}, the 1-PI generating functional, what we actually computed is the 
eigenvalue $\Lambda_{3/2}$ associated with the perturbation around the fixed point. The odd feature pointed out 
by KRT is that, contrary to what is usually found for {\it analytic} fixed-point theories (see, {\it e.g.}, [\onlinecite{cardy_textbook}]), the 
two-point correlation function of the nonanalytic (fluctuating) operator $\mathcal O_{3/2}(x)$ {\it a priori} corresponding to the perturbation 
does not show the power-law spatial dependence at long distance with an exponent $2(d-\Lambda_{3/2})$,  {\it i.e.}, 
$\langle \mathcal O_{3/2}(x)\mathcal O_{3/2}(y)\rangle_c\sim 1/\vert x-y\vert^{2(d-\Lambda_{3/2})}$, which one would anticipate. 

As we have already stressed, the connection between eigenvalues around the fixed point, scaling dimensions, and large-distance spatial 
of correlation functions is rather subtle in the RFIM at criticality. In our approach, this is due to the zero-temperature nature of the fixed point. 
Difficulties already arise in the treatment of analytical (polynomial) operators, as discussed in Sec.~\ref{sub_dangerous}, but it appears even 
more striking when considering nonanalytical perturbations of the SUSY/DR fixed point. More explicitly, in our perturbative FRG treatment in 
$d=6-\epsilon$, we considered a perturbation in the renormalized 1-PI second  cumulant of the form
\begin{equation}
-\frac{w_k}4 \int_x \sum_{a,b}\varphi_a(x)\varphi_b(x) \vert \varphi_a(x)-\varphi_b(x)\vert, \nonumber
\end{equation}
which is associated with a cusp in the second cumulant of the renormalized random field (obtained by deriving with respect to $\varphi_a$ 
and $\varphi_b$). We showed through a 2-loop calculation that $w_k$ is an irrelevant coupling constant going as $k^{\Lambda_{3/2}}$ when 
$k\to 0$ with $\Lambda_{3/2}=1-\epsilon/2-5\epsilon^2/36 + {\rm O}(\epsilon^3)$ (see Sec.~\ref{sub_RFIM_perturb} and 
Fig.~\ref{fig_lambdas_RFIM}). However, the pair correlation function of the corresponding fluctuating operator, 
$\langle \sum_{a,b}\vert \varphi_a(x)-\varphi_b(x)\vert^3\sum_{c,d} \vert \varphi_c(y)-\varphi_d(y)\vert^3 \rangle_c$, even after having taken 
care of the factors of $n$ coming from the number of replicas, does not seem to have a leading behavior at large separation in 
$1/\vert x-y\vert^{2(d-\Lambda_{3/2})}$ as is usually found from the relation between scaling dimension and eigenvalue.\cite{cardy_textbook} 
This was illustrated by KRT on a toy model.\cite{rychkov_RFIM-II} We have repeated the calculation around the Gaussian fixed point (at one 
loop so that there is no anomalous dimension of the field) by using a different treatment of the nonanalyticity at zero temperature that was 
introduced by Chauve {\it et al.} for the FRG of a disordered elastic manifold.\cite{FRGledoussal-chauve}  We have obtained the very same 
result as KRT.\cite{rychkov_RFIM-II} The nonanalytical operator seems to be decomposed at long distance into powers of analytical operators 
with no sign of the anomalous eigenvalue $\Lambda_{3/2}$. The conditions for relating the eigenvalue around the fixed point with the power-law 
spatial decay as in the usual (heuristic) derivation of  [\onlinecite{cardy_textbook}] are manifestly not satisfied in the presence of  
nonanalyticities in a zero-temperature fixed-point theory.

This observation points to the rather unique features associated with such nonanalyticities. The peculiar behavior already appears  at the 
UV/Gaussian level: One cannot derive an effective action with a nonanalyticity such as that in the above expression of the 1-PI second cumulant 
by simply adding to the bare action a non-Gaussian term with a nonanalytic functional dependence and doing a perturbation expansion around 
the Gaussian theory. Any naive approach of this kind gives back the DR Gaussian theory. This points to a nontrivial connection between 
1-PI quantities in the effective action and fluctuating operators in the action whenever nonanalyticities are present in the former. 

These specificities have been discussed in great detail by Chauve {\it et al.} in the context of the FRG for a disordered elastic 
manifold.\cite{FRGledoussal-chauve} This is clearly unusual compared to more standard cases such as the pure $\phi^4$ theory and comes 
from the fact that the long-distance physics of disordered systems such as the RFIM and random elastic manifolds is affected by the presence 
of singular collective phenomena in the form of avalanches.

\subsection{Unusual mechanism of the change of fixed point at $d_{\rm DR}$}

In his lectures on the RFIM\cite{rychkov_lectures}  Rychkov raises additional concerns about our nonperturbative FRG approach.

One question is about the unusual mechanism by which the new cuspy fixed point emerges from the two merging DR/SUSY fixed points when 
$d<d_{\rm DR}$. For the lowest eigenvalue governing the stability of the fixed point, this leads to a square-root singularity in 
$\sqrt{\vert d_{\rm DR}-d\vert}$ associated with a discontinuity at $d_{\rm DR}$. This peculiar behavior is indeed unseen in other models. It 
results from the intrinsically functional nature of the RG description of the RFIM at criticality and from the fact that the limit $d\to d_{\rm DR}^-$ 
is nonuniform in the field dependence of the FRG functions associated with the second and higher-order cumulants of the renormalized disorder. 
The new fixed point actually appears through a boundary-layer mechanism in these functions. This is unconventional (as is the zero-temperature 
and functional nature of the RFIM fixed points\cite{villain84,fisher86,cardy_textbook}), but the phenomenon has been well characterized 
in our previous papers: see [\onlinecite{FPbalog,balog20}]. 

A related issue is the existence and the nature of the unstable SUSY/DR fixed point which we predict to exist above $d_{DR}$ and which 
then coalesces with the stable SUSY/DR fixed point in $d=d_{DR}$, leading to its disappearance. The main critical scaling behavior, which 
corresponds to the SUSY/DR sector for equal replica fields, is identical for the stable and the unstable fixed points. But the two fixed points 
differ when considering the functions that appear in the second and higher-order cumulants of the renormalized random field when the 
replica-field arguments are different, {\it e.g.}, $\delta_{*,2}(\varphi)$ in Eq.~(\ref{eq_cuspless_delta}). From the comparison with the pattern 
as a function of $N$ in the RFO($N$)M in $d=4+\epsilon$ (see Fig.~\ref{fig_alpha_RFON} bottom), we expect that the unstable fixed point 
is unobservable because somehow ``infinitely unstable'' close to $d=6$. This would explain why a previous numerical search for 
finding this fixed point directly in $d=6$ was unsuccessful.\cite{tissier_pertFRG} (Note also that the fixed point is not perturbatively reachable in $d=6$.) 
It only becomes ``accessibly unstable'' when approaching $d_{\rm DR}$. It remains that this, admittedly physically odd, unstable fixed point is 
mathematically well defined as another solution of the fixed-point equations near $d_{\DR}$ when $\Lambda_2 \to 0$: see Fig.~\ref{fig_RFIM_dDR}.
\\


\begin{thebibliography}{15}
\bibitem{footnote_SUSY_FP}
It should be stressed that, even in the domain of dimensions where SUSY and DR apply, the RFIM fixed point and its 
vicinity contain more than just the fixed point of the pure $\phi^4$ theory in 2 dimensions less. Indeed, the latter corresponds to a sector 
of the full RFIM fixed-point effective action in which the 1-PI cumulants are restricted to equal field arguments (which can also be thought 
of as associated with a single replica). Cumulants of order 2 and higher when considered for nonequal field arguments should also reach 
a fixed point, and the approach to this fixed point then involve eigenperturbations that do not appear in the spectrum of the pure $\phi^4$ 
theory in 2 dimensions less. In particular, scale-free avalanches are present in the RFIM at criticality for $d\geq d_{\rm DR}$, a feature that  
does not alter the SUSY/DR sector but is not described by it. The fixed point which disappears in $d<d_{\rm DR}$ is the full RFIM fixed point 
and this is due to the sector associated with nonequal field arguments in the cumulants; the fixed point of the pure $\phi^4$ theory in 2 
dimensions less of course continues to exist.
\bibitem{footnote_marginal}
In some systems, such as frustrated magnets,\cite{frustrated_magnets} there may be several coupling constants that are marginal at the 
upper critical dimension and are then treated through nonlinear equations in the perturbative RG below this dimension. In the RFIM, and 
more generally RFO($N$)M, the only marginal coupling constant at the upper critical dimension $d=6$ is that associated with the $\phi^4$ 
term in the first cumulant.
\bibitem{footnote_NPRG}
Reviews on the general approach based on the exact FRG for the effective action in a wide variety of situations can be found in [\onlinecite{berges02}] and 
[\onlinecite{dupuis_review}].
\bibitem{footnote_perturb-fit}
Note that polynomial fits using 2-loop perturbative results and a postulated square-root behavior have given 
$d_{\rm DR}\approx 5.03-5.06$,\cite{tissier_pertFRG} which is again in the above predicted range of Eq.~(\ref{eq_error-bar}).
\end{thebibliography}
\end{document}